\documentclass[11pt,a4paper]{article}
\usepackage{jheppub}
\usepackage{dsdshorthand}
\usepackage{amsmath,amssymb,amsfonts,mathtools}
\usepackage{booktabs}
\usepackage{enumitem}
\usepackage{xcolor}
\usepackage{tikz}
\usepackage{graphicx}
\usepackage{cleveref}
\usepackage{slashed}

\numberwithin{equation}{section}

\newcommand{\bea}{\begin{equation}\begin{aligned}}
\newcommand{\eea}[1]{\label{#1}\end{aligned}\end{equation}}
\newcommand{\beq}{\begin{equation}}
\newcommand{\eeq}{\end{equation}}

\newcommand{\ads}{AdS_5\times S^5}
\newcommand{\lb}{\bar\lambda}
\newcommand{\Lb}{\bar\Lambda}
\newcommand{\vev}[1]{\langle #1\rangle}
\newcommand{\fc}{f}
\newcommand{\Qc}{Q_{\rm cov}}
\newcommand{\cV}{\mathcal V}
\newcommand{\cN}{\mathcal N}
\newcommand{\tr}{\operatorname{tr}}
\newcommand{\Str}{\operatorname{Str}}
\newcommand{\Fp}{\mathbb F_p}
\newcommand{\ch}[1]{[#1]}

\title{Scattering amplitudes in \texorpdfstring{$AdS_5\times S^5$}{AdS5xS5} from the pure spinor superstring}
\author{Luis F. Alday$^\lambda$, Thiago Fleury$^\theta$, Tobias Hansen$^x$, Lucas N. S. Martins$^\theta$ and Dennis Zavaleta$^\theta$}
\affiliation{${}^\lambda$Mathematical Institute, University of Oxford,
Woodstock Road, Oxford, OX2 6GG, UK\\
${}^\theta$Instituto Cienovus,
Av. Pres. Juscelino Kubitschek, 1327, 4 andar, Conj 41,
CEP: 04543-011, São Paulo, Brazil.\\
$^x$Department of Mathematical Sciences, Durham University,
Stockton Road, Durham, DH1 3LE, UK}
\emailAdd{luis.alday@maths.ox.ac.uk}
\emailAdd{tsi.fleury@gmail.com}
\emailAdd{tobias.p.hansen@durham.ac.uk}
\emailAdd{lucas\_nmartins@hotmail.com}
\emailAdd{dennis.zavaleta@unesp.br}

\abstract{
We formulate a BRST-invariant prescription for the integration of zero modes in the pure spinor superstring on $AdS_5\times S^5$. The measure includes contributions absent in flat space and admits a description in terms of picture changing operators or a non-minimal regulator. The prescription is valid for genus-zero amplitudes with arbitrary multiplicity and external states and removes a central obstacle to amplitude computations in this background. We use it to perform the first direct worldsheet computation of two and three-point amplitudes of half-BPS supergravity states. The properly normalised three-point amplitudes exactly reproduce the results of planar ${\cal N}=4$ super Yang–Mills theory.
}

\begin{document}
\maketitle

\section{Introduction}\label{sec:intro}

The S-matrix is a fundamental observable of quantum gravity. The RNS  worldsheet formalism of superstrings provides a recipe for an ultraviolet-finite and perturbatively unitary string S-matrix in asymptotically flat spacetime. On the other hand the formalism has a number of limitations on curved backgrounds. Some of these limitations can in principle be overcome in the string field theory framework, but this remains technically very challenging in practice. One outstanding difficulty is incorporating finite Ramond–Ramond (RR) fluxes into a conventional local RNS worldsheet description. The GS formalism does naturally incorporate the RR flux, but in curved backgrounds its covariant quantization is difficult. 

A natural setting to address this issue is type IIB string theory on $AdS_5 \times S^5$. This background supports a self-dual RR five-form flux and despite being maximally supersymmetric, a first principles worldsheet computation of even the simplest tree-level amplitudes has been thus far unavailable. A crucial advantage of choosing this background is that the string theory is holographically dual to ${\cal N}=4$ SYM. The supergravity sector of the theory gives rise to a tower of Kaluza-Klein supermultiplets on $AdS_5$, dual to half-BPS multiplets of ${\cal N}=4$ SYM. String amplitudes of these states map to correlation functions of the dual operators through AdS/CFT. Much is known about these correlation functions, providing precise predictions for the worldsheet computation. For example, two and three point amplitudes should be protected from receiving quantum (worldsheet) corrections and their form is explicitly known. For four point amplitudes at tree-level, the location of their poles should be dictated by the spectrum of planar ${\cal N}=4$ SYM, in principle determined by integrability techniques. In fact, this was shown to be  
highly constraining when supplemented with a certain single-valuedness property. Namely, only certain mathematical constants can appear in the low energy expansion of string amplitudes. This has been established in flat spacetime, and is conjectured to hold in curved spacetimes. This has led to a very specific prediction for the four-point amplitude on $AdS_5 \times S^5$ in a curvature expansion around flat space \cite{Alday:2022uxp,Alday:2022xwz,Alday:2023jdk,Alday:2023mvu}. 

A direct worldsheet approach to these observables would provide a practical framework for computing string interactions on $AdS_5 \times S^5$, beyond the supergravity approximation and semiclassical string calculations. Such a framework would also lead to significant progress in our understanding of string theory in the presence of RR fluxes for other cases, which are ubiquitous in holographic dualities. 

The most promising formalism for this endeavour is the pure spinor superstring \cite{Berkovits:2000fe}. The pure spinor formalism uses target-space superspace coordinates together with bosonic spinor ghosts, satisfying the pure spinor constraint. This allows the RR flux to be incorporated into a local worldsheet action while keeping the full $PSU(2,2|4)$ superisometry manifest. The model can be shown to be BRST invariant and conformally invariant, hence consistent, at the classical and quantum level to all orders in perturbation theory. Furthermore, physical vertex operators are described by BRST cohomology, which provides a covariant way to study string interactions in this background. Genus zero scattering amplitudes take the schematic form
\begin{equation}
{\cal A}^{(0)}_n \sim \int \prod_{i=4}^n d^2 z_i \langle V_1(0) V_2(1)V_3(\infty) \prod_{i=4}^n U_i(z_i,\bar z_i) \rangle
\end{equation}
where $V_i$ are unintegrated vertices and $U_i$ are integrated vertices, and we have used the $SL(2,C)$ symmetry to fix the locations of the three unintegrated vertices. In order to compute this amplitude we proceed as follows. 

The first step is the expression for the vertex operators. Following \cite{Berkovits:2019ulm,Berkovits:2019rwq,Martins:2019wwi,Mikhailov:2020nlg} the unintegrated vertex operators for the supergravity states have been constructed explicitly, first at a special point \cite{Fleury:2021ieo} and then in full generality \cite{Fleury:2026xpg}. They are labelled by a boundary point $x\in\partial AdS_5$, a nonzero complex null $SO(6)$ polarization vector $y$, and their protected dimension $k=2,3,\ldots$. These labels specify the external data of the amplitude. Each vertex is a local function of the worldsheet fields evaluated at its worldsheet insertion point. These fields comprise the bosonic embedding coordinates $(X,Y)\in AdS_5\times S^5$, the fermionic coordinates of the target superspace, both encoded in the supercoset representative $g(z,\bar z)$, and the left- and right-moving pure-spinor ghosts
\begin{equation}
V_k(z,\bar z ; x,y) = {\cal V}_k(g(z,\bar z),\lambda_L(z,\bar z),\lambda_R(z,\bar z) |x,y).
\end{equation}
The next two steps are encoded in the definition of the brackets $\langle \cdot \rangle$. The second step is to integrate out the non-zero modes by using their OPE. The two dimensional sigma-model is a consistent theory at the quantum level \cite{Berkovits:2004xu}, with $\alpha'$ the loop expansion parameter. The OPE can in principle be computed to any desired order in $\alpha'$, see \cite{Puletti:2006vb} for the computation at leading order. In this paper we will assume that two and three point worldsheet correlators of supergravity states receive no $\alpha'$ corrections. Similar assumptions were made in \cite{Berkovits:2008qc,Berkovits:2008ga}. This is expected from the corresponding non-renormalisation theorems in ${\cal N}=4$ SYM, see \cite{Lee:1998bxa,Eden:1999gh,Howe:1999hz,Heslop:2001gp,Baggio:2012rr}, but we leave a direct proof within the pure spinor formalism for future work. Of course, this will not be the case for four-point correlators, or correlators involving non-protected operators. 

The last step is the integration over the zero modes, parametrised by a common bulk point, 32 fermionic coordinates and the left and right moving pure spinors. This step is subtle: the pure spinor space contains non-compact directions, which introduce divergences, while fermionic integrals vanish unless they are saturated. Hence a suitable measure and a regulator need to be introduced. A prescription for the integration of zero modes on $AdS_5 \times S^5$ was proposed in \cite{Berkovits:2008ga}. In that proposal, picture lowering operators are not used, and the superspace integration and the pure spinor integration decouple. We show, however, that this proposal is inconsistent, in a subtle way, with the BRST symmetry of the pure spinor formalism. Furthermore it was observed in \cite{Fleury:2021ieo} that this prescription leads to vanishing
three-point amplitudes for massless states, in clear contradiction with the AdS/CFT duality. 

In this paper we develop a zero-mode prescription adapted from flat space and compatible with BRST symmetry. We formulate it using picture changing operators and give an equivalent description in terms of a regulator in the non-minimal formalism. Crucially, adapting the construction to $AdS_5 \times S^5$ requires additional terms in the picture-changing operators which are essential for obtaining the amplitudes computed below. A correct prescription to treat zero modes was the key missing ingredient for the computation of amplitudes in the pure spinor formalism on $AdS_5 \times S^5$. Let us stress that, while in this paper we will only consider two and three point amplitudes of  supergravity states, the prescription for the integration of zero modes is valid for genus zero amplitudes of arbitrary multiplicity and external states. 

With the expression for the supergravity vertex operators and the prescription for the integration of zero modes at hand, we then turn to the computation of scattering amplitudes. In this paper we consider two and three point amplitudes ${\cal A}^{k,k}_2$, ${\cal A}^{k_1,k_2,k_3}_3$. The first observation is that the space-time dependence is the correct one
\begin{eqnarray}
{\cal A}^{k,k}_2(x_1,y_1,x_2,y_2) &=& N_k \frac{(y_1 \cdot y_2)^k}{(x_1 \cdot x_2)^k} ,\\
{\cal A}^{k_1,k_2,k_3}_3(x_1,y_1,x_2,y_2,x_3,y_3) &=&C_{k_1 k_2 k_3} \frac{(y_1 \cdot y_2)^{\alpha_3}(y_1 \cdot y_3)^{\alpha_2}(y_2 \cdot y_3)^{\alpha_1}}{(x_1 \cdot x_2)^{\alpha_3}(x_1 \cdot x_3)^{\alpha_2}(x_2 \cdot x_3)^{\alpha_1}} 
\end{eqnarray}
with 
$2\alpha_1=(k_2+k_3-k_1),2\alpha_2=(k_1+k_3-k_2),2\alpha_3=(k_1+k_2-k_3)$. This is to be expected, since this dependence is fixed by $PSU(2,2|4)$. More remarkable is the fact that the normalised three point amplitudes give
\begin{equation}
\frac{C_{k_1 k_2 k_3}^2}{N_{k_1}N_{k_2}N_{k_3}} =  (4\pi g_s)^2 \left( \frac{\alpha'}{R^2}\right)^4 k_1 k_2 k_3.  \end{equation}
Upon using the AdS/CFT dictionary, this agrees precisely with the known planar ${\cal N}=4$ SYM result!

The paper is organised as follows. Section~\ref{sec:review} reviews the pure spinor formalism, first in flat space and then on $AdS_5 \times S^5$, and establishes our notation. In section~\ref{sec:PCOs}, we introduce our prescription for the zero-mode integration on $AdS_5 \times S^5$, after reviewing the corresponding prescriptions in flat space. In section~\ref{sec:2pt}, we apply our prescription to two-point amplitudes. This computation is subtle because fixing two insertion points leaves a non-compact subgroup of the worldsheet conformal Killing group unfixed. We address this residual symmetry by inserting a mostly BRST-exact operator. Section~\ref{sec:3pt} presents the computation of three-point amplitudes. In a way, this is a conceptually simpler computation. We then compare the normalised three-point amplitude with the corresponding result in planar ${\cal N}=4$ SYM, finding exact agreement. The conclusions outline open problems and future directions, while technical details are collected in the appendices. In particular, in appendix~\ref{sec:previous_proposal}, we discuss a previous proposal for the zero-mode integration on $AdS_5 \times S^5$ and show that it is not consistent with BRST symmetry.
 
\section{Review of the pure spinor formalism}\label{sec:review}

In this section we briefly review the pure spinor formalism. We start, in subsection \ref{sec: The pure spinor superstring}, developing the formalism in flat background and focusing on the left-movers, where most equations are readily applicable for scattering amplitudes involving open string states. Then, we proceed, in subsection \ref{sec: The type II backgrounds}, with the formalism in type II backgrounds, including both flat and curved backgrounds. The goal is to prepare for the most important subsection \ref{sec: The type IIB ads background} where the focus is $\ads$. There we  introduce our notation and review established results that will be useful for the present paper.

\subsection{The pure spinor superstring}
\label{sec: The pure spinor superstring}

In flat background, focusing on the left-movers only, the worldsheet action in pure spinor minimal formalism is given by
\begin{equation}
S = \frac{1}{2\pi\alpha'}\int d^2z\left(\frac{1}{2}\partial x^{m}\overline\partial x_{m} + p_{\alpha} \overline\partial\theta^{\alpha} + w_{\alpha}\overline\partial\lambda^{\alpha}\right)\,,    
\label{eq: flat_worldsheet_action}
\end{equation} 
where $(x^{m},\theta^{\alpha}, p_{\alpha})$ correspond to spacetime superspace coordinates and $(\lambda^{\alpha},w_{\alpha})$ are the pure spinor variables. Here $\alpha=1,\dots,16$ are the ten-dimensional Majorana-Weyl spinorial indices, while $m=0,\dots,9$ are the vector indices. The variables $\lambda^{\alpha}$ are required to satisfy the pure spinor constraints
\begin{equation}
\lambda^{\alpha}\gamma_{\alpha\beta}^{m}\lambda^{\beta} = 0\,,
\label{eq: pure_spinor_constraint_so19}
\end{equation}
where $\gamma_{\alpha\beta}^{m}=\gamma_{\beta\alpha}^{m}$ are the symmetric chiral off-diagonal blocks of the ten-dimensional gamma matrices. Together with the symmetric anti-chiral off-diagonal blocks $\gamma^{m\alpha\beta}=\gamma^{m\beta\alpha}$ they satisfy
\begin{equation}
\gamma_{\alpha\beta}^{(m}\gamma^{n)\beta\gamma} = 2 \eta^{mn}\delta^{\gamma}_{\alpha}\,,\qquad \gamma^{m}_{(\alpha\beta}\gamma_{m\gamma)\delta}=0\,. 
\label{eq: 10d_gamma_matrices}
\end{equation}
Combining the gamma matrices in (\ref{eq: 10d_gamma_matrices}) into creation and annihilation operators, we obtain anti-commutation relations
\begin{equation}
\{c_{I},c_{J}\} = 0\,,\qquad \{c^{I},c^{J}\}=0\,,\qquad \{c_{I},c^{J}\}=\delta_I^J\,,  
\label{eq: u5 notation}
\end{equation}
\begin{equation}
c^{I} = \frac{\gamma^{2I}+i^{[I]}\gamma^{2I+1}}{2}\,,\qquad c_{I} = \frac{\gamma^{2I}-i^{[I]}\gamma^{2I+1}}{2}\,,\qquad I=0,\dots, 4\,,  \nonumber
\end{equation}
where the $i^ {[I]}$ switch between holomorphic and light-cone combinations, through
\begin{equation}
[I]=\begin{cases}
   0\,,\qquad \text{if }I=0 \\
   1\,,\qquad \text{otherwise.}
\end{cases}    
\end{equation}
The ground state $|0\rangle$ annihilated by all annihilators of  (\ref{eq: u5 notation}), $c_{I}|0\rangle=0$, is a distinguished pure spinor. Any spinor can be obtained through linear combinations of products of creation operators acting on this pure spinor. The chirality condition imposes that these products must involve an even number of creation operators. Therefore we can write
\begin{equation}
\lambda = \lambda^{+}|0\rangle + \frac{1}{2}\lambda^{+}_{IJ}c^{I}c^{J}|0\rangle + \frac{1}{4!}\varepsilon_{IJKLM}^{++}\lambda^{I}_{+}c^{J}c^{K}c^{L}c^{M}|0\rangle\,.    
\label{eq: decomposition_chiral_spinor_ground_state}
\end{equation}
where the $16$ components $(\lambda^{+}, \lambda^{+}_{IJ}, \lambda_{+}^{I})$ correspond to the decomposition of a chiral spinor into linear combinations of products of creation operators acting on the reference pure spinor $|0\rangle$. In this basis the pure spinor constraint becomes 
\begin{equation}
 \lambda^{+}   \lambda_{+}^{I} +\frac{1}{8}\varepsilon^{IJKLM}_{++}\lambda^{+}_{JK} \lambda^{+}_{LM} = 0\,,\qquad \lambda_{+}^{I} \lambda^{+}_{IJ} = 0\,.
 \label{eq: pure_spinor_constraint_in_u5}
\end{equation}
In the patch $\lambda^{+}\neq 0$, one can solve the pure spinor constraint by expressing the components $\lambda_{+}^{I}$ in terms of the $11$ independent components $(\lambda^{+}, \lambda^{+}_{IJ})$,
\begin{equation}
\lambda_{+}^{I} = -\frac{1}{8\lambda^{+}} \varepsilon^{IJKLM}_{++}\lambda^{+}_{JK}  \lambda^{+}_{LM}\,. 
\label{eq: pure_spinor_sol_u5}
\end{equation}
One can cover the space of pure spinors by $16$ patches, each corresponding to a particular choice of a ground state $|0\rangle$, by requiring that the corresponding component $\lambda^{+}$ in the (\ref{eq: decomposition_chiral_spinor_ground_state}) decomposition 
is non-zero. For example, one can consider the ground state $c_{1,2,3}|0\rangle = 0$ and $c^{0,4}|0\rangle =0$, decompose chiral spinors just as in (\ref{eq: decomposition_chiral_spinor_ground_state}), write the pure spinor constraint as in (\ref{eq: pure_spinor_constraint_in_u5}) and then impose $\lambda^{+}\neq 0$ so that it can be solved as in (\ref{eq: pure_spinor_sol_u5}). By switching the roles of the creation and annihilation operators in deciding the ground state, this results in $2^4$ patches. These $16$ patches provide a cover for the pure spinor space.

The pure spinor constraints (\ref{eq: pure_spinor_constraint_so19}) generate gauge symmetries in the worldsheet action (\ref{eq: flat_worldsheet_action}). These gauge transformations act only on the canonical conjugates $w_{\alpha}$ and are given by
\begin{equation}\label{ps-gauge-transform}
\delta w_{\alpha} =\Lambda_{m}(\gamma^{m}\lambda)_{\alpha}\,,
\end{equation}
where $\Lambda_{m}$ is a local parameter on the worldsheet. Notice that not all the components of $\Lambda_{m}$ couple to the gauge transformations of $w_{\alpha}$, as for a pure spinor $\lambda^{\alpha}$, there are five linearly independent vectors, $\Lambda^{(i,\lambda)}_{m}$, with $i=1,\dots,5$, that once contracted with gamma matrices annihilate the pure spinor $\Lambda^{(i, {\lambda})}_{m}\gamma^{m}_{\alpha\beta}\lambda^{\beta} = 0$. Therefore, gauge-fixing this symmetry would break Lorentz covariance. For this reason, in the pure spinor formalism we usually do not gauge fix these symmetries, but instead restrict the operators to depend on $w_{\alpha}$ only through the gauge invariant quantities
\begin{equation}
\alpha' N= \lambda^{\alpha}w_{\alpha}   \,\qquad \alpha'N_{mn} =-\frac{1}{2}\lambda^{\alpha}(\gamma^{mn})_{\alpha}\,^{\beta}w_{\beta}\,.
\end{equation}
Later we will assign ghost number $+1$ to $\lambda^\alpha$, and ghost number $-1$ to $w_\alpha$. Notice then that it is not possible to construct objects with negative ghost number via polynomials in $(\lambda^{\alpha}, N, N_{mn})$. 

\subsubsection{Equations of motion and OPEs}
\label{sec: Equations of motion and OPEs}

The equations of motion obtained from the worldsheet action in (\ref{eq: flat_worldsheet_action}) are harmonic and holomorphic conditions
\begin{equation}
\overline\partial\partial x^{m} = 0\,,\qquad \overline\partial\theta^{\alpha}= 0\,,\qquad \overline\partial p_{\alpha} = 0\,,\nonumber  
\end{equation}
\begin{equation}
\overline\partial\lambda^{\alpha} = 0\,,\qquad \overline \partial N =0\,,\qquad \overline\partial N_{mn} = 0\,.     
\end{equation}
Since the above equations are linear, one can readily compute the singular OPEs:
\begin{equation}
x^{m}(z',\overline z') x^{n}(z,\overline z)\approx -\alpha'\eta^{mn}\log |z'-z|^2+\mathcal{O}(1)\,,\qquad p_{\alpha}(z')\theta^{\beta}(z)\approx \frac{\alpha'\delta^{\beta}_{\alpha}}{z'-z} +\mathcal{O}(1)\,,\nonumber 
\end{equation}
\begin{equation}
N(z')\lambda^{\alpha}(z)\approx \frac{\lambda^{\alpha}(z)}{z'-z}+\mathcal{O}(1)\,,\qquad  N_{mn}(z')\lambda^{\alpha}(z)\approx \frac{\frac{1}{2}(\gamma_{mn}\lambda)^{\alpha}(z)}{z'-z}+\mathcal{O}(1) \,,
\end{equation}
and
\begin{equation}
N_{mn}(z')N_{pq}(z)\approx \frac{+3\eta_{p[m}\eta_{n]q}}{(z'-z)^2}+\frac{\eta_{p[m}N_{n]q}-\eta_{q[m}N_{n]p}}{(z'-z)}+\mathcal{O}(1)\,,  \nonumber  
\end{equation}
\begin{equation}
N(z')N(z)\approx \frac{-4}{(z'-z)^2}+\mathcal{O}(1)\,.
\end{equation}
Here $\approx$ denotes equality up to terms that vanish upon using the equations of motion. 
It follows from the equations of motion and OPEs above that the gauge invariant quantities, $N$ and $N_{mn}$, are holomorphic currents whose corresponding charges generate, respectively, Weyl and Lorentz transformations on the pure spinor variables $\lambda^{\alpha}$.

\subsubsection{Stress-energy tensor}
\label{sec: Stress-energy tensor}

The worldsheet action in (\ref{eq: flat_worldsheet_action}) couples with the worldsheet conformal structure through the holomorphic stress-energy tensor
\begin{equation}
\alpha' T(z)= -\frac{1}{2}:\partial x^{m}\partial x_{m}: - :p_{\alpha}\partial \theta^{\alpha}: -\frac{\alpha'}{20}:N_{mn}N^{mn}: - \frac{\alpha'}{8}:NN: + \alpha'\partial N\,,
\end{equation}
where we used the Sugawara construction to stay within polynomials of $(\lambda^{\alpha}, N, N_{mn})$. By noticing that
\begin{equation}
:w_{\alpha}\partial\lambda^{\alpha}: \ \approx -\frac{\alpha'}{20}:N_{mn}N^{mn}: - \frac{\alpha'}{8}:NN: + \ \alpha'\partial N\,,
\end{equation}
it becomes evident that the pure spinor variables $\lambda^\alpha$ are worldsheet variables of zero conformal weight. Therefore, there are $11$ zero modes that must be integrated in worldsheet correlators for any Riemann surface. The conformal anomaly is absent as it can be shown, using the above OPEs, that
\begin{equation}
T(z')T(z)\approx \frac{0}{(z'-z)^4}+\frac{2T(z)}{(z'-z)^2} + \frac{\partial T(z)}{(z'-z)}+\mathcal{O}(1)\,,
\end{equation}
with vanishing central charge. Therefore, the worldsheet theory is conformally invariant at the quantum level. On the other hand, when we compute the OPE between the holomorphic stress-energy tensor and the holomorphic current $N$ corresponding to Weyl scaling of the pure spinor variables, there is a triple pole anomaly equal to $+8$:
\begin{equation}
T(z')N(z) \approx \frac{8}{(z'-z)^3} + \frac{N}{(z'-z)^2}  + \frac{\partial N}{(z'-z)} +\mathcal{O}(1)\,.   
\label{eq: ghost number anomaly}
\end{equation}
Therefore, the measure in which we compute worldsheet correlators must integrate over $11$ zero modes, coming from the pure spinors, but it should scale with charge $+8$ under Weyl scaling.  We will have more to say about this in section \ref{sec:PCOs}.

\subsubsection{Spacetime supersymmetry}

In the pure spinor formalism, the spacetime supersymmetry transformations are generated by the charge operators
\begin{equation}
q_{\alpha} =  \oint \frac{dz}{2\pi i\alpha' } \left(  p_{\alpha} +\frac{1}{2}\partial x^{m}(\gamma_{m}\theta)_{\alpha} -\frac{1}{24}(\partial\theta\gamma^{m}\theta)(\gamma_{m}\theta)_{\alpha}\right)\,.    
\label{eq: spacetime_susy_flat_brackground}
\end{equation}
Using the OPEs presented in subsection \ref{sec: Equations of motion and OPEs}, one can show that these supercharges act on local operators corresponding to normal-ordered superfields of $(x,\theta)$ as $Q_{\alpha} =  \frac{\partial}{\partial\theta^{\alpha}} - \frac{1}{2}(\gamma^{m}\theta)_{\alpha}\frac{\partial}{\partial x^{m}}$. 
They satisfy the super-Poincaré algebra $\{q_{\alpha}, q_{\beta}\}\approx \gamma^{m}_{\alpha\beta}\oint \frac{dz}{2\pi i\alpha'}\partial x_{m}$, just as $\{Q_{\alpha},Q_{\beta}\}=\gamma^{m}_{\alpha\beta}\frac{\partial}{\partial x^{m}}$. They do not act on the pure spinor sector $(\lambda^{\alpha}, N, N_{mn})$.

Although the worldsheet of (\ref{eq: flat_worldsheet_action}) is supersymmetric, this is not manifest. To make the worldsheet action manifestly supersymmetric, we must perform a field redefinition. Define
\begin{equation}
d_{\alpha} = p_{\alpha} - \frac{1}{2}\partial x^{m}(\gamma_{m}\theta)_{\alpha} + \frac{1}{8}(\partial\theta\gamma^{m}\theta)(\gamma_{m}\theta)_{\alpha}  \,.
\label{eq: shift_d_alpha}
\end{equation}
The action can then be written as
\begin{equation}
S = \frac{1}{2\pi\alpha'}\int d^2z\left(\frac{1}{2}\pi^{m}\overline\pi_{m} + B(\partial,\overline\partial)+ d_{\alpha} \overline\pi^{\alpha} + w_{\alpha}\overline\partial\lambda^{\alpha}\right)\,,    
\label{eq:flat_worldsheet_action_susy}
\end{equation}
where
\begin{equation}
\pi^{m} = \partial x^{m} - \frac{1}{2}  (\partial\theta\gamma^m\theta  )\,,\qquad \overline \pi^{m} = \overline \partial x^{m} - \frac{1}{2}  (\overline \partial\theta\gamma^m\theta  )\,,\qquad \overline\pi^{\alpha} = \overline\partial\theta^{\alpha}\,,
\end{equation}
are, together with $d_{\alpha}$, supersymmetric. $B(\partial,\overline\partial)$ is the worldsheet pull-back of the flat-background Kalb-Ramond field potential
\begin{equation}
B(\partial,\overline\partial)= \frac{1}{4}\left(\partial x^{m}(\theta\gamma_{m}\overline\partial\theta) - \overline\partial x^{m}(\theta\gamma_{m}\partial\theta)\right)
  = \frac{1}{4}\left(\pi^{m}(\theta\gamma_{m}\overline\partial\theta) - \overline\pi^{m}(\theta\gamma_{m}\partial\theta)\right)\,,
\end{equation}
and it is supersymmetric up to total derivatives. To make it fully manifestly supersymmetric, without the need for total derivatives, we use Stokes theorem, by writing $\int d^2zB(\partial,\overline\partial)=\int d^2zdy H(\partial_{y},\partial,\overline\partial)$, where $H(\partial_{y},\partial,\overline\partial)$ is the pull-back of the Kalb-Ramond field strength on a three-dimensional world-volume whose boundary correspond to the worldsheet. Notice that the first two terms in (\ref{eq:flat_worldsheet_action_susy}), $\frac{1}{2}\pi^{m}\overline\pi_{m} + B$, are nothing but the usual Green-Schwarz Lagrangian. The pure spinor formalism adds to the GS worldsheet action the terms $d_{\alpha}\overline\pi^{\alpha}+w_{\alpha}\overline\partial\lambda^{\alpha}$. The term $d_{\alpha}\overline\pi^{\alpha}$, in particular, breaks the $\kappa$-symmetry of the GS formalism.

\subsubsection{BRST symmetry}
\label{sec:brst_symmetry}

The pure spinor formalism, besides adding the term $d_{\alpha}\overline\pi^{\alpha}+w_{\alpha}\overline\partial\lambda^{\alpha}$ to the GS worldsheet action, which breaks $\kappa$-symmetry, also replaces $\kappa$-symmetry with a BRST symmetry. Physical states are required to be in a BRST cohomology whose BRST invariance induces the linearized spacetime equations of motion and decoupling from BRST exact states induces the linearized spacetime gauge-invariances. The BRST charge operator is given by the holomorphic and nilpotent operator
\begin{equation}
Q = \oint \frac{dz}{2\pi i} \lambda^{\alpha}d_{\alpha}\,,
\label{eq: BRST_charge_flat_background}
\end{equation}
where its holomorphicity follows from the worldsheet equation of motion, while its nilpotency follows from the singular OPE
$d_{\alpha}(z')d_{\beta}(z) \approx -\frac{1}{2}\frac{\gamma_{\alpha\beta}^{m}\partial x_{m}}{z'-z} + \mathcal{O}(1)$ and the pure spinor constraints $\lambda^{\alpha}\gamma^{m}_{\alpha\beta}\lambda^{\beta}=0$. Since $d_{\alpha}$ is spacetime supersymmetric and $\lambda^{\alpha}$ does not transform under spacetime supersymmetry it follows that the BRST charge (\ref{eq: BRST_charge_flat_background}) is also spacetime supersymmetric. Therefore, the spacetime supersymmetry passes to the BRST cohomology.

Because the superspace sector variables $(x^{m},\theta^{\alpha},p_{\alpha})$ are required to carry zero ghost number, we must assign $+1$ ghost number to the pure spinor variables to make the BRST charge operator $Q$ in (\ref{eq: BRST_charge_flat_background}) a ghost number $+1$ operator. As we saw in subsection \ref{sec: The pure spinor superstring}, because of the gauge-invariances generated by the pure spinor constraint, we restrict the pure spinor sector to involve only the gauge-invariant operators $(\lambda^{\alpha}, N, N_{mn})$. However, if we restrict the pure spinor dependence of our operators to be polynomial in these gauge-invariant operators, then it is not possible to construct operators with strictly negative ghost number. Naively, one would try to construct operators with negative powers of $\lambda^{\alpha}$, but this is problematic since it would trivialize the cohomology, due to $Q(\frac{\theta^{+}}{\lambda^{+}})=1$ implying that any $V$ obeying $QV= 0$ can be written as $V = Q\left(\frac{\theta^{+}}{\lambda^{+}}V\right) = 0$.

There are two ways to circumvent this obstruction. The first one is to consider the picture-lowering operators $\theta^{+} \delta (Q(\theta^{+}))= \theta^{+} \delta (\lambda^{+})$ where $\theta^{+}$ is one of the components of $\theta^{\alpha}$. These operators are BRST invariant and carry $-1$ ghost number but necessarily spoil Lorentz covariance. The other approach is to introduce the non-minimal variables which essentially covariantize the picture-lowering operators. We will see this in section \ref{sec:PCOs}.

\subsubsection{Unintegrated vertex operators}

Using the BRST charge operator (\ref{eq: BRST_charge_flat_background}), the unintegrated vertex operators $V$ representing physical states are required to obey the condition $Q V \approx 0$ and the equivalence relation $V \sim V + Q W$. For concreteness, let us focus on the unintegrated vertex operators for massless open string states. These states have ghost number $+1$ so they take the following form
\begin{equation}
V = \lambda^{\alpha}A_{\alpha}(x,\theta)\,,
\label{eq: massless_open_string_state_unintegrated}
\end{equation}
where the above expression is the most general Lorentz invariant expression of ghost number $+1$ with zero conformal weight. Requiring these states to be in BRST cohomology implies the linearized equations of motion and gauge invariance of the super-Maxwell supermultiplet
\begin{equation}
\lambda^{\alpha}\lambda^{\beta} D_{\alpha}A_{\beta}(x,\theta) = 0\,,    
\qquad 
A_{\alpha}(x,\theta) \sim     A_{\alpha}(x,\theta)  + D_{\alpha}\Omega(x,\theta)\,.
\label{eq: A_alpha_beta_on-shell}
\end{equation}
Here, $D_{\alpha} = \frac{\partial}{\partial\theta^{\alpha}} + \frac{1}{2}(\gamma^{m}\theta)_{\alpha}\frac{\partial}{\partial x^{m}} $ is the supercovariant derivative, satisfying $\{D_{\alpha},Q_{\alpha}\} = 0$. Constructing the superfields
\begin{equation}
A_{m}= \frac{1}{8}(\gamma_{m}D)^{\alpha}A_{\alpha}\,,\qquad W^{\alpha} = (DD)^{\alpha\beta}A_{\beta}\,,\qquad F_{mn}=\frac{1}{8}\partial_{[m}(\gamma_{n]}D)^{\alpha}A_{\alpha}\,,
\label{eq: super_maxwell}
\end{equation}
where $DD^{\alpha\beta} = \frac{1}{80}(\gamma^{m}D)^{\alpha}(\gamma_{n}D)^{\beta} -\frac{1}{10}\gamma_{m}^{\alpha\beta}\partial^{m}$, we obtain that the first components of the $\theta$-expansion of $A_{m}$, $W^{\alpha}$ and $F_{mn}$ correspond, respectively, to the Maxwell potential, the photino and the Maxwell field-strength.

\subsubsection{Integrated vertex operators}

One complication of the formalism is the difficulty in constructing objects with negative ghost number. In particular, it is not possible to construct the $b$-ghost, satisfying $\{Q, b\} = T$, where $T$ is the stress-energy tensor, using only polynomials of $(\lambda^{\alpha}, N, N_{mn})$, worldsheet derivatives $(\partial, \overline\partial)$, and spacetime variables $(x^{m}, \theta^{\alpha})$. Therefore, to construct integrated vertex operators from unintegrated ones, we use the descent relations. For an open string state, the integrated vertex operator $\int d z U$ corresponding to the unintegrated $V$ is obtained by solving
\begin{equation}
\partial V \approx \{Q, W\}\qquad [Q ,U ]\approx \overline \partial W\,,
\label{eq: descent_relations_open}
\end{equation}
for $(W,U)$. For example, for the massless open string state, represented by the unintegrated vertex operator in (\ref{eq: massless_open_string_state_unintegrated}), the corresponding
integrated vertex operator is
\begin{equation}
U = \int dz \left(\partial\theta^{\alpha}A_{\alpha} + \pi^{m}A_{m}+ W^{\alpha}d_{\alpha} + \frac{1}{2}F_{mn}N^{mn}\right)\,,    
\end{equation}
where $(A_{\alpha}, A_{m}, W^{\alpha}, F_{mn})$ are spacetime superfields appearing in (\ref{eq: super_maxwell}).

\subsubsection{Tree-level scattering amplitudes}
\label{treelevelflat}

In the last two subsections we have discussed what characterizes unintegrated vertex operators, and how to obtain integrated ones from these. Concretely, we restrict our attention to open string states, since in this subsection we do not wish to specify the right-movers content. Closed strings will be discussed in subsection \ref{sec: The type II backgrounds}, where we focus on type II superstrings.

Having the unintegrated and integrated vertex operators for open strings, $V$ and $\int dz U$, our next step is to compute the scattering amplitudes in the pure spinor formalism through the worldsheet correlator. Consider the three-point amplitude between three massless open strings states at genus zero. This corresponds to a disk correlator with three unintegrated vertices inserted along the boundary
\begin{equation}
\mathcal{A}^{(123)} = \left\langle V^{(1)}(0)V^{(2)}(1)V^{(3)}(\infty)\right\rangle  \,,\qquad V^{(i)} = \lambda^{\alpha}A_{\alpha}^{(i)}(x,\theta)\,,
\end{equation}
where Chan-Paton factors will be suppressed. Solving the equations (\ref{eq: A_alpha_beta_on-shell}) determining the superfield $A_{\alpha}(x,\theta)$, its $\theta$-expansion results in
\begin{equation}
A_\alpha^{(i)}=\Big[\tfrac12(\gamma^m\theta)_\alpha e^{(i)}_m+\tfrac13(\gamma^m\theta)_\alpha(\theta\gamma_m\xi^{(i)})-\tfrac1{32}(\gamma_p\theta)_\alpha(\theta\gamma^{mnp}\theta)\partial_{[m}e_{n]}^{(i)}+\dots\Big]e^{ik^{(i)}\cdot X}\,,
\label{TheAvertex}
\end{equation}
where $e_{m}^{(i)}$ correspond to the gluon polarizations and $\xi^{(i)\alpha}$ correspond to the gluino polarizations. For example, suppose we want to compute the scattering amplitudes between three gluons. In this case, we would set $\xi^{(i)\alpha}=0$ for $i=1,2,3$. Another example would be the computation involving two gluinos and one gluon, which would correspond to $e^{(1)}_{m} = e^{(2)}_{m} = 0$ and $\xi^{(3)\alpha}=0$.

Through their OPEs, we can eliminate any non-zero mode dependence, so what is left in the correlators are the zero modes. We will discuss the zero-mode measure of the worldsheet correlators more carefully in section \ref{sec:PCOs}. For now, let us use a short cut which consists of taking the disk zero-mode measure to satisfy\footnote{We use the convention of e.g.\ \cite{Mafra:2022wml} because it implies canonically normalised amplitudes in flat space.}
\begin{equation}
\langle (\lambda\gamma^{m}\theta)(\lambda\gamma^{n}\theta)(\lambda\gamma^{p}\theta)(\theta\gamma_{mnp}\theta)\rangle = 2880\,,
\label{eq:shortcut_measure_flat}
\end{equation}
while terms with different number of $\lambda$s and $\theta$s, other than three and five, vanish. This is consistent with the anomaly we saw in \ref{sec: Stress-energy tensor}. There we concluded the measure must have charge $+8$ under scaling of the pure spinor variables, but is integrated over $11$ zero modes of those. Exactly three more pure spinors need to be supplied by the vertex operators. Furthermore, when integrating over the 11 spinor zero modes, this also cancels 11 $\theta$s, due to Lorentz invariance. Five more  $\theta$s need to be provided by the vertex operators. The zero-mode measure (\ref{eq:shortcut_measure_flat}) completely determines the worldsheet correlators to be
\begin{equation}
 \mathcal{A}^{(123)}=   (\xi^{(1)}\gamma^m\xi^{(2)})\,e^{(3)}_m\,,
\end{equation}
for the amplitude of two gluinos and one gluon, and
\begin{equation}
\mathcal{A}^{(123)}=i\Big[(e^{(1)}\!\cdot\! e^{(2)})(e^{(3)}\!\cdot\! k^{(1)})+(e^{(2)}\!\cdot\! e^{(3)})(e^{(1)}\!\cdot\! k^{(2)})+(e^{(3)}\!\cdot\! e^{(1)})(e^{(2)}\!\cdot\! k^{(3)})\Big]\;
\label{eq:3g}
\end{equation}
for the three-gluon amplitude, where the following identities have been used
\begin{equation}
\vev{(\lambda\gamma^m\theta)(\lambda\gamma^n\theta)(\lambda\gamma^p\theta)(\theta\gamma_{qrs}\theta)}=24\,\delta^{[m}_q\delta^n_r\delta^{p]}_s.
\end{equation}
More generally, one can introduce a tensor $T^{\alpha\beta\gamma;\delta_1\delta_2\delta_3\delta_4\delta_5}$ satisfying
\begin{equation}
\vev{\lambda^\alpha\lambda^\beta\lambda^\gamma\theta^{\delta_1}\theta^{\delta_2}\theta^{\delta_3}\theta^{\delta_4}\theta^{\delta_5}} = T^{\alpha\beta\gamma;\delta_1\delta_2\delta_3\delta_4\delta_5}    \,.
\label{TheTupindices}
\end{equation}

\subsection{The type II backgrounds}
\label{sec: The type II backgrounds}

In the previous subsection we saw how the formalism works in flat background by focusing on the left-movers only. This describes open superstrings or, with a few adjustments, closed strings in heterotic backgrounds. From now on, we will focus on closed superstrings on type II backgrounds.

\subsubsection{Flat backgrounds}

The worldsheet action for the pure spinor superstring in type II flat background is obtained simply by combining left and right copies of pure spinors
\begin{equation}
S = \int d^2z \left(\partial x^{m}\overline\partial x_{m} + p_{L\alpha} \overline\partial\theta_L^{\alpha} + w_{L\alpha}\overline\partial\lambda^{\alpha}_L +  p_{R\widehat\alpha} \partial \theta_R^{\widehat\alpha} +  w_{R\widehat\alpha}\partial \lambda_R^{\widehat\alpha}\right),    
\end{equation}
where the spinorial indices $(\alpha,\widehat\alpha)$ have same chirality in type IIB and opposite chirality in type IIA. The equations of motion are
\begin{equation}
\overline\partial\theta^\alpha_{L}=0,\quad \overline\partial  p_{L\alpha}=0,\quad \partial\overline\partial x_m=0,\quad \overline\partial\lambda_L^\alpha=0,\quad \overline\partial w_{L\alpha}=0\,,\nonumber
\end{equation}
\begin{equation}
\partial\theta^{\widehat\alpha}_{R}=0,\quad \partial  p_{R\widehat\alpha}=0\,\quad \partial\lambda_R^{\widehat\alpha}=0,\quad \partial w_{R\widehat\alpha}=0\,.
\end{equation}
The spacetime supersymmetries are given by
\begin{equation}
q_{L\alpha}=\oint\frac{dz}{2\pi i\alpha'}\left[p_{L\alpha} +\frac{1}{2}\partial x^{m}(\gamma_{m}\theta_{L})_{\alpha} -\frac{1}{24}(\partial\theta_{L}\gamma^{m}\theta_{L})(\gamma_{m}\theta_{L})_{\alpha}\right]   \,,\nonumber
\end{equation}
\begin{equation}
q_{R\widehat\alpha}=\oint\frac{d\overline z}{2\pi i\alpha'} \left[p_{R\widehat\alpha} +\frac{1}{2}\overline\partial x^{m}(\gamma_{m}\theta_{R})_{\widehat\alpha} -\frac{1}{24}(\overline\partial\theta_{R}\gamma^{m}\theta_{R})(\gamma_{m}\theta_{R})_{\widehat\alpha}\right]\,.   
\end{equation}
As before, OPEs can be easily obtained from this worldsheet action. In addition, to make the action manifestly supersymmetric we perform, just as in (\ref{eq: shift_d_alpha}), the shifts
\begin{equation}
d_{L\alpha}=p_{L\alpha}-\tfrac12\big(\partial x^m+\tfrac14\theta_{L}\gamma^m\partial\theta_{L}+\tfrac14\theta_{R}\gamma^m\partial\theta_{R}\big)(\gamma_m\theta_L)_{\alpha}\,,\nonumber
\end{equation}
\begin{equation}
d_{R\widehat\alpha}=p_{R\widehat\alpha}-\tfrac12\big(\overline\partial x^m+\tfrac14\theta_{L}\gamma^m\overline\partial\theta_{L}+\tfrac14\theta_{R}\gamma^m\overline\partial\theta_{R}\big)(\gamma_m\theta_R)_{\widehat\alpha}\,,
\end{equation}
and define the supersymmetric quantities
\begin{equation}
\pi^{\alpha} = \partial \theta_{L}^{\alpha}\,,\qquad \overline\pi^{\alpha}=\overline\partial\theta^{\alpha}_{L}\,,\qquad \pi^{\widehat\alpha} = \partial \theta_{R}^{\widehat\alpha}\,,\qquad \overline\pi^{\widehat\alpha}=\overline\partial\theta^{\widehat\alpha}_{R}    \nonumber 
\end{equation}
\begin{equation}
\pi^{m}=\partial x^{m}+\frac{1}{2}(\partial\theta_{L}\gamma^{m}\theta_{L})+\frac{1}{2}(\partial\theta_{R}\gamma^{m}\theta_{R})\,,\nonumber
\end{equation}
\begin{equation}
\overline\pi^{m} = \overline\partial x^{m}+\frac{1}{2}(\overline\partial\theta_{L}\gamma^{m}\theta_{L})+\frac{1}{2}(\overline\partial\theta_{R}\gamma^{m}\theta_{R})  \,,    
\end{equation}
in terms of which the worldsheet action takes the form
\begin{equation}
S = \int d^2z \left(\pi^{m}\overline\pi_{m} + B(\partial,\overline\partial)+d_{L\alpha} \overline\partial\theta_L^{\alpha} + w_{L\alpha}\overline\partial\lambda^{\alpha}_L +  d_{R\widehat\alpha} \partial \theta_R^{\widehat\alpha} +  w_{R\widehat\alpha}\partial \lambda_R^{\widehat\alpha}\right)\,,
\label{eq: worldsheet_action_type_ii_susy}
\end{equation}
where $B(\partial,\overline\partial)$ is the worldsheet pull-back 
\begin{equation}
B(\partial,\overline\partial)=\tfrac14\left[\partial x^m\,(\theta_L\gamma_m\overline\partial\theta_L)-\overline\partial x^m\,(\theta_L\gamma_m\partial\theta_L)\right]-\tfrac14\left[\partial x^m\,(\theta_R\gamma_m\overline\partial\theta_R)-\overline\partial x^m\,(\theta_R\gamma_m\partial\theta_R)\right]\nonumber    
\end{equation}
\begin{equation}
-\tfrac{1}{8}\Big[(\theta_{L}\gamma^m\partial\theta_{L})(\theta_{R}\gamma_m\overline\partial\theta_{R})-(\theta_{L}\gamma^m\overline\partial\theta_{L})(\theta_{R}\gamma_m\partial\theta_{R})\Big]\,.
\label{eq:Bsigma}
\end{equation}
of the Kalb-Ramond potential 
\begin{equation}
B = \frac{1}{4}dx^{m}\wedge\left[(\theta_{L}\gamma^{m}d\theta_{L}) - (\theta_{R}\gamma^{m}d\theta_{R})\right]  - \frac{1}{8}  (\theta_{L}\gamma^{m}d\theta_{L})\wedge (\theta_{R}\gamma^{m}d\theta_{R})\,.
\end{equation}
The BRST symmetry is generated by the sum of the holomorphic and antiholomorphic pieces, corresponding to the left and right movers
\begin{equation}
Q =\oint \frac{dz}{2\pi i} \lambda_{L}^{\alpha}d_{L\alpha} - \oint\frac{d\overline z}{2\pi i} \lambda_{R}^{\widehat\alpha}d_{R\widehat\alpha}\,.
\end{equation}
Nilpotency follows from the OPE, the pure spinor constraints and the equations of motion.

\subsubsection{Unintegrated vertex operators}

The vertex operators corresponding to closed massless states in type II flat background are given by
\begin{equation}
V =\lambda^{\alpha}_{L}\lambda_{R}^{\widehat\alpha}A_{\alpha\widehat\alpha}(x,\theta_L,\theta_R)\,,    
\end{equation}
where BRST closedness imposes the linearized supergravity equations of motion
\begin{equation}
\lambda^{\alpha}_{L}\lambda_{L}^{\beta}\lambda_{R}^{\widehat\alpha}D_{\alpha}A_{\beta\widehat\alpha}+ \lambda^{\alpha}_{L}\lambda_{R}^{\widehat\alpha}\lambda_{R}^{\widehat\beta}D_{\widehat\alpha}A_{\alpha\widehat\beta}=0\,,\qquad 
\end{equation}
and BRST exactness leads to the gauge invariances
\begin{equation}
A_{\alpha\widehat\alpha}  \cong A_{\alpha\widehat\alpha} + D_{\alpha}\Omega_{\widehat\alpha} + D_{\widehat\alpha}\Omega_{\alpha}\,.
\end{equation}
By acting with the covariant derivatives $D_{\alpha}$ and $D_{\widehat\alpha}$, we obtain the tower of superfields
\bea
  A_{m\widehat\alpha} &= \frac{1}{8}(\gamma_{m}D)^{\alpha}A_{\alpha\widehat\alpha}\,,\quad & A_{\alpha m} &=  -\frac{1}{8}(\gamma_{m}D)^{\widehat\alpha}A_{\alpha\widehat\alpha}\,,\\
 A_{mn}&=\frac{1}{64}(\gamma_{n}D)^{\widehat\alpha}(\gamma_{m}D)^{\alpha}A_{\alpha\widehat\alpha}\mathrlap{{}=   -\frac{1}{64}(\gamma_{m}D)^{\alpha}(\gamma_{n}D)^{\widehat\alpha} A_{\alpha\widehat\alpha},} \\
A_{\alpha}\,^{\widehat\alpha}&=\frac{1}{8}(DD)^{\widehat\alpha\widehat\beta} A_{\alpha \widehat\beta}\,,\quad& A^{\alpha}\,_{\widehat\alpha} &= \frac{1}{8}(DD)^{\alpha\beta}A_{\beta \widehat\alpha}\,, \\
A^{\alpha\widehat\alpha} &= (DD)^{\alpha\beta}(DD)^{\widehat\alpha\widehat\beta}A_{\beta \widehat\beta}\mathrlap{{}= (DD)^{\widehat\alpha\widehat\beta}(DD)^{\alpha\beta}A_{\beta \widehat\beta} \,,} \\
A_{m}\,^{\widehat\alpha}&=\frac{1}{8}(DD)^{\widehat\alpha\widehat\beta}(\gamma_{m}D)^{\alpha} A_{\alpha \widehat\beta}\,,\quad & A^{\alpha}\,_{m} &= \frac{1}{8}(DD)^{\alpha\beta}(\gamma_{m}D)^{\widehat\alpha}A_{\beta \widehat\alpha}\,, \\
 A_{mn \widehat\alpha}&=\frac{1}{8}\partial_{[m}(\gamma_{n]}D)^{\alpha}A_{\alpha\widehat\alpha}\,,\quad &    A_{\alpha mn} &=  -\frac{1}{8}\partial_{[m}(\gamma_{n]}D)^{\widehat\alpha}A_{\alpha\widehat\alpha},\\
 A_{mn}\,^{\widehat\alpha}&=\frac{1}{8}\partial_{[m}(\gamma_{n]}D)^{\alpha}(DD)^{\widehat\alpha\widehat\beta} A_{\alpha \widehat\beta}\,,\quad &    A^{\alpha}\,_{ mn} &=  -\frac{1}{8}\partial_{[m}(\gamma_{n]}D)^{\widehat\alpha}(DD)^{\alpha\beta}A_{\beta \widehat\alpha},\\
A_{m,pq} &=  \frac{1}{64}\partial_{[p}(\gamma_{q]}D)^{\widehat\alpha}(\gamma_{m}D)^{\alpha}A_{\alpha\widehat\alpha} \,,\quad& A_{mn,p}&= \frac{1}{64}\partial_{[p}(\gamma_{q]}D)^{\alpha}(\gamma_{m}D)^{\widehat\alpha}A_{\alpha\widehat\alpha} \,,\\
  A_{mn pq} &=  - \frac{1}{64}\partial_{[m}(\gamma_{n]}D)^{\alpha}\partial_{[p}(\gamma_{q]}D)^{\widehat\alpha}A_{\alpha\widehat\alpha}\mathrlap{{}= \frac{1}{64}\partial_{[p}(\gamma_{q]}D)^{\widehat\alpha}\partial_{[m}(\gamma_{n]}D)^{\alpha}A_{\alpha\widehat\alpha}.}
\eea{eq: tower_superfields_type_ii}
The graviton $h_{mn}$, the B-field $b_{mn}$ and the dilaton $\phi$ sit in the first component of the superfield expansion 
\begin{equation}
 A_{mn}|_{\theta=0}=h_{mn}+b_{mn}+\eta_{mn}\phi   \,,
\end{equation}
while the gravitini $\left(\xi_{Lm}\,^{\alpha},\xi_{Rm}\,^{\widehat\alpha}\right)$ and dilatini $\left(\chi_{L\alpha},\chi_{R\widehat\alpha}\right)$ sit in the first component of the superfield expansion
\begin{equation}
 A^{\alpha}\,_{m}|_{\theta=0} = (\gamma_{m}\chi)^{\alpha} + \xi_{m}\,^{\alpha}\,,\qquad   A_{m}\,^{\widehat\alpha}|_{\theta=0} = (\gamma_{m}\chi)^{\widehat\alpha} + \xi_{m}\,^{\widehat\alpha}\,. 
\end{equation}
Finally, the Ramond-Ramond field strength sits in the first component of the superfield expansion
\begin{equation}
 A^{\alpha\widehat\alpha}|_{\theta=0} = \begin{cases} 
 F\delta^{\alpha\widehat\alpha}+F_{mn}(\gamma^{mn})^{\alpha\widehat\alpha}+F_{mnpq}(\gamma^{mnpq})^{\alpha\widehat\alpha} \,,&\text{for type IIA}\\ 
 F_{m}(\gamma^{m})^{\alpha\widehat\alpha}+F_{mnp}(\gamma^{mnp})^{\alpha\widehat\alpha} +F_{mnpqr}(\gamma^{mnpqr})^{\alpha\widehat\alpha} \,,&\text{for type IIB}.
 \end{cases}  
\end{equation}
\subsubsection{Integrated vertex operators}
As for the open string case, to obtain the integrated vertex operators $\int d^2zU$ corresponding to the closed string unintegrated vertex $V$ we solve a set of descent equations. In this case they become
\begin{equation}
\partial V \approx \{Q,W\}\,,\qquad \overline \partial V \approx    \{Q,\overline  W\}\,,\qquad [Q,U] \approx \overline\partial W - \partial\overline W,
\label{eq: descent_relations_closed}
\end{equation}
for $(W, \overline W, U)$. The corresponding integrated vertex operator for type II massless states is given by
\begin{equation}
U=\begin{bmatrix}
   \pi^{\alpha} \\
    \pi^{m}\\
    d_{L\alpha}\\
    \frac{1}{2}N^{mn}_{L}
\end{bmatrix}  ^{T}\cdot
\begin{bmatrix}
 A_{\alpha\widehat\beta}& A_{\alpha p}& A_{\alpha}\,^{\widehat\beta}& A_{\alpha pq}\\  
  A_{m\widehat\beta}& A_{mp}& A_{m}\,^{\widehat\beta}& A_{m,pq}\\  
   A^{\alpha}\,_{\widehat\beta}& A^{\alpha}\,_{p}& A^{\alpha\widehat\beta}& A^{\alpha}\,_{pq}\\  
    A_{mn\widehat\beta}& A_{mn,p}& A_{mn}\,^{\widehat\beta}& A_{mnpq}
\end{bmatrix}  \cdot
\begin{bmatrix}
   \overline\pi^{\widehat\beta} \\
    \overline\pi^{p}\\
    d_{R\widehat\beta}\\
    -\frac{1}{2}N^{pq}_{R}
\end{bmatrix}.
\label{integrated-vo}\end{equation}
Besides for computing scattering amplitudes involving type II massless states, these integrated vertex operators are useful in that they provide deformations for the worldsheet action, as in
\begin{equation}
S_{\varepsilon} = S^{(0)} +\varepsilon S^{(1)} + \frac{\varepsilon^2}{2} S^{(2)}  +\mathcal{O}(\varepsilon^3).  
\end{equation}
where $S^{(0)}=S$ is the type II flat background worldsheet action (\ref{eq: worldsheet_action_type_ii_susy}) and $S^{(1)} = \int d^2z U$ is the integrated vertex operator. To maintain BRST symmetry, we must also deform the BRST transformations
\begin{equation}
Q_{\varepsilon} = Q^{(0)} + \varepsilon Q^{(1)} + \frac{\varepsilon^2}{2} Q^{(2)}+\mathcal{O}(\varepsilon^3)\,,  
\end{equation}
so that $Q_{\varepsilon}S_{\varepsilon} = 0$. For example, at leading order in $\varepsilon$, we obtain
\begin{equation}
Q^{(0)} S^{(1)} + Q^{(1)}S^{(0)} =0\,.
\end{equation}
The second term, $Q^{(1)}S^{(0)}$, is a linear combination of equations of motion, so the above equation has solutions since the integrated vertex operator is BRST invariant up to equations of motion. Keeping track of these equations of motion provides the correction $Q^{(1)}$ to the BRST transformations. At the next order, we have
\begin{equation}
Q^{(0)}S^{(2)} + Q^{(1)}S^{(1)}+ Q^{(2)}S^{(0)}  =0\,,
\end{equation}
such that if the integrated vertex operator is not BRST invariant after the first order correction,  i.e. $Q^{(1)}S^{(1)}$ does not vanish up to equations of motion, the worldsheet action back-reacts on $S^{(2)}$, so that $Q^{(0)}S^{(2)}$ cancels $Q^{(1)}S^{(1)}$ up to equations of motion. Then, these equations of motion provide the second-order correction $Q^{(2)}$ to the BRST transformations. This continues at higher orders.

\subsubsection{Curved backgrounds}

As we saw in subsection \ref{sec:brst_symmetry}, the BRST symmetry is characterized by its nilpotency and holomorphicity. In \cite{Berkovits:2001ue}, the pure spinor formalism was generalized to curved backgrounds by allowing worldsheet action deformations that preserve BRST nilpotency and holomorphicity, where the BRST charge is maintained to be
\begin{equation}
Q =\oint \frac{dz}{2\pi i} \lambda_{L}^{\alpha}d_{L\alpha} - \oint\frac{d\overline z}{2\pi i} \lambda_{R}^{\widehat\alpha}d_{R\widehat\alpha}\,.
\end{equation}
At leading order in $\alpha'$, it was shown that imposing these conditions results in the supergravity equations for the background fields parametrizing the worldsheet action deformations. In what follows, we describe this in more detail.

We obtain the action of the pure spinor string in curved backgrounds by promoting the matter worldsheet fields $(x^m, \theta_L^\alpha , \theta_R^{\widehat\alpha})$ to describe local coordinates $(Z^M) = (x^m,\theta_L^\mu, \theta_R^{\widehat\mu})$ of some supermanifold and then covariantizing the flat space action deformed by the integrated vertex operator $S^{(0)} + \int d^2z\ U$ w.r.t.\ super-diffeomorphisms. This produces the worldsheet action of the pure spinor formalism in a general type II background
\bea
S_{\text{Type II}} = \int \frac{d^2z}{2\pi\alpha'} \bigg[& \frac{1}{2}\pi^{a}\overline\pi^{b}\eta_{ab} + B(\partial,\overline\partial)+ \overline \pi^{\alpha} d_{L\alpha} + \pi^{\widehat\alpha}d_{R\widehat\alpha} \\
  &+ w_{L\alpha}\overline\nabla\lambda_{L}^{\alpha} +w_{R\widehat\alpha}\nabla\lambda_{R}^{\widehat\alpha}+S_{\alpha}\,^{\beta}\,_{\widehat\alpha}\,^{\widehat\beta} w_{L\beta}\lambda_{L}^{\alpha} w_{R\widehat\beta}\lambda_{R}^{\widehat\alpha} \\
  & + P^{\alpha\widehat\alpha}d_{L\alpha}d_{R\widehat\alpha}+C_{\alpha}\,^{\beta\widehat\alpha}\lambda_{L}^{\alpha}w_{L\beta}d_{R\widehat\alpha}+C_{\widehat\alpha}\,^{\widehat\beta\alpha}\lambda_{R}^{\widehat\alpha}w_{R\widehat\beta}d_{L\alpha}  \bigg]
\eea{eq: curved_background_worldsheet_action}
where $P^{\alpha\widehat\alpha}(Z)$, $C_{\alpha}\,^{\beta\widehat\alpha}(Z)$, $C_{\widehat\alpha}\,^{\widehat\beta\alpha}(Z)$ and $S_{\alpha}\,^{\beta}\,_{\widehat\alpha}\,^{\widehat\beta}(Z)$ are superfields associated to the field-strength of RR fields, gravitini, dilatini, and the curvature of the supermanifold. Also,
\begin{equation}
\pi^{a} = \partial Z^{M}E_{M}\,^{a}\,,\qquad \pi^{\alpha} =  \partial Z^{M}E_{M}\,^{\alpha}\,,\qquad \pi^{\widehat\alpha} =  \partial Z^{M}E_{M}\,^{\widehat\alpha},\nonumber   
\end{equation}
\begin{equation}
\overline \pi^{a} = \overline \partial Z^{M}E_{M}\,^{a}\,,\qquad \overline \pi^{\alpha} =  \overline \partial Z^{M}E_{M}\,^{\alpha}\,,\qquad \overline \pi^{\widehat\alpha} =  \overline \partial Z^{M}E_{M}\,^{\widehat\alpha}   ,
\end{equation}
are the worldsheet pullbacks of the superspace vielbein $E^A_M(Z)$ with $A = a, \alpha,\widehat\alpha$,
\begin{equation}
\overline\nabla\lambda_{L}^{\alpha} = \overline\partial\lambda_{L}^{\alpha} + \overline\partial Z^{M}\Omega_{M}\,^{\alpha}\,_{\beta}\lambda_{L}^{\beta}\,,\qquad \nabla\lambda_{R}^{\widehat\alpha} = \partial\lambda_{R}^{\widehat\alpha} + \overline\partial Z^{M}\Omega_{M}\,^{\widehat\alpha}\,_{\widehat\beta}\lambda_{R}^{\widehat\beta},
\end{equation}
are the covariant derivatives of $\lambda_L^\alpha$ and $\lambda_R^{\widehat\alpha}$ with respect to the worldsheet pullbacks of target space connections, and
\begin{equation}
B(\partial,\overline\partial) = \partial Z^{M}\overline\partial Z^{N}B_{MN}\,,  
\end{equation}
is the worldsheet pullback of the Kalb-Ramond potential.

The worldsheet action (\ref{eq: curved_background_worldsheet_action}) is invariant under superspace reparametrizations, provided that background fields transform appropriately. Besides, the spacetime structure group, which is composed of Weyl and Lorentz transformations acting on the flat indices $(a,\alpha,\widehat\alpha)$ of the background fields, are symmetries of the worldsheet action provided that the pure spinor variables also transform, and that the background fields appearing in
\begin{equation}
\Omega_{M}\,^{\alpha}\,_{\beta}\lambda_{L}^{\beta}w_{L\alpha} =  \frac{1}{2}\Omega_{M}^{(L)}N_{L}  + \frac{1}{2}\Omega_{M}^{(L)}\,^{ab}N_{Lab}\,,\nonumber  
\end{equation}
\begin{equation}
\Omega_{M}\,^{\widehat\alpha}\,_{\widehat\beta}\lambda_{R}^{\widehat\beta}w_{R\widehat\alpha} =  \frac{1}{2}\Omega_{M}^{(R)}N_{R}  + \frac{1}{2}\Omega_{M}^{(R)}\,^{ab}N_{Rab}\,,  
\end{equation}
transform as connections. Notice that the above decomposition involves only the gauge invariant operators of the pure spinor sector, and this imposes restrictions to the structure group.

\subsection{The type IIB \texorpdfstring{$\ads$}{AdS5xS5} background}
\label{sec: The type IIB ads background}

Plugging in the spacetime background superfields corresponding to the type IIB $\ads$ background into the worldsheet action for type II curved background (\ref{eq: curved_background_worldsheet_action}), we obtain
\bea
S= R^2\int\frac{d^2z}{2\pi\alpha'}\bigg[&\frac{1}{2}\pi^{a}\overline \pi_{a}+ B(\partial,\overline\partial)+\overline \pi^{\alpha}d_{L\alpha} +\pi^{\widehat\alpha}d_{R\widehat\alpha}+\overline\nabla\lambda_{L}^{\alpha}w_{L\alpha}+w_{ R\widehat\alpha}\nabla\lambda_{R}^{\widehat\alpha}   \\
& +\eta^{\alpha\widehat\alpha}d_{L\alpha}d_{R\widehat\alpha}-\frac{1}{4}R_{abcd}N^{ab}_{L}N^{cd}_{R})\bigg]\,,
\eea{S_AdS}
where the Kalb-Ramond potential is given by
\begin{eqnarray}
B(\partial,\overline\partial)= -\frac{1}{2}(\pi^{\alpha} \overline \pi\,^{\widehat\alpha}-\overline \pi\,^{\alpha} \pi^{\widehat\alpha})\eta_{\alpha\widehat\alpha}\,.    
\end{eqnarray}
Unlike in flat background, in $\ads$ the worldsheet action has a term $\eta^{\alpha\widehat\alpha}d_{L\alpha}d_{R\widehat\alpha}$ that allows us to eliminate the worldsheet variables $d_{\alpha}$ and $d_{\widehat\alpha}$ by their equations of motion. This results in
\bea
S=R^2\int\frac{d^2z}{2\pi\alpha'}\bigg[&\frac{1}{2}\pi^{a}\overline \pi_{a}+(\pi^{\alpha} \overline \pi^{\widehat\alpha}+\overline \pi^{\alpha}\pi^{\widehat\alpha})\eta_{\alpha\widehat\alpha}-\frac{1}{2}(\pi^{\alpha} \overline \pi\,^{\widehat\alpha}-\overline \pi\,^{\alpha} \pi^{\widehat\alpha})\eta_{\alpha\widehat\alpha}  \\
&+w_{L\alpha}\overline\nabla\lambda_{L}^{\alpha}+w_{ R\widehat\alpha}\nabla\lambda_{R}^{\widehat\alpha} -\frac{1}{4}R_{abcd}N_{L}^{ab}N_{R}^{cd} \bigg]\,.
\eea{eq: ads_worldsheet_action}
Here $R$ is the common radius of $AdS_5$ and $S^5$. 
Note that the first line of the worldsheet action above contains only the worldsheet variables corresponding to the superspace coordinates, while the second line contain the pure spinor variables. The first line is different to the GS worldsheet action in that it contains the term $(\pi^{\alpha} \overline \pi^{\widehat\alpha}+\overline \pi^{\alpha}\pi^{\widehat\alpha})\eta_{\alpha\widehat\alpha}$, which is real, thus providing a non-degenerate metric for the fermionic directions. The second line contains the spacetime spin connection coupling with the pure spinor variables, via $w_{L\alpha}\overline\nabla\lambda_{L}^{\alpha}+w_{ R\widehat\alpha}\nabla\lambda_{R}^{\widehat\alpha}$, which implies that the pure spinor variables function as place holders for the spin group. The term $\frac{1}{4}R_{abcd}N_{L}^{ab}N_{R}^{cd} $ is required for BRST invariance.

Another special feature of the $\ads$ background is that, because we are able to eliminate the auxiliary variables $d_{L\alpha}$ and $d_{R\widehat\alpha}$ through their equations of motion, the BRST transformations, now generated by
\begin{equation}
Q =\oint \frac{dz}{2\pi i} \lambda_{L}^{\alpha}\pi^{\widehat\alpha}\eta_{\alpha\widehat\alpha} - \oint\frac{d\overline z}{2\pi i} \lambda_{R}^{\widehat\alpha}\overline\pi^{\alpha}\eta_{\alpha\widehat\alpha}\,,
\end{equation}
act geometrically in the spacetime superspace worldsheet variables:
\begin{eqnarray}
Q Z^{M} = \lambda^{\alpha}_{L}E_{\alpha}\,^{M} + \lambda^{\widehat\alpha}_{R}E_{\widehat\alpha}\,^{M} \,.
\label{eq: brst_is_geom_in_ads}
\end{eqnarray}
The non-geometrical ones, involving $d_{L \alpha}$ and $d_{R \alpha}$, are not required once we eliminate these variables by their equations of motions.

\subsubsection{The superspace as a supercoset}
\label{sec: The superspace as a supercoset}

The $\ads$ superspace can be viewed as the supercoset $ \frac{PSU(2,2|4)}{USp(2,2)\times USp(4)}$. Therefore, the worldsheet variables corresponding to the superspace coordinates, $Z^{M}(z,\overline z)$, can be replaced by the worldsheet variables
\begin{equation}
g(z,\overline z) \in \frac{PSU(2,2|4)}{USp(2,2)\times USp(4)} ,
\end{equation}
where the $USp(2,2)\times USp(4)$ are gauge invariances
\begin{equation}
g(z,\overline z) \sim g(z,\overline z)h(z,\overline z)\,,\qquad h(z,\overline z)\in  USp(2,2)\times USp(4)\,. 
\end{equation}
The BRST transformations in (\ref{eq: brst_is_geom_in_ads}) thus acts as fermionic right-translation of the coset variables whose parameters are the worldsheet pure spinors:
\begin{equation}
Q g \sim g \left(\lambda_{L}^{\alpha}t_{\alpha}+\lambda_{R}^{\widehat\alpha}t_{\widehat\alpha} \right) \,,  
\end{equation}
where $\sim$ denotes equal up to $USp(2,2)\times USp(4)$ gauge transformations.

The worldsheet currents that appear in the worldsheet action (\ref{eq: ads_worldsheet_action}) are written in terms of these $ \frac{PSU(2,2|4)}{USp(2,2)\times USp(4)}$ worldsheet variables as
\begin{equation}
 \pi^{a} = \left(g^{-1}\partial g\right)^{a} \,,\qquad  \pi^{\alpha} = \left(g^{-1}\partial g\right)^{\alpha}\,,\qquad  \pi^{\widehat\alpha} = \left(g^{-1}\partial g\right)^{\widehat\alpha}, \nonumber 
\end{equation}
\begin{equation}
 \overline \pi^{a} = \left(g^{-1}\overline \partial g\right)^{a} \,,\qquad  \overline \pi^{\alpha} = \left(g^{-1}\overline \partial g\right)^{\alpha}\,,\qquad  \overline \pi^{\widehat\alpha} = \left(g^{-1}\overline \partial g\right)^{\widehat\alpha}  ,
\end{equation}
where
\begin{equation}
g^{-1}\partial g = \pi^{a}t_{a} + \pi^{\alpha} t_{\alpha} +  \pi^{\widehat\alpha} t_{\widehat\alpha} - \Omega^{ab}t_{ab}\,,\qquad g^{-1}\overline \partial g = \overline \pi^{a}t_{a} + \overline \pi^{\alpha} t_{\alpha} +  \overline \pi^{\widehat\alpha} t_{\widehat\alpha} - \overline \Omega^{ab}t_{ab}  ,  
\label{vielbeinconnection}
\end{equation}
is the decomposition of $g^{-1}\partial g$ into the $PSU(2,2|4)$ generators $(t_a, t_{\alpha}, t_{\widehat\alpha}, t_{ab})$ decomposed by the $\mathbb{Z}_4$ grading 
\begin{equation}
\mathfrak{g} = \mathfrak{g}_0 +\mathfrak{g}_1 +\mathfrak{g}_2 +\mathfrak{g}_3 \nonumber   
\end{equation}
where
\begin{equation}
t_{ab}\in\mathfrak{g}_0\,,\qquad     t_{\alpha}\in\mathfrak{g}_1\,,\qquad t_{a}\in\mathfrak{g}_2\,,\qquad t_{\widehat\alpha}\in\mathfrak{g}_3\,.
\end{equation}
The $PSU(2,2|4)$ algebra, in terms of these generators, is
\begin{equation}
\{t_{\alpha},t_{\beta}\} = \gamma_{\alpha\beta}^{a}t_{a}\,,\qquad   \{t_{\widehat \alpha},t_{\widehat \beta}\} = \gamma_{\widehat \alpha\widehat \beta}^{a}t_{a}\,,\qquad \{t_{\alpha},t_{\widehat\alpha}\}= R_{\alpha\widehat\alpha}\,^{ab}t_{ab},  \nonumber
\end{equation}
\begin{equation}
[t_{a},t_{\alpha}] = -\frac{1}{2}\gamma_{a\alpha\beta}P^{\beta\widehat\alpha}t_{\widehat\alpha}\,,\qquad [t_{a},t_{\widehat\alpha}] = +\frac{1}{2}\gamma_{a\widehat\alpha\widehat\beta}P^{\alpha\widehat\beta}t_{\alpha}\,,\qquad [t_{a},t_{b}]= R_{ab}\,^{cd}t_{cd} ,
\label{eq: rr_flux_acting_algebra}
\end{equation}
with the $USp(2,2)\times USp(4)$, the spin group of $SO(4,1)\times SO(5)$, given by
\begin{equation}
[t_{ab},t_{cd}]= \eta_{c[a} t_{b]d} - \eta_{d[a} t_{b]c}\,,
\end{equation}
\begin{equation}
  [t_{ab},t_{\alpha}]=\frac{1}{2}\gamma_{ab\alpha}\,^{\beta}t_{\beta}\,,\qquad [t_{ab},t_{c}] = \eta_{c[a}t_{b]}\,,\qquad [t_{ab},t_{\widehat\alpha}]=\frac{1}{2}\gamma_{ab\widehat\alpha}\,^{\widehat\beta}t_{\widehat\beta}  \,.\nonumber 
\end{equation}
The background superfields take the constant values
\begin{equation}
P^{\alpha\widehat\alpha} = -\frac{1}{R}\,\eta^{\alpha\widehat\alpha}\,,\qquad
R_{\alpha\widehat\alpha}{}^{ab} = \mp\frac{1}{2R}(\gamma^{ab}\eta)_{\alpha\widehat\alpha}\,,\qquad
R_{ab}{}^{cd} = \mp\frac{1}{2R^2}\left(\delta_{a}^{c}\delta_{b}^{d}-\delta_{a}^{d}\delta_{b}^{c}\right)\,,
\end{equation}
where $\eta_{\alpha\widehat\beta}=(\gamma^{01234})_{\alpha\widehat\beta}$ and $\eta^{\alpha\widehat\beta}=(\gamma^{01234})^{\alpha\widehat\beta}$, so that $\eta_{\alpha\widehat\beta}\eta^{\widehat\beta\gamma}=-\delta_{\alpha}^{\gamma}$, are the normalized Ramond-Ramond bi-spinors, and the upper (lower) sign are for the $AdS_5$ ($S^5$) indices.  If we consider the worldsheet currents to be valued at the $PSU(2,2|4)$ algebra
\begin{equation}
\pi^{(1)} = \pi^{\alpha} t_{\alpha} \,,\qquad \pi^{(2)}=  \pi^{a}t_{a} \,,\qquad \pi^{(3)} =  \pi^{\widehat\alpha} t_{\widehat\alpha}\,, \nonumber
\end{equation}
\begin{equation}
\overline \pi^{(1)} = \overline \pi^{\alpha} t_{\alpha} \,,\qquad \overline \pi^{(2)}=  \overline \pi^{a}t_{a} \,,\qquad \overline \pi^{(3)} =  \overline \pi^{\widehat\alpha} t_{\widehat\alpha}\,, 
\end{equation}
and similar for the pure spinor variables
\begin{equation}
\lambda_{L} = \lambda_{L}^{\alpha}t_{\alpha}\,,\qquad w_{L}= w_{L\alpha}\eta^{\alpha\widehat\alpha}t_{\widehat\alpha}\,,\qquad   \lambda_{R} = \lambda^{\widehat\alpha}_{R}t_{\widehat\alpha}\,,\qquad w_{R} = w_{R\widehat\alpha}\eta^{\alpha\widehat\alpha}t_{\alpha},  
\end{equation}
the worldsheet action becomes
\bea
S = R^2\int\frac{d^2z}{2\pi\alpha'} \text{Str}\bigg[&\frac{1}{2}\pi^{(2)}\pi^{(2)} + \frac{1}{4}\pi^{(1)}\overline\pi^{(3)} + \frac{3}{4}\overline\pi^{(1)}\pi^{(3)}\\
&+ w_{L}\overline\nabla\lambda_{L}+w_{R}\nabla \lambda_{R}-\frac{1}{4}\{w_{L},\lambda_{L}\}\{w_{R},\lambda_{R}\}\bigg],
\eea{SS}
where
\begin{equation}
\text{Str}\left[t_{a}t_{b}\right] = \eta_{ab}\,,\qquad \text{Str}\left[t_{\alpha}t_{\widehat\beta}\right] = \eta_{\alpha\widehat\beta}\,,\qquad \text{Str}\left[t_{ab}t_{cd}\right] = \mp  \eta_{a[c}\eta_{d]b}\,.  
\end{equation}
The pure spinor constraints become $\{\lambda_{L},\lambda_{L}\} = 0$ and $\{\lambda_{R},\lambda_{R}\} = 0$. As we can see from the (anti-)commutation relations in (\ref{eq: rr_flux_acting_algebra}), the spacetime translation mixes the two Majorana-Weyl supercharges $t_{\alpha}$ and $t_{\widehat\alpha}$. Therefore, if we want to express the spinors using $SU(2,2)\times SU(4)$ indices, we must diagonalize the spacetime translations by considering linear combinations of $t_{\alpha}$ and $t_{\widehat\alpha}$. The result is
\begin{equation}
\varepsilon=\frac{\varepsilon_L-i\varepsilon_R}{2},\qquad
\overline\varepsilon =\frac{\varepsilon_L+i\varepsilon_R}{2},
\label{eq:l diag_of_translation}
\end{equation}
where $\varepsilon = \varepsilon^{I}_{R}t^{R}_{I}$, $\overline\varepsilon = \varepsilon_{I}^{R}t_{R}^{I}$, $\varepsilon_{L}=\varepsilon^{\alpha}t_{\alpha} $ and $\varepsilon_{R}=\varepsilon^{\widehat\alpha}t_{\widehat\alpha}$, and $(t^{R}_{S}, t^{I}_{J})$ are the spacetime supercharges anti-commuting into the $SU(2,2)\times SU(4)$ generators (see appendix \ref{app: algebra}).

\subsubsection{Parametrizations of the \texorpdfstring{$\ads$}{AdS5xS5} superspace}

In \cite{Berkovits:2019rwq, Berkovits:2019ulm}, an $\ads$ superspace parametrization was proposed using the $\frac{PSU(2,2|4)}{USp(2,2)\times USp(4)}$ supercoset (see appendix \ref{app: algebra})
\begin{equation}
g_{\text{old}}= e^{(\psi^{\alpha}_i q^i_{\alpha} + \theta^{\bar{i}}_{\dot\alpha} q^{\dot\alpha}_{\bar{i}} + x^{\alpha}_{\dot\alpha} K^{\dot\alpha}_{\alpha} + y^{\bar{i}}_{i} K^{i}_{\bar{i}} )} 
e^{(\psi^{\dot\alpha}_i q^i_{\dot\alpha} + \theta^{\bar{i}}_{\alpha} q^{\alpha}_{\bar{i}})}
e^{(\psi^{\alpha}_{\bar{i}} q^{\bar{i}}_{\alpha}+\theta^i_{\dot\alpha} q^{\dot\alpha}_i) }
e^{(\psi^{\dot\alpha}_{\bar{i}} q^{\bar{i}}_{\dot\alpha}+\theta^i_{\alpha} q^{\alpha}_i )}
e^{ z \Delta + w J} \, , 
\label{eq: OldCosetRepresentative}
\end{equation}
where $(x^{\alpha}_{\dot\alpha},z)$ are $AdS_5$ coordinates, $(y^{\overline i}_{i}, w)$ are $S^5$ coordinates, and all the other coordinates, $(\theta_{\alpha}^{i}, \theta^{\overline i}_{\alpha}, \theta_{\dot\alpha}^{\overline i},\theta^{\overline i}_{\dot\alpha})$ and $(\psi^{\alpha}_{i},\psi^{\dot\alpha}_{i},\psi^{\alpha}_{\overline i},\psi^{\dot\alpha}_{\overline i})$, are fermionic. Although the $SU(2,2)\times SU(4)$ transformations are complicated in these coordinates, as they mix bosons with fermions, the half-BPS vertex operators annihilated by the $24$ supercharges
\begin{equation}
q_{\alpha}^{i}\,,\qquad    q_{\overline i}^{\dot\alpha}\,,\qquad  q_{\alpha}^{\overline i} \,,\qquad q_{\dot\alpha}^{i}\,,\qquad q_{i}^{\dot\alpha}\,,\qquad q_{\overline i}^{\alpha}\,,  
\end{equation}
simplify in terms of these coordinates. In particular, these vertex operators depend only on the coordinates $(z+w)$, $\theta^{i}_{\alpha}$ and $\psi^{\dot\alpha}_{\overline i}$. They were explicitly constructed in \cite{Fleury:2021ieo}. Their structure is schematically given by
\begin{equation}
V =  \left(\langle\overline\lambda^2_{\pm}\rangle + 
\langle\overline\lambda^2_{\pm} \theta_+^2\rangle P_1(\ell)
+\langle\overline\lambda^2_{\pm} \theta_+^4\rangle P_2(\ell)
+ \langle\overline\lambda^2_{\pm} \theta_+^6\rangle P_3 (\ell)+
\langle\overline\lambda^2_{\pm} \theta_+^8\rangle P_4(\ell)\right) e^{-\ell(z+w)}\,   \, .
\label{schematically} 
\end{equation}
To compute non-trivial scattering amplitudes, however, we need to consider more general half-BPS states, not annihilated by these $24$ supersymmetries. One way to obtain these states is through $SU(2,2)\times SU(4)$ transformations.

Since the $SU(2,2)\times SU(4)$ transformations are computationally heavy in the coordinates (\ref{eq: OldCosetRepresentative}), \cite{Fleury:2026xpg} performed a superspace reparameterization bringing the vertex operator of \cite{Fleury:2021ieo} to the coordinates
 \begin{equation}
g_{\text{new}} = e^{\Psi^S_I q^I_S} e^{\Theta_S^I q_I^S} U V  \, , 
\label{eq: NewCosetRepresentative}
\end{equation} 
where
\begin{equation}
U= e^{X^{\alpha}_{\dot\alpha} K^{\dot\alpha}_{\alpha}} e^{Z \Delta} \, , \qquad  V = 
e^{Y^{\bar{i}}_i K^i_{\bar{i}}} e^{W J} \, , 
\label{eq: CosetIdeal}    
\end{equation}
where $SU(2,2)\times SU(4)$ acts in a  straightforward way.
 This allows for the construction of generic half-BPS vertex operators simply by acting with the $PSU(2,2|4)$ transformation on the vertex obtained in \cite{Fleury:2021ieo}, now written using the coordinates (\ref{eq: NewCosetRepresentative}).

\subsubsection{\texorpdfstring{$SU(2,2)\times SU(4)$}{SU(2,2)xSU(4)} covariance}

Although the fermionic coordinates in the parametrization (\ref{eq: CosetIdeal}) are $SU(2,2)\times SU(4)$ covariant, the spacetime coordinates are not. Making all coordinates $SU(2,2)\times SU(4)$ covariant will prove useful, as we will see later in the paper. To achieve this, consider the $SU(2,2)\times SU(4)$ covariant variables $(X_{RS}, Y_{IJ}, \Psi^S_I, \Theta_S^I)$ 
\begin{equation}
  \epsilon^{RSTP} X_{RS} X_{TP} = -8\,,\qquad  \epsilon^{IJKL} Y_{IJ} Y_{KL} = -8\,,
  \label{eq: constraints_charac_x_and_y}
\end{equation}
 \begin{equation}
g_{\text{fermionic}} = e^{\Psi^S_I q^I_S} e^{\Theta_S^I q_I^S} \, ,
\label{gfermionic}
\end{equation} 
where $\epsilon^{RSTP}$ and $\epsilon^{IJKL}$ are the  $SU(2,2)$ and $SU(4)$ invariants. We will now show that the above $SU(2,2)\times SU(4)$ covariant variables are field redefinitions of the $\ads$ superspace coordinates.

We first introduce an additional $SU(2,2)\times SU(4)$ gauge group through the local compensators $U_{R}\,^{\widetilde R}$ and $V_{I}\,^{\widetilde I}$, where the tilded indices are the gauge-transforming ones. One can construct the following gauge-transforming variables by dressing the covariant ones with compensators
\begin{equation}
 \widetilde X_{\widetilde{R}\widetilde{S}} = U_{\widetilde R}\,^{{R}}\,X_{RS} U^{{S}}\,_{\widetilde S}\,, \qquad 
  \widetilde Y_{\widetilde{I}\widetilde{J}}= V_{\widetilde I}\,^{{I}}\,Y_{IJ}  \, V^{{J}}\,_{\widetilde J}\,,    
  \label{XYrotation}
\end{equation}
 \begin{equation}
\widetilde g_{\text{fermionic}} = e^{\Psi^S_I q^I_S} e^{\Theta_S^I q_I^S} U V  \, .
\nonumber
\end{equation}
Of course, one can gauge away the compensators, returning to our starting point, but let us instead consider the gauge
\begin{equation}
  \widetilde X_{\widetilde R\widetilde S} = \varepsilon_{\widetilde{R}\widetilde{S}}\,, \qquad 
  \widetilde Y_{\widetilde I\widetilde J} =  \varepsilon_{\widetilde{I}\widetilde{J}}\,,
\label{eq: X and Y def}
\end{equation}
where
\begin{equation}
\varepsilon_{\widetilde R\widetilde S} = \begin{pmatrix}
   \varepsilon_{\alpha\beta}& 0\\
   0 & -\varepsilon_{\dot\alpha\dot\beta}
\end{pmatrix}\,,\qquad   \varepsilon_{\widetilde I\widetilde J}  = \begin{pmatrix}
   \varepsilon_{ij}& 0\\
   0 & -\varepsilon_{\overline i\overline j}
\end{pmatrix}\,,
\end{equation}
with  $(\varepsilon_{\alpha\beta},\varepsilon_{\dot\alpha\dot\beta},\varepsilon_{ij},\varepsilon_{\overline i\overline j})$ denoting the two-dimensional Levi-Civita symbols, with $\varepsilon_{1 2}=\varepsilon_{34}=1$. This completely solves the constraints (\ref{eq: constraints_charac_x_and_y}) and reduces the gauge group to $USp(2,2)\times USp(4)$. 

Note that in this gauge the bosonic degrees of freedom are absorbed into the compensators, such that all degrees of freedom are now in $\widetilde g_{\text{fermionic}}$. Since the residual gauge group is $USp(2,2)\times USp(4)$, which acts by right multiplication on $\widetilde g_{\text{fermionic}}$, the degrees of freedom becomes an element of the $\frac{PSU(2,2|4)}{USp(2,2)\times USp(4)}$ supercoset, but as we saw in sub-subsection \ref{sec: The superspace as a supercoset}, this corresponds precisely to the $\ads$ superspace. Therefore, we have established a field redefinition relating the covariant variables $(X_{RS}, Y_{IJ}, \Psi^S_I, \Theta_S^I)$ introduced above to the $\ads$ superspace coordinates. Alternatively, one can work with the full gauge group in which the fermionic variables are described by the fermionic supercoset
\begin{equation}
\widetilde g_{\text{fermionic}}(z,\overline z) \in \frac{PSU(2,2|4)}{SU(2,2)\times SU(4)}\,,
\label{gferm}
\end{equation}
while the bosonic variables are described by the gauge invariant $X_{RS}$ and $Y_{IJ}$ obeying the constraints (\ref{eq: constraints_charac_x_and_y}). 

This would be the whole story if we neglect the pure spinors variables. Because the pure spinors transform under the $USp(2,2)\times USp(4)$ gauge group, the above manipulation must be re-examined. In the partially fixed gauge (\ref{eq: X and Y def}), where the residual gauge group is $USp(2,2)\times USp(4)$, the pure spinor constraints are given by
\begin{equation}
(\lambda_{L})^{\widetilde{R}}_{\widetilde{I}} \varepsilon_{\widetilde{R} 
\widetilde{S}} (\lambda_{L})^{\widetilde{S}}_
{\widetilde{J}} = -\frac{1}{4} (\lambda_{L} \lambda_{L})  
\varepsilon_{\widetilde{I} \widetilde{J}} \,,\qquad  (\lambda_L)^{\widetilde{R}}_{\widetilde{I}} \varepsilon^{\widetilde{I} \widetilde{J}} (\lambda_L)^{\widetilde{S}}_{\widetilde{J}} = -\frac{1}{4} (\lambda_L \lambda_L)  
\varepsilon^{\widetilde{R} \widetilde{S}}  
\label{eq: RLvinculos}
\end{equation}
\begin{equation}
(\lambda_{R})^{\widetilde{R}}_{\widetilde{I}} \varepsilon_{\widetilde{R} 
\widetilde{S}} (\lambda_{R})^{\widetilde{S}}_
{\widetilde{J}} = -\frac{1}{4} (\lambda_{R} \lambda_{R})  
\varepsilon_{\widetilde{I} \widetilde{J}} \,,\qquad  (\lambda_R)^{\widetilde{R}}_{\widetilde{I}} \varepsilon^{\widetilde{I} \widetilde{J}} (\lambda_R)^{\widetilde{S}}_{\widetilde{J}} = -\frac{1}{4} (\lambda_R \lambda_R)  
\varepsilon^{\widetilde{R} \widetilde{S}} \,. 
\nonumber
\end{equation} 
It is natural to construct gauge invariant variables by dressing the pure spinors with the compensators
\begin{equation}
(\Lambda_L)_I^R = V_I^{\widetilde{I}} (\lambda_L)_{\widetilde{I}}^{\widetilde{R}} (U^{-1})_{\widetilde{R}}^R \, , \quad \quad \quad 
(\Lambda_R)_I^R = V_I^{\widetilde{I}} (\lambda_R)_{\widetilde{I}}^{\widetilde{R}} (U^{-1})_{\widetilde{R}}^R\, ,
\label{defLambda}
\end{equation}
resulting in the constraints
\bea
(\Lambda_{L})^{R}_{I} X_{R 
{S}} (\Lambda_{L})^{{S}}_
{{J}} &= -\frac{1}{4} (\Lambda_{L} \Lambda_{L})  
Y_{IJ} \,,&\qquad  (\Lambda_L)^{R}_{I} Y^{I J} (\Lambda_L)^{S}_{J} &= -\frac{1}{4} (\Lambda_L \Lambda_L)  
X^{R S}  ,\\
(\Lambda_{R})^{R}_{I} X_{R 
{S}} (\Lambda_{R})^{S}_
{{J}} &= -\frac{1}{4} (\Lambda_{R} \Lambda_{R})  
Y_{I J} \,,&\qquad  (\Lambda_R)^{R}_{I} Y^{I J} (\Lambda_R)^{S}_{J} &= -\frac{1}{4} (\Lambda_R \Lambda_R)  
X^{R S}  \,,
\eea{eq: Constraints_carach_ps_cov}
where $X_{RS}$ and $Y_{IJ}$ are the gauge invariant variables obeying the constraints (\ref{eq: constraints_charac_x_and_y}), and they are used to lower and raise indices of $(\Lambda_{L})_{I}^{R}$ and $ (\Lambda_{R})_{I}^{R}$. Therefore, the $SU(2,2)\times SU(4)$ covariant variables for the pure spinor formalism in $\ads$ are $X_{RS}$, $Y_{IJ}$, $\Psi^S_I$, $\Theta_S^I, (\Lambda_{L})_{I}^{R}$ and $(\Lambda_{R})_{I}^{R}$ obeying the constraints (\ref{eq: constraints_charac_x_and_y}) and (\ref{eq: Constraints_carach_ps_cov}).

Of course, we must keep track of the BRST transformations as we perform these field redefinitions mapping the pure spinors and the $\ads$ supercoset into the $SU(2,2)\times SU(4)$ covariant variables of the previous sub-subsection. These BRST transformations are given by
\begin{equation}
Q \cdot \Psi^R_I  = \Lambda^R_I  \, , \quad \quad Q \cdot \Theta^I_R = \bar{\Lambda}^I_R - \Theta^I_S \Lambda^S_J \Theta^J_R \, , 
\label{BRSThere}
\end{equation}
and
\bea
 Q \cdot X^{RS} &= - \Lambda^R_{I} \Theta^I_{P} X^{PS} - \Lambda^S_{I} \Theta^I_{P} X^{RP} + \frac{1}{2} (\Lambda^P_{J} \Theta^J_{P}) X^{RS} \, ,\\
 Q \cdot Y_{IJ} &=  \Lambda^R_{I} \Theta^K_{R} Y_{KJ} + \Lambda^R_{J} \Theta^K_{R} Y_{IK} - \frac{1}{2} (\Lambda^P_{K} \Theta^K_{P}) Y_{IJ} \, . 
\eea{Qbulkpoints}
where $\Lambda^R_{I}$ and $\overline \Lambda_{I}^{R}$ are linear combinations of the gauge-invariant pure spinors $(\Lambda_{L})_{I}^{R}$ and $(\Lambda_{R})_{I}^{R}$ similar to the ones in (\ref{eq:l diag_of_translation}), satisfying
\begin{equation}
(\Lambda_R)^R_I = i (\Lambda^R_I + X^{R S} Y_{I J} \bar\Lambda^J_S) \, , \quad \quad 
(\Lambda_L)^R_I =  (\Lambda^R_I - X^{R S} Y_{I J} \bar\Lambda^J_S) \, . 
\label{LambdaLR}
\end{equation}
One can also consider similar linear combinations of the fermionic parameters:
\begin{equation}
(\Theta_R)^R_I = i (\Psi^R_I + X^{R S} Y_{I J} \Theta^J_S) \, , \quad \quad 
(\Theta_L)^R_I =  (\Psi^R_I - X^{R S} Y_{I J} \Theta^J_S) \, ,
\label{thetaLR}
\end{equation}
where corresponding BRST transformations have as leading order the gauge-invariant pure spinors $(\Lambda_{L})_{I}^{R}$ and $(\Lambda_{R})_{I}^{R}$, and then fermionic corrections.

\subsubsection{Half-BPS unintegrated vertex operator}

Next, we present the unintegrated half-BPS vertex operator constructed in \cite{Fleury:2026xpg}, which uses the $SU(2,2)\times SU(4)$ covariant variables introduced in the previous sub-subsection. It describes a
half-BPS state, labelled by a point of the boundary of $AdS_5$, given by a future-directed null vector $x_i\in\mathbb R^{1,5}$ (so that $x_i\cdot X>0$), a
null polarisation $y_i\in\mathbb C^6$ of the $R$-symmetry, and the weight $k_i$, which is the dimension and the $R$-charge of the dual chiral primary operator
\begin{equation}
\mathcal O_{k}(x_i,y_i)=\tr\big(y_i\cdot\Phi(x_i)\big)^{k} .
\end{equation}
The unintegrated vertex operator for this state has the form
\begin{equation}
V(i) = V(x_i,y_i,k_i)=\frac{(y_i\cdot Y)^{k_i}}{(x_i\cdot X)^{k_i}}\sum_{m=0,2,\ldots,12}V_m (i),
\label{eq:vertex}
\end{equation}
where the prefactor is the product of the bulk-to-boundary propagators on $AdS_5$ and $S^5$, and $V_m$ is a
polynomial in the fermions $\Psi,\Theta$ ($m$ of them), the pure spinors $\Lambda,\Lb$ (two), $X,Y$, and normalised polarisations $x_i/(2\,x_i\cdot X)$ and $y_i/(2\,y_i\cdot Y)$ inside traces of products of $4\times4$ matrices.
The vertex is manifestly homogeneous of degree $k_i$ in $y_i$ and of degree $-k_i$ in $x_i$. It is $Q$-closed and only determined up to the addition of $Q$-exact expressions. 
The coefficients are polynomials in $k_i$ of degree at most $m/2$.
 The bottom component is the dilaton-like operator
\begin{equation}
V_0= \tfrac i2\,\ch{\Lambda_LX\Lambda_RY} ,
\label{eq:V0}
\end{equation}
where $[ \cdot ]$ denotes a trace and matrix products are performed in the obvious way, e.g.\ $\ch{\Lambda_LX\Lambda_RY} = (\Lambda_L)^R_I X_{RS} (\Lambda_R)^S_J Y^{JI}$. For $k=0$ all parts $V_m$ with $m\ge2$ vanish and $V=V_0$, which is $Q$-closed by itself.
The full vertex has 3469 terms in the variables $(\Psi,\Theta,\Lambda,\Lb,X,Y,x_i,y_i)$ (4, 16, 148, 538, 1139, 1116 and 508 terms with $0,2,\ldots,12$ fermions).
In the present paper, we will use \eqref{LambdaLR} and \eqref{thetaLR} to change to the sector variables $(\Theta_L,\Theta_R,\Lambda_L,\Lambda_R,X,Y,x_i,y_i)$, which will lead to a factorised integration measure for the zero modes.
\cite{Fleury:2026xpg} also constructed the corresponding integrated vertex operators for these unintegrated ones, by solving the descent relations. 
With these results at hand we can consider insertions of both unintegrated and integrated vertex operators for half-BPS closed string states on the worldsheet. The next step towards computing scattering amplitudes in $\ads$ is to address the zero-mode integration. This will be done in the next section.

\section{Zero-mode integration}\label{sec:PCOs}

In this section, we define our prescription for the zero-mode integration for the pure spinor 
formalism in $AdS_5 \times S^5$.
At a fixed bulk point $(X,Y)$ (i.e.\ in the local frame defined by $(X,Y)$), $\Lambda_L$ and $\Lambda_R$ are ordinary ten-dimensional pure spinors, and $\Theta_L$, $\Theta_R$ are
the corresponding Weyl spinors of the same chirality (the precise translation will be given in section \ref{sec:translation} below) and the prescription for integrating their zero modes is invariant under $SO(10)$, although the integrand is only $SO(1,4) \times SO(5)$ invariant.
Therefore, we begin by reviewing the known flat space prescriptions, 
written using picture changing operators (PCOs) in the minimal formalism  or equivalently a regulator when using non-minimal variables.
The prescription factorises, so we work in one sector and drop the $L$ or $R$ subscript. 

We will show in subsection \ref{sec:Qcov} below that the action of the BRST operator in AdS produces an additional term compared to flat space
\bea
{}&\text{flat space} && \ads\\
{}&Q(C\cdot \theta) = C \cdot \lambda \qquad &\to \qquad &Q(C\cdot \theta) = C \cdot (\lambda + f)\,,\\
{}&Q(\bar\lambda \cdot \theta) = \bar\lambda \cdot \lambda \qquad &\to \qquad &Q(\bar\lambda\cdot \theta) = \bar\lambda \cdot (\lambda + f)\,,
\eea{Qtheta_AdS}
where $C_\alpha$ and $\bar\lambda_\alpha$ are extra spinor variables that appear in the picture changing operators or in the non-minimal formalism.
We will see that the extra term $f$ leads to corrections to the usual flat space zero mode measure. We will show that both the minimal formalism with picture changing operators and the non-minimal formalism with its regulator define a linear functional, both with the same properties. In appendix \ref{app:prescription} we show that these properties uniquely fix the functional and derive a differential operator that satisfies the properties. In subsection \ref{sec:translation} we explain how to translate this prescription to $AdS_5 \times S^5$ variables, and subsection \ref{sec:prescription} states the final prescription.

\subsection{Minimal formalism with picture changing operators}\label{sec:minimal}

The pure spinor variable $\lambda^{\alpha}$ and 
its conjugate $w_{\alpha}$ 
are bosonic. 
As in the $(\beta, \gamma)$ system 
of the RNS formalism, they require the introduction of delta functions 
for defining their zero-mode path integration. For the $(\beta, \gamma)$ system it is convenient to bosonize the fields as follows 
\begin{equation}
\beta = \partial \xi e^{- \phi} \, , \quad \quad 
\gamma = \eta e^{\phi}
\, , 
\end{equation}
in terms of which the delta function has the simple 
expression
\begin{equation}
\delta(\gamma)= e^{-\phi} \, .
\end{equation}
This operator is called the picture-lowering operator.
It is possible to formally 
bosonise the pure spinor ghosts in a similar way by making the identification $(w, \lambda) \to (\beta, \gamma)$. However, 
the $\lambda^{\alpha}$ is not a chiral spinor since it is constrained and 
the system is a curved $(\beta, \gamma)$ system. 
Nevertheless, 
it is possible to absorb the eleven zero-modes of $\lambda^{\alpha}$ in the minimal formalism   
by using picture changing operators 
\cite{Berkovits:2005bt,Berkovits:2004px,Craps_2005,Hoogeveen_2007}. In order to construct 
these operators it is necessary to introduce constant spacetime spinors 
(this is always the case when the bosonic ghosts are not Lorentz scalars) or tensors. This in principle breaks the Lorentz covariance, however, as discussed below, it is possible to recover the full $SO(10)$ Lorentz symmetry by averaging over all the possible such tensors \cite{Berkovits_2009} or defining the operators using multi-variable residue theorems \cite{Bedoya:2011kv,GomezZuniga:2011tla}. 
A proposed family of flat space PCOs is
\begin{equation}
Y_{C} = C_{\alpha} \theta^{\alpha} \delta( Q(C_{\beta}  \theta^{\beta}) ) \, , 
\label{flatspacep}
\end{equation}  
with $C_{\alpha}$ a constant spinor labelling the PCO. In flat space
\beq
Q(C \cdot \theta) = C \cdot \lambda.
\label{deltaarg_flat}
\eeq 
The computation of an open string amplitude will involve eleven PCOs. The computation of a closed string 
amplitude will involve twenty-two 
PCOs, since there are two independent pure spinors $\lambda_L$, $\lambda_R$.   
The PCO
$Y_{C}$ has several important properties. It is $Q$-closed in a distributional sense 
\begin{equation}
Q \cdot Y_{C} = (C \lambda) \delta(C \lambda) \, .
\end{equation}
Its derivative is $Q$-exact, 
\begin{equation}
\partial Y_{C} = (C \partial \theta) \delta(C \lambda) + (C \theta) 
(C \partial \lambda) \partial \delta(C \lambda) = Q \cdot [(C \partial \theta)(C \theta)  
\partial \delta(C \lambda)] \, . 
\end{equation}
In addition, both its supersymmetry and Lorentz variations are $Q$-exact.  
The prescription for $N$-point tree-level amplitudes for open-strings 
is then 
\begin{equation}
\mathcal{A}_N = \left\langle V_{1}(z_1)   V_{2}(z_2)   V_{3}(z_3)
\int dz_4 U_4(z_4) \ldots  \int dz_N U_N(z_N) 
\right\rangle^\text{minimal} \, , 
\end{equation}
where $V_i$ and $U_i$ are unintegrated and integrated vertex operators respectively.
The $\langle . \rangle^\text{minimal}$ includes integration over the zero-modes of the sixteen $\theta$'s and the pure spinors after all the non-zero modes have been eliminated by their OPE's.
This includes the non-zero modes from the picture changing operators because we chose their positions to be at $z_\ell = \infty$ so that their OPE's don't contribute  
\begin{equation}
\langle W(\lambda,\theta) \rangle^\text{minimal}=
\int [DC] d^{16}\theta[D\lambda] \;W(\lambda,\theta) \prod_{\ell=1}^{11}  Y_{C_\ell}   , \qquad [DC] = \prod_{\ell=1}^{11}  [d C_\ell] .
\label{eq:A}
\end{equation}
When there are only zero-modes left, we have to perform the integrations
\begin{equation}
\begin{aligned}
\mathcal{A} & = 
\int [DC] d^{16} \theta [D\lambda]  \lambda^{\alpha} \lambda^{\beta} \lambda^{\gamma} 
f_{\alpha \beta \gamma}(\theta) (C_1 \theta) \ldots 
 (C_{11} \theta) \delta(C_1 \lambda)  \ldots \delta(C_{11} \lambda) \\
 & = \int [DC] d^{16} \theta \, (\epsilon T)^{\alpha \beta \gamma}_{ \rho_1 \ldots \rho_{11}} 
 \, d \lambda^{\rho_1} \ldots d \lambda^{\rho_{11}}  f_{\alpha \beta \gamma}(\theta) 
 (C_1 \theta) \ldots 
 (C_{11} \theta) \delta(C_1 \lambda)  \ldots \delta(C_{11} \lambda) \, , 
\label{flatzeromodes}
\end{aligned} 
\end{equation}
where the tensor $T$ of \eqref{TheTupindices} contracted with $\epsilon$ is given by  
\begin{equation}
(\epsilon T)^{\alpha \beta \gamma}_{\beta_1 \ldots \beta_{11}}
\propto \, \epsilon_{\beta_1 \ldots \beta_{16}} \gamma_m^{\alpha \beta_{12}} 
 \gamma_n^{\beta \beta_{13}}  \gamma_p^{\gamma \beta_{14}} (\gamma^{mnp})^{\beta_{15} \beta_{16}} \, . 
 \label{epsT}
\end{equation}
As explained in subsection \ref{sec: Stress-energy tensor} the measure 
$[D \lambda]$ carries ghost number +8 
and together with the extra three $\lambda$'s  
the ghost number is +11 
which is the number 
of independent components of a pure spinor giving the differentials in the second line of \eqref{flatzeromodes}.  
The tensor above has the property of being completely antisymmetric in its lower indices and symmetric and traceless in its upper indices. 
This tensor also has the following 
integral representation  
\beq
\int [DC] [D \lambda]
\lambda^\alpha \lambda^\beta \lambda^\gamma
\prod_{\ell=1}^{11} (C_\ell)_{\beta_\ell} 
\delta(C_1 \lambda) \cdots \delta(C_{11} \lambda)
\propto (\epsilon T)^{\alpha \beta \gamma}_{\beta_1 \ldots \beta_{11}} \, , 
\label{PCO_integral}
\eeq
which follows from Lorentz invariance.
Performing the integration directly 
is complicated because of the Jacobian 
for changing the arguments of the $\delta$ functions 
is involved. 
Recently integrations over orthogonal Grassmannians were considered, see for example
\cite{orthogonalgrassmannian}. This construction 
could be adapted to formally define the 
$C_\ell$ integrals. 

It was demonstrated in \cite{Hoogeveen_2010} that the integration over the $C$'s  in \eqref{flatzeromodes} is necessary.
Otherwise, the zero-mode integration 
is singular enough 
that neither $Q$-exact states may decouple nor distributional identities may hold.
Another solution which works for fixed independent $C$'s having an eleven dimensional tangent space everywhere when contracted with the $\lambda$'s is by using multi-variable residue theorems and trading the delta functions for poles and contours, see \cite{griffiths1978principles, Bedoya:2011kv,GomezZuniga:2011tla}.
By contracting the $\beta_\ell$ indices in \eqref{PCO_integral} with an integral over 16 $\theta$'s, which evaluates to the epsilon tensor, one shows that the result of the amplitude using the prescription just described is equivalent to the one described in 
subsection \ref{treelevelflat}, see 
also \cite{Mafra:2008gkx} for details. 

Let us now consider what happens in AdS, when \eqref{deltaarg_flat} is replaced by \eqref{Qtheta_AdS}. In this case the PCO \eqref{flatspacep} becomes
\begin{equation}
Y_C=(C\cdot\theta)\,\delta\big(Q(C\cdot\theta)\big)
=(C\cdot\theta)\,\delta\big(C\cdot(\lambda+\fc)\big).
\label{eq:Y}
\end{equation}
For each PCO we can Taylor expand the arguments around $C_\ell\cdot\lambda$
\bea
\prod_{\ell=1}^{11}\delta\big(C_\ell\cdot(\lambda+\fc)\big)
={}&\sum_{a_1,\ldots,a_{11}\ge0}\prod_\ell\frac{(C_\ell\cdot\fc)^{a_\ell}}{a_\ell!}\,\delta^{(a_\ell)}(C_\ell\cdot\lambda)\\
={}&\sum_{j\ge0}\frac1{j!}\,\fc^{\mu_1}\cdots\fc^{\mu_j}\,\partial_{\mu_1}\cdots\partial_{\mu_j}\prod_\ell \delta(C_\ell\cdot\lambda) ,
\eea{eq:taylor}
so that the integral can be written
\beq
\langle W(\lambda,\theta) \rangle^\text{minimal}=
\sum_{j\ge0}\frac1{j!} F_j^\text{minimal}[\fc^{\mu_1}\cdots\fc^{\mu_j} W(\lambda,\theta)]
\eeq
in terms of the functional
\beq
F_j^\text{minimal}[h^{\mu_1\ldots\mu_j}] = 
\int [DC] d^{16}\theta [D\lambda]\,
h^{\mu_1\ldots\mu_j}(\lambda,\theta)\,\prod_\ell (C_\ell\cdot\theta)\;\partial_{\mu_1}\cdots\partial_{\mu_j}\prod_\ell \delta(C_\ell\cdot\lambda) .
\label{Fminimal}
\eeq
As the delta function is homogeneous of degree $-1$, we have that
\beq
F_j^\text{minimal}[\lambda^{(\mu_1} H^{\mu_2\ldots\mu_j)}]
= - (10+j)  F_{j-1}^\text{minimal}[H^{\mu_1\ldots\mu_{j-1}}].
\label{eq:m_scaling}
\eeq

\subsection{Non-minimal formalism with regulator}\label{sec:non-minimal}

The minimal pure spinor formalism does not contain the 
complex conjugate of the pure spinor $\lambda^{\alpha}$ among its worldsheet fields. 
In 10d the complex conjugate of a chiral spinor is an antichiral spinor
\begin{equation}
(\lambda^{\alpha})^{\dag} = \bar{\lambda}_{\alpha} \, .  
\end{equation} 
It is possible to add a non-minimal set of variables which includes the bosonic spinor $\bar{\lambda}_{\alpha}$ 
together with its conjugate $\bar{\omega}^{\alpha}$ and the fermionic spinor pair $(r_{\alpha},s^{\alpha})$. We will briefly review the non-minimal formalism because our prescription for the zero-mode integration in $\ads$ admits an equivalent description 
in terms of these variables. 
The constraints of the new variables are  
\begin{equation}
\bar{\lambda}_{\alpha} \gamma_m^{\alpha \beta} \bar{\lambda}_{\beta} =0 \, , \quad \quad 
\bar{\lambda}_{\alpha} \gamma_m^{\alpha \beta} r_{\beta} =0 \, ,
\end{equation}
and the gauge transformations are 
\begin{equation}
\delta \bar{\omega}^{\alpha} =
\bar{\phi}^m (\gamma_m \bar{\lambda})^{\alpha} 
- \phi^m (\gamma_m r)^{\alpha} \, , \quad 
\delta s^{\alpha} = 
\phi^m (\gamma_m \bar{\lambda})^{\alpha} \, , 
\end{equation}
for any $\bar{\phi}^m$ and 
$\phi^m$. 
The BRST operator receives an additional contribution 
\begin{equation}
Q_{\rm{minimal}} = \int dz  
\lambda^{\alpha} d_{\alpha} \quad \rightarrow \quad 
Q_{\rm{non-minimal}}
=\int dz  
(\lambda^{\alpha} d_{\alpha} 
+ \bar{\omega}^{\alpha} r_{\alpha}) \, . 
\end{equation}
Importantly, this extra term does not change the cohomology of the original BRST operator because of the quartet argument \cite{Berkovits:2005bt}.  
The amplitude prescription \cite{Berkovits:2005bt,Berkovits:2006vi,GomezZuniga:2011tla} that gives the same result for the amplitudes in the minimal formalism is
\beq
\langle W(\lambda,\theta) \rangle^\text{non-minimal} =\int[d\lambda]\,[d\lb]\,[dr]\,d^{16}\theta\;\mathcal{N}\;W(\lambda,\theta) ,
\label{eq:momdef}
\eeq
with the regulator for the non-compact integrations 
\begin{equation}
\cN=e^{-Q(\lb\cdot\theta)} .
\label{eq:Nflat}
\end{equation}
In flat space it is
\beq
e^{-Q(\lb\cdot\theta)}=e^{-\lb\cdot\lambda-r\cdot\theta}.
\label{regulator_flat}
\eeq
Note that $\cN$ is BRST closed by
construction, and since $\cN=1+Q(\ldots)$, inserting it does not change the value of BRST-closed integrands. The integrations are performed using the rules 
\begin{align}
[d\lambda]\lambda^\beta \lambda^\gamma \lambda^\delta &= (\epsilon T)^{\beta\gamma\delta}_{\alpha_1 \dots \alpha_{11}} d\lambda^{\alpha_1} \dots d\lambda^{\alpha_{11}} \, , \\
[d\bar{\lambda}]\bar{\lambda}_\beta \bar{\lambda}_\gamma \bar{\lambda}_\delta &= (\epsilon T^{-1})^{\alpha_1 \dots \alpha_{11}}_{\beta\gamma\delta} d\bar{\lambda}_{\alpha_1} \dots d\bar{\lambda}_{\alpha_{11}}\, , 
\end{align} 
and 
\begin{equation}
  [dr] = (\epsilon T)^{\beta\gamma\delta}_{\alpha_1\ldots\alpha_{11}}\,
  \overline{\lambda}_\beta\, \overline{\lambda}_\gamma\, \overline{\lambda}_\delta
  \left(\frac{\partial}{\partial r_{\alpha_1}}\right)\ldots
  \left(\frac{\partial}{\partial r_{\alpha_{11}}}\right) \, . 
\end{equation}

In AdS, again using \eqref{Qtheta_AdS}, we have to replace \eqref{regulator_flat} by
$e^{-Q(\lb\cdot \theta)} = e^{-\lb\cdot\lambda-r\cdot\theta}\,e^{-\lb\cdot f}$, so that we can write
\beq
\langle W(\lambda,\theta) \rangle^\text{non-minimal}=
\sum_{j\ge0}\frac1{j!} F_{j,1}^\text{non-minimal}[\fc^{\mu_1}\cdots\fc^{\mu_j} W(\lambda,\theta)]
\eeq
in terms of the functional
\begin{equation}
F_{j,\sigma}^\text{non-minimal}[h^{\mu_1\ldots\mu_j}]
=(-1)^j \int[d\lambda]\,[d\lb]\,[dr]\,d^{16}\theta\;e^{-\sigma \lb\cdot\lambda-r\cdot\theta}\;
\lb_{\mu_1}\cdots\lb_{\mu_j}\,h^{\mu_1\ldots\mu_j}(\lambda,\theta) .
\label{Fnonminimal}
\end{equation}
Here we defined the parameter $\sigma$ to show that 
\bea
{}&F_{j,1}^\text{non-minimal}[\lambda^{(\mu_1} H^{\mu_2\ldots\mu_j)}]
= \left[\partial_\sigma F_{j-1,\sigma}^\text{non-minimal}[H^{\mu_1\ldots\mu_{j-1}}]  \right]_{\sigma = 1}\\
={}& \left[\partial_\sigma \sigma^{-10-j} F_{j-1,1}^\text{non-minimal}[H^{\mu_1\ldots\mu_{j-1}}]  \right]_{\sigma = 1}
= - (10+j)  F_{j-1,1}^\text{non-minimal}[H^{\mu_1\ldots\mu_{j-1}}] ,
\eea{eq:nm_scaling}
where we rescaled $\lambda \to \lambda / \sigma$ in the second step and used that $[d\lambda]$ scales with $\lambda^{8}$ and $H^{\mu_1\ldots\mu_{j-1}}$ with $\lambda^{3+(j-1)}$.

\subsection{The functional and change to AdS variables}\label{sec:translation}

The functionals emerging from the minimal \eqref{Fminimal} and non-minimal \eqref{Fnonminimal} formalisms both have to be Lorentz invariant, well defined on the pure spinor cone, and satisfy the scaling relation in \eqref{eq:m_scaling} or \eqref{eq:nm_scaling}. In appendix \ref{app:prescription} we show that this uniquely fixes the functional, and that it can be expressed as
\beq
F_j[h^{\mu_1\ldots\mu_j}] = \kappa_j Z^{(4)}_{\mu_j}\cdots Z^{(j+2)}_{\mu_2}Z^{(j+3)}_{\mu_1}\,h^{\mu_1\ldots\mu_j} \Big|_{\lambda^\alpha \lambda^\beta \lambda^\gamma \theta^{\delta_1} \theta^{\delta_2} \theta^{\delta_3} \theta^{\delta_4} \theta^{\delta_5} \to T^{\alpha \beta \gamma; \delta_1 \ldots \delta_5}}\,,
\label{functional_flat}
\eeq
with $\kappa_j$ given in \eqref{kappadef}. The functional first acts with the differential operator
\begin{equation}
Z^{(d)}[h^\mu]=\partial_\mu h^\mu-\frac{1}{4(d+2)}\lambda^\nu\gamma^m_{\mu\nu}\gamma_m^{\alpha\beta}\partial_\alpha\partial_\beta h^\mu ,
\qquad \partial_\alpha=\frac{\partial}{\partial\lambda^\alpha} ,
\label{eq:Zd_flat}
\end{equation}
followed by the insertion of the tensor $T$ of \eqref{TheTupindices}
\beq
T^{\alpha\beta\gamma;\delta_1\ldots\delta_5}
=\frac{1}{\mathcal N_T}\;\gamma_m^{\alpha[\delta_1}\,\gamma_n^{|\beta|\delta_2}\,\gamma_p^{|\gamma|\delta_3}\,(\gamma^{mnp})^{\delta_4\delta_5]} ,\qquad
\mathcal N_T=2^{8}\cdot7 ,
\label{eq:Tflat}
\eeq
where the normalisation is fixed by \eqref{eq:shortcut_measure_flat}.
Both the functional and the tensor $T$ are written with $10d$ spinor indices and gamma matrices. Our integrands are instead written in terms of the $4\times4$ matrices $(\Lambda_s)^R{}_I$ and
$(\Theta_s)^R{}_I$, $s=L,R$, of (\ref{LambdaLR},\ref{thetaLR}) and of the bulk point $(X,Y)$. In this subsection we translate $T$ and $Z^{(d)}$
to these variables.

To translate we need the gamma matrices and the sums over vector indices in the AdS
variables.
A spinor index is a pair of an $SU(2,2)$ index and an $SU(4)$ index\footnote{We reinstate the sector index $s=L,R$ that we dropped when discussing the measure for one sector.},
\beq
\lambda_s^\alpha\;\to\;(\Lambda_s)^R{}_I ,\qquad \theta_s^\alpha\;\to\;(\Theta_s)^R{}_I ,\qquad
\partial_\alpha\;\to\;\partial^I{}_R=\frac{\partial}{\partial(\Lambda_s)^R{}_I} ,
\eeq
and the ten vector directions $m=(a,a')$ are the tangent directions of $AdS_5$ and $S^5$ at the bulk point. Let
$\sigma^{\mathsf I}_{RS}$ and $\bar\sigma^{\mathsf I\,RS}=\frac12\epsilon^{RSTU}\sigma^{\mathsf I}_{TU}$,
$\mathsf I=1,\ldots,6$, be the six-dimensional Pauli matrices, with
$\sigma^{\mathsf I}\bar\sigma^{\mathsf J}+\sigma^{\mathsf J}\bar\sigma^{\mathsf I}=2\delta^{\mathsf{IJ}}$, so that
$X_{RS}=X_{\mathsf I}\sigma^{\mathsf I}_{RS}$ and $X^{RS}=X_{\mathsf I}\bar\sigma^{\mathsf I\,RS}$ with
$X_{RS}X^{ST}=\delta_R^T$ by \eqref{eq: constraints_charac_x_and_y}. We denote by $\sigma^a$, $a=1,\ldots,5$, the Pauli
matrices of an orthonormal basis of the directions orthogonal to $X$, and by $\sigma^{a'}$ those of the directions
orthogonal to $Y$. The gamma matrices are then 
\bea
\gamma^a_{\alpha\beta}&=\sigma^a_{RS}\,Y^{IJ} ,&\qquad \gamma^{a'}_{\alpha\beta}&=i\,X_{RS}\,\bar\sigma^{a'\,IJ} ,\\
\gamma_a^{\alpha\beta}&=\bar\sigma_a^{RS}\,Y_{IJ} ,&\qquad \gamma_{a'}^{\alpha\beta}&=-i\,X^{RS}\,\sigma_{a'\,IJ} ,
\eea{eq:gammaAdS}
with $\alpha=(R,I)$ and $\beta=(S,J)$. In the frame \eqref{eq: X and Y def} the sixth direction is along $\sigma_6$ or $\bar\sigma_6$ which is rotated to the form used here with \eqref{XYrotation}. They satisfy the Clifford algebra with metric $\delta^{mn}$, and with them the
constraint $\lambda\gamma^m\lambda=0$ is \eqref{eq: Constraints_carach_ps_cov}: at a fixed bulk point $\Lambda_L$ and
$\Lambda_R$ are ten-dimensional pure spinors in the local frame \eqref{eq:gammaAdS}.

The sums over vector indices follow from the completeness relations of the six Pauli matrices, by subtracting the
direction of $X$ and using $\epsilon^{RSTU}=-(X^{RS}X^{TU}+X^{RT}X^{US}+X^{RU}X^{ST})$:
\bea
\sum_{a=1}^5\sigma^a_{RS}\,\bar\sigma_a^{TU}&=2\big(\delta_R^U\delta_S^T-\delta_R^T\delta_S^U\big)-X_{RS}X^{TU} ,\\
\sum_{a=1}^5\bar\sigma_a^{RS}\,\bar\sigma_a^{TU}&=X^{RS}X^{TU}+2\,X^{RT}X^{US}+2\,X^{RU}X^{ST} ,
\eea{eq:complete5}
and the same with $Y$ for the sphere. The second relation also holds with all indices lowered. After the sums no Pauli matrices are left, only $\delta$, $X$ and $Y$. As an
example, with $\gamma=(T,K)$ and $\delta=(U,L)$,
\bea
\gamma_m^{\alpha\beta}\gamma^{m\,\gamma\delta}
&=\Big(\sum_a\bar\sigma_a^{RS}\bar\sigma_a^{TU}\Big)Y_{IJ}Y_{KL}
-X^{RS}X^{TU}\Big(\sum_{a'}\sigma_{a'\,IJ}\,\sigma_{a'\,KL}\Big)\\
&=2\big(X^{RT}X^{US}+X^{RU}X^{ST}\big)Y_{IJ}Y_{KL}-2\,X^{RS}X^{TU}\big(Y_{IK}Y_{LJ}+Y_{IL}Y_{JK}\big) ,
\eea{eq:gammagamma}
where the first term comes from $AdS_5$ and the second from $S^5$.

The tensor \eqref{eq:Tflat} is translated in the same way. It has eight $SU(2,2)$ indices $R_1,\ldots,R_8$ and eight
$SU(4)$ indices $I_1,\ldots,I_8$, where the slots $1,2,3$ belong to the $\Lambda_s$'s and $4,\ldots,8$ to the
$\Theta_s$'s, and each set can be paired in $105$ ways. The result is
\beq
\vev{(\Lambda_s)^{R_1}{}_{I_1}(\Lambda_s)^{R_2}{}_{I_2}(\Lambda_s)^{R_3}{}_{I_3}\,(\Theta_s)^{R_4}{}_{I_4}\cdots(\Theta_s)^{R_8}{}_{I_8}}
=\frac1{6\mathcal N_T}\sum_{k=1}^{105}\sum_{m=1}^{105}c_{km}\,\mathcal R_k^{R_1\ldots R_8}\,\mathcal J_{m\,I_1\ldots I_8} ,
\label{eq:T2k}
\eeq
where $\mathcal R_k$ are the products of four $X$'s that pair the $SU(2,2)$ indices and $\mathcal J_m$ the products of
four $Y$'s that pair the $SU(4)$ indices. Of the coefficients 2685 are nonzero, with values
$c_{km}\in\{\pm\frac85,\pm\frac{16}5,\pm\frac{24}5,\pm\frac{32}5,\pm8\}$.

The operator $Z^{(d)}[h^\mu]$ in \eqref{eq:Zd_flat}
acts on a function $h^\mu$ with one spinor index, which is now a matrix $h^R{}_I$. With $\mu=(R,I)$, $\nu=(S,J)$,
$\alpha=(T,K)$, $\beta=(U,L)$ and the first line of \eqref{eq:complete5},
\beq
\gamma^m_{\mu\nu}\gamma_m^{\alpha\beta}=
\Big[2\big(\delta_R^U\delta_S^T-\delta_R^T\delta_S^U\big)-X_{RS}X^{TU}\Big]Y^{IJ}Y_{KL}
+X_{RS}X^{TU}\Big[2\big(\delta^I_L\delta^J_K-\delta^I_K\delta^J_L\big)-Y^{IJ}Y_{KL}\Big] .
\label{eq:Kins}
\eeq
In the contraction with the two derivatives, which commute, the two terms with Kronecker deltas in each bracket are
equal. Hence
\beq
\begin{aligned}
Z^{(d)}[h^R{}_I]={}&\partial^I{}_R\,h^R{}_I-\frac{1}{d+2}\,(\Lambda_s)^S{}_J\Big[Y^{IJ}Y_{KL}\,\partial^K{}_S\partial^L{}_R
-X_{RS}X^{TU}\,\partial^I{}_T\partial^J{}_U\\
&\hspace{4.2cm}-\frac12\,X_{RS}X^{TU}\,Y^{IJ}Y_{KL}\,\partial^K{}_T\partial^L{}_U\Big]h^R{}_I .
\end{aligned}
\label{eq:ZAdS}
\eeq

\subsection{BRST invariance of the measure}\label{sec:Qcov}

In this subsection, we compute 
$f_L$ and $f_R$ of \eqref{Qtheta_AdS} and in this way we fix the argument of the delta functions in the $\ads$ picture changing operators, or equivalently, the corrections to the non-minimal regulators. 
To this end
we need to 
fix the 
transformation 
properties of the $C$'s.  
The $C$'s carry the indices 
$C^I_R$ and they contract 
as $C^I_R (\Theta)^R_I$. 
There are two ways of viewing
\eqref{eq:Y}.
The variables $\Theta_R$ and $\Theta_L$ defined in 
\eqref{thetaLR}
have the property that 
their BRST transformations at leading order in the fermionic variables are 
\begin{equation}
\begin{aligned}
Q \cdot (\Theta_R)^R_I = (\Lambda_R)^R_I + \ldots \, , \quad \quad 
Q \cdot (\Theta_L)^R_I = (\Lambda_L)^R_I + \ldots \, . 
\end{aligned} 
\end{equation}
This implies that 
the argument of the delta functions
will contain the following term at lowest order in the fermionic variables, 
\begin{equation}
C^I_R (\Lambda_L)^R_I + \ldots =
C^I_R V_I^{\widetilde{I}} (\lambda_L)_{\widetilde{I}}^{\widetilde{R}} (U^{-1})_{\widetilde{R}}^R + \ldots = 
C^{\widetilde{I}}_{\widetilde{R}}
(\lambda_L)_{\widetilde{I}}^{\widetilde{R}} + \ldots \, . 
\label{Cleadingtermnew}
\end{equation}
Note that the expression above is independent of
$U$ and $V$ if the $C^I_R$'s are of the form 
\begin{equation}
C^I_R = C^{\widetilde{I}}_{\widetilde{R}}
(V^{-1})^I_{\widetilde{I}} (U)^{\widetilde{R}}_R \, ,
\label{Cwithtildes}
\end{equation}
in terms of $C^{\widetilde{I}}_{\widetilde{R}}$ which are independent of
$U$ and $V$.
This is required 
for \eqref{Cleadingtermnew}
to have the expected
argument of the flat space picture changing operators of \eqref{flatspacep}. 
This also explains why the tensor of \eqref{eq:T2k} 
depends on $(X,Y)$ as this follows from these additional factors of $U$ and $V$.
In defining the $C$'s as
in \eqref{Cwithtildes}, it is clear that $Q \cdot C$ is different from zero because it acts on the 
$V^{-1}$ and $U$ factors. 
One verifies that 
\begin{equation}
Q \cdot C=F\cdot C \, , 
\label{rotationC}
\end{equation}
where $F$ is the same 
$SU(2,2) \times SU(4)$ rotation 
appearing on \eqref{Qbulkpoints}, 
\begin{equation}
F := [\Theta, \Lambda] \, , \quad \quad
[\Theta, \Lambda]
= \Theta^I_R \Lambda^S_J
\left(\delta^R_S M^J_I + \delta^J_I M^R_S\right) \, . 
\end{equation} 

Equivalently, we can work directly with the $C$'s 
with the untilded indices $C^I_R$. 
In this case,
all the bosonic variables with indices $R$ and $I$ rotate with the same $F$ above, which includes the bulk points. 
Moreover, the spinors $\Lambda_L$ and $\Lambda_R$ satisfy the constraints given in \eqref{eq: Constraints_carach_ps_cov} which now depend on $(X,Y)$. 
Since we are integrating over $X$ and $Y$ at the end of the computation, the $C$-average, or analogously the $\lb$ integral, is defined with a measure that is
invariant \emph{in the local frame}
at $(X,Y)$. 
Thus, it is convenient to define the operator 
$\Qc=Q-Q_F$ which keeps the frame fixed. 
The operator $Q_F$ is the rotation by $F$ acting on every object according to its indices. The main property of $Q_{\rm{cov}}$ is 
\begin{equation}
Q \cdot(C \cdot \Theta_s) =
C \cdot (Q_{\rm{cov}} \cdot \Theta_s) \, . 
\end{equation}
It remains to compute 
$Q_{\rm{cov}} \cdot \Theta_s$.
Note that $Q_F$ agrees with $Q$ on $X$, $Y$,
$\Lambda$ and $\Lb$, which transform only by the compensating rotation, and
\begin{equation}
Q_F \cdot \Psi=-\Psi\Theta\Lambda-\Lambda\Theta\Psi \, , \quad  Q_F \cdot \Theta=-2\,\Theta\Lambda\Theta. 
\end{equation}
So $\Qc$ is the pure right translation
of the fermionic coset,
\begin{equation}
\begin{aligned}
&\Qc\Psi=\Lambda+\Psi\Theta\Lambda+\Lambda\Theta\Psi ,\qquad 
\Qc\Theta=\Lb+\Theta\Lambda\Theta \, , \\ 
&\Qc X=\Qc Y=\Qc\Lambda=\Qc\Lb=0 .
\end{aligned}
\end{equation}
Importantly, acting with $Q_{\rm{cov}}$ on the fermionic variables 
\eqref{thetaLR} gives 
the desired $f_L$ and $f_R$,
\begin{equation}
\boxed{\;\begin{aligned}
&\Qc \cdot \Theta_L=\Lambda_L+\fc_L ,&\qquad &\fc_L=\Psi\Theta\Lambda+\Lambda\Theta\Psi+X \Theta\Lambda\Theta Y ,\\
&\Qc \cdot \Theta_R=\Lambda_R+\fc_R ,&\qquad &\fc_R=i\big(\Psi\Theta\Lambda+\Lambda\Theta\Psi- X \Theta\Lambda\Theta Y \big) .
\end{aligned}\;}
\label{eq:fcov}
\end{equation}
They 
can be written in terms of
$\Lambda_L$, $\Lambda_R$, $\Theta_L$ and $\Theta_R$ by using the inverse of the maps \eqref{LambdaLR} and \eqref{thetaLR}.

This discussion 
can be extended 
to the non-minimal variables. 
The computation of the $f$
in this case is motivated by keeping the local frame  
\begin{equation}
Q \cdot \big(\lb\gamma^m(X,Y)\lb\big)=0 \, , 
\end{equation}
and this forces the following transformation  
\begin{equation}
Q \cdot \lb=r+F\cdot\lb \, , 
\end{equation}
and one defines $Q \cdot r$ such
that $Q^2 \cdot \lb=0$. Collecting the relevant transformations 
\begin{equation}
Q \cdot (\lb\cdot\theta_L)=r\cdot\theta_L+\lb\cdot\Qc \cdot \theta_L ,
\end{equation}
which gives the same previously computed $f_L$ and $f_R$. 

The full prescription for computing string amplitudes will be given in the next subsection. 
The first tests we have performed are the decoupling of several $Q$-exact sample states. 
In flat-space the proof of decoupling of $Q$-exact states can be found for example in \cite{Berkovits_2009}.
It is possible to adapt the same arguments here. In other words,
as in flat space our picture changing operators are $Q$-exact in a distributional sense, as
we are using the same pure spinor measure.

\subsection{The prescription}\label{sec:prescription}

The zero modes are the bulk point $(X,Y)$, the fermions and the pure spinors. At a fixed bulk point the integral over
the fermions and pure spinors of each sector is the functional \eqref{functional_flat}, with $f$ given by $\fc_L$ and
$\fc_R$ of \eqref{eq:fcov}. For an integrand $W$ written in terms of $\Lambda_{L,R}$ and $\Theta_{L,R}$\footnote{The integration over $dX$ and $dY$ absorbs the superdeterminant of the supervielbein of the Haar supermeasure, see the Appendix \ref{Haarsupermeasure} for details.},
\beq
\boxed{\;\begin{gathered}
\vev{W}=\int_{AdS_5}\!dX\int_{S^5}\!dY\sum_{j_L,j_R\ge0}\frac{\kappa_{j_L}\kappa_{j_R}}{j_L!\,j_R!}\;
Z^L_{j_L}Z^R_{j_R}\Big[W\,(\fc_L)^{j_L}(\fc_R)^{j_R}\Big]
\bigg|_{\Lambda_s^3\Theta_s^5\to T_s} ,\\[4pt]
\kappa_j=\frac{(-1)^j\,336}{(j+6)(j+7)(j+8)} ,\qquad
Z^s_j=Z^{(4)}_{\mu_j}\cdots Z^{(j+2)}_{\mu_2}Z^{(j+3)}_{\mu_1} ,
\end{gathered}\;}
\label{eq:presc}
\eeq
where $Z^s_j$ is a product of $j$ operators \eqref{eq:ZAdS} of sector $s=L,R$, which contract the indices of the $j$
factors of $\fc_s$, and in the last step the terms with three $\Lambda_s$ and five $\Theta_s$ in each sector are
replaced by the tensor \eqref{eq:T2k}, while all other terms are set to zero. Since the fermions anticommute, they have to be brought into the order $\Theta_L^5\Theta_R^5$ before this replacement. The term $j_L=j_R=0$ is the flat space prescription in the local frame. The term
$(j_L,j_R)$ gets contributions from the part of $W$ with $10-2(j_L+j_R)$ fermions, so the sum ends at $j_L+j_R=5$.
While the integration prescription factorises, both the vertex operators which are integrated and each $f_s$ mix the variables from the $L$ and $R$ sectors.

We tested the prescription with BRST-exact integrands $Q(E)$, where $E$ is invariant under $SU(2,2)\times SU(4)$. Their
integral over the fermions and pure spinors must vanish at each bulk point. If $E$ has nine fermions,
$Q(E)=E_{10}+E_8$ has parts with ten and eight fermions and
\beq
\vev{Q(E)}=E_{10}\Big|_{\Lambda_s^3\Theta_s^5\to T_s}+\kappa_1\sum_{s=L,R}Z^s_1[E_8\,\fc_s]\Big|_{\Lambda_s^3\Theta_s^5\to T_s} .
\eeq
The first term is the flat space prescription and does not
vanish by itself. For example, for
$E=\ch{\Theta\Lambda}\ch{\Lb X\Theta Y}\ch{\Psi\Theta\Psi\Lb}\ch{\Psi\Theta\Lambda\Lb}\ch{\Psi\Theta}$ the two terms
are, in units of $\mathcal N_T^{-2}$,
\beq
\vev{Q(E)}=-30-\tfrac23\cdot(-45)=0 .
\eeq
If $E$ has seven fermions there is no flat space term, and the terms of first order cancel against those of second
order, which tests the terms with $\fc_s\fc_s$ and with $\fc_L\fc_R$. With five
fermions the second and third orders cancel. Finally we checked that $\vev{V(1)\,V(2)\,Q(E)}$ with two half-BPS vertex operators
vanishes after the integration over the bulk point, which tests the vertex operators, the functional and the bulk integrals of section \ref{sec:3pt-integrals} all together.

\section{The two-point amplitude}\label{sec:2pt}

The computation of two-point string amplitudes needs a regularization. 
For closed strings, sphere string amplitudes are integrals over the positions of the vertex operators modulo the conformal Killing
group $SL(2,\mathbb C)$. 
In a two-point amplitude, once the two insertion points on the worldsheet have been fixed,
the residual symmetry
includes a non-compact dilatation direction, and hence has infinite
volume. 
In \cite{Erbin:2019uiz}
Erbin, Maldacena and Skliros pointed out  
that one of the delta
functions of momentum conservation is redundant, $\delta(k_1^0+k_2^0)=\delta(0)$, and this volume of time translations
cancels the volume of the residual group.
After properly taking into account all these effects the correctly normalised 
two-point function can be computed 
\cite{Erbin:2019uiz,Seki:2019ycg}.

In a BRST formalism without reparametrisation ghosts, such as the pure spinor superstring, this cancellation or regularization requires two
ingredients \cite{Kashyap:2020tgx,Kashyap:2022mpr,Kishimoto:2024aig,Kishimoto:2024yuw,Kashyap:2025rvf}: a \emph{mostly BRST
exact operator} for the non-compact part and, for closed strings, a vertex of \emph{ghost number three} for the compact
part. We first recall the computations and the definitions of these operators in flat space and then consider $\ads$.

\subsection{Flat space}

\paragraph{Open superstring: 
the mostly BRST exact operator}

The 
purpose of the \emph{mostly BRST exact operator} 
is to regularize the worldsheet correlator for the two-point function 
computation.
This regulator for the pure spinor formalism is given by \cite{Kashyap:2020tgx}
\begin{equation}
\mathcal{V}^{+1} \equiv  \frac12\int_{-\infty}^{\infty} dq\, (\lambda\gamma^{0}\theta)\,e^{iqX^{0}} \,.
\label{MostlyBRST}
\end{equation}
The name, \emph{mostly BRST exact operator}, stems from the fact that the integrand is BRST exact for all values of  $q\neq 0$, which follows from
\begin{equation}
 Q\left(\frac{e^{iqX^0}}{iq}\right) = \frac12(\lambda\gamma^0\theta)\, e^{iq X^{0}}\,.
\label{usualMostly}
\end{equation}
More generally, let $\phi$ be a target-space function that is shifted by a constant under the residual non-compact symmetry mentioned above. The mostly BRST exact operator is \cite{Kashyap:2020tgx,Kashyap:2022mpr}
\begin{equation}
\cV^{+1}=\int_{-\infty}^{\infty}\frac{dq}{iq}\,Q\big(e^{iq\phi}\big)=\int dq\,(Q\phi)\,e^{iq\phi}=2\pi\,\delta(\phi)\,Q\phi \, , 
\label{eq:mbrst}
\end{equation}
and the previous expression \eqref{MostlyBRST} is recovered for the particular choice
$\phi = X^0$. 
The last form of \eqref{eq:mbrst} shows that $\cV$ is a Faddeev--Popov insertion:
$\delta(\phi)$ restricts the target space to a gauge slice of the residual symmetry, and $Q\phi$, which has ghost number one,
is the BRST completion of the Jacobian. For open strings $\vev{V^{+1}V^{+1}}$ would naively vanish, since the ghost number cannot be saturated. Instead, one should consider the 'regulated' two point function $\vev{V^{+1}V^{+1}\cV^{+1}}$, where now the ghost number is saturated. The
gluon two-point function was computed in this way in the pure spinor formalism \cite{Kashyap:2020tgx}.
Let us review the computation. 
For concreteness consider the photon's two-point function. The two vertex operators are given by
\begin{equation}
V^{+1}_{i}= c_1 (\lambda\gamma^{a}\theta)A_{a}^{(i)}(x) + c_2(\lambda\gamma^{a}\theta)(\theta\gamma_{abc}\theta)F^{(i)bc}(x) + \ldots\,,\qquad i=1,2 \,,  
\end{equation}
where $F_{ab}=\partial_aA_b-\partial_bA_a$, the fields are evaluated at the insertion points $z_i$, and $c_1=\frac12$, $c_2=-\frac1{32}$ for the superfield \eqref{TheAvertex}. The proposal is that the two-point function amplitude is given by
\begin{equation}
\mathcal{A} = \langle V_{1}^{+1} V_{2}^{+1}\mathcal{V}_{3}^{+1}\rangle \,,
\end{equation}
where 
$\mathcal{V}_{3}^{+1}$ is the regulator given in 
\eqref{MostlyBRST} inserted at $z_3$. The terms with three pure spinors and five $\theta$'s have the potential of one vertex operator and the field strength of the other. Since $(\lambda\gamma^a\theta)$ is anticommuting, integrating over the zero modes of $\theta$ and $\lambda$ gives
\begin{equation}
\mathcal{A} =  -\frac{c_1c_2 }{2}\int_{-\infty}^{\infty} dq \left\langle e^{iqx^{0}(z_3)}\left(A_{c}^{(1)}F_{ab}^{(2)} - A_{c}^{(2)}F_{ab}^{(1)}\right)\right\rangle
\langle (\lambda\gamma^{c}\theta)(\lambda\gamma^{0}\theta)(\lambda\gamma_{e}\theta)(\theta\gamma^{eab}\theta)\rangle\,,
\end{equation}
where 
\begin{equation}
\langle(\lambda\gamma^{c}\theta)(\lambda\gamma^{d}\theta)(\lambda\gamma_{e}\theta)(\theta\gamma^{eab}\theta)\rangle  = 64\,\delta^{[c}_{a}\delta^{d]}_{b}=32\big(\delta^c_a\delta^d_b-\delta^d_a\delta^c_b\big)\,. 
\end{equation}
Contracting the above deltas, we obtain
\begin{equation}
\mathcal{A} = - 32\,c_1c_2\int_{-\infty}^{\infty} dq \left\langle e^{iqx^{0}(z_3)}\left(A^{(1)}_{a}F^{(2)a0} - A^{(2)}_{a}F^{(1)a0}\right)\right\rangle \,.
\end{equation}
In order to understand the role of $e^{iqx^{0}(z_3)}$ in the regularization, let us consider plane waves:
\begin{equation}
A^{(i)}_{a}(x) = e^{(i)}_{a}\, e^{ik^{(i)}\cdot x}\implies F^{(i)}_{ab}(x)=i\big(k^{(i)}_{a}e^{(i)}_{b}-k^{(i)}_{b}e^{(i)}_{a}\big)e^{ik^{(i)}\cdot x}\,.    
\end{equation}
For $k^{(2)}=-k^{(1)}$ and transverse polarisations, the zero-mode integral with the insertion of $(\lambda\gamma^m\theta)$ is 
\begin{equation}
\vev{(\lambda\gamma^m\theta)\,V_1^{+1}V_2^{+1}}=-128i\,c_1c_2\,k^{(1)m}\,\big(e^{(1)}\!\cdot e^{(2)}\big)=2i\,k^{(1)m}\,\big(e^{(1)}\!\cdot e^{(2)}\big) ,
\label{eq:Fm}
\end{equation}
without the plane wave factors, so that the two-point scattering amplitude is 
\begin{equation}
\mathcal{A} 
 = -64i\,c_1c_2\,k^{(1)0}\,\big(e^{(1)}\!\cdot e^{(2)}\big)
 \int_{-\infty}^{\infty} dq \left\langle e^{iqx^{0}(z_3)}e^{ik^{(1)}\cdot x(z_{1})}e^{ik^{(2)}\cdot x(z_{2})}\right\rangle \,,    
\end{equation}
where
\begin{equation}
\begin{aligned}
{}& \left\langle e^{iqx^{0}(z_3)}e^{ik^{(1)}\cdot x(z_{1})}e^{ik^{(2)}\cdot x(z_{2})}\right\rangle \\
&=  \delta^{d}(k^{(1)}+k^{(2)}+q)\, |z_3 - z_1|^{2\alpha' q\cdot k^{(1)}} |z_3-z_2|^{2\alpha' q\cdot k^{(2)}} |z_1-z_2|^{2\alpha' k^{(1)}\cdot k^{(2)}}\,,  \end{aligned} 
\end{equation}
and $q$ only has a time component. Because of the conservation of the spatial momenta, $\vec{k}^{(1)}+\vec{k}^{(2)} = 0$, imposed by $d-1$ of the Dirac deltas, the on-shell constraints coming from BRST invariance of the vertex operators imply that $k^{(1)0} + k^{(2)0} = 0$, so that we end up with
\begin{equation*}
\left\langle e^{iqx^{0}(z_3)}e^{ik^{(1)}\cdot x(z_{1})}e^{ik^{(2)}\cdot x(z_{2})}\right\rangle  =  \delta^{d-1}(\vec{k}^{(1)}+\vec{k}^{(2)})\,\delta(q)\, |z_3 - z_1|^{2\alpha' q\cdot k^{(1)}} |z_3-z_2|^{2\alpha' q\cdot k^{(2)}} \,.   
\end{equation*}
Plugging this back to the amplitude we obtain
\begin{equation}
\mathcal{A} =i\,k^{(1)0}\,\big(e^{(1)}\!\cdot e^{(2)}\big)\, \delta^{d-1}(\vec{k}^{(1)}+\vec{k}^{(2)})\,.   
\label{eq:2pt_open_flat}
\end{equation}
The role of the $e^{iqx^0(z_3)}$, and the integration over $q$, is to absorb one redundant delta function from the intersection between the momentum conservation and on-shell conditions on physical vertex operators.

\paragraph{Closed superstrings}

In \cite{Seki:2021ivm}, it was noted that generalizing the mostly BRST exact operator to the closed string is not straightforward due to a ghost number mismatch. Each vertex operator contributes ghost number $+2$ while the mostly BRST exact operator has ghost number $+1$. The total ghost number is $+5$, missing one unit to saturate the anomaly. This issue was resolved for the bosonic string in \cite{Kishimoto:2024aig}, by replacing one of the vertex operators of ghost number $+2$ with one of ghost number $+3$, constructed in \cite{Kishimoto:2024yuw}. If $V^{+2}$ denotes the bosonic vertex operator of ghost number $+2$, the vertex operator of ghost number $+3$ proposed in \cite{Kishimoto:2024aig} and \cite{Kishimoto:2024yuw} is
\begin{equation}
V^{+3} = \frac{1}{4\pi}(\overline\partial \overline c - \partial c) V^{+2}  \,. 
\label{V3c}
\end{equation}
In the pure spinor formalism the $b$-ghost is not fundamental and it 
is constructed in terms of the other fields \cite{Jusinskas:2013sha}. 
Also it is not clear how to properly define 
a $c$ ghost \cite{Jusinskas:2014jtq}.
Thus the construction 
\eqref{V3c}
is difficult to pursue in the pure spinor formalism. Nevertheless, 
it is possible to compute the ghost number three cohomology directly.
This was done in \cite{Mikhailov:2014qka} where it was also shown that they transform 
under supersymmetry as a linearized Type IIB supergravity supermultiplet. They are given by 
\begin{equation}
V^{+3}_a=\frac12\,a_m\big[(\lambda\gamma^m\theta)-(\hat\lambda\gamma^m\hat\theta)\big]\,V^{+2} ,
\label{eq:Wmikh}
\end{equation}
with a constant vector $a_m$. With the plus sign the operator is exact,
$\frac12a_m[(\lambda\gamma^m\theta)+(\hat\lambda\gamma^m\hat\theta)]V^{+2}=Q\big((a\cdot x)V^{+2}\big)$; with the minus sign it is not, and the
obstruction to exactness is the scalar $a\cdot k$, the charge of the state under the translation $a$.

In flat background, we have factorization between left and right movers, so the computation of expectation values between closed vertex operators is essentially a product of left and right computations involving open vertex operators. For example, the vertex operator of ghost number $+2$ factorizes,
\begin{equation}
V^{+2} = V^{+1}\,\widehat V^{+1}\,.    
\end{equation}
The mostly BRST exact operator \eqref{eq:mbrst} with $\phi=x^0$ is now
\begin{equation}
\mathcal{V}^{+1} = \int_{-\infty}^{+\infty} \frac{dq}{iq}\, Q\big(e^{iqx^{0}}\big) = \frac12\int_{-\infty}^{\infty} dq \left[(\lambda\gamma^{0}\theta)+(\hat\lambda\gamma^{0}\hat\theta)\right]e^{iqx^{0}} \,,  
\end{equation}
and the two-point function is computed as \cite{Kashyap:2025rvf}
\begin{equation}
\mathcal{A} = \frac{1}{a\cdot k}\left\langle V^{+2}_{1}\,V^{+3}_{a,2}\,\mathcal{V}^{+1}\right\rangle ,
\label{eq:2ptpresc}
\end{equation}
where we divide by the charge $a\cdot k$, to which the bracket is proportional. In the zero-mode integral the left-moving part of $\cV^{+1}$ pairs with the right-moving part of $V^{+3}_a$ and vice versa. Each of the two terms is a product of two open-string brackets \eqref{eq:Fm}, and they add. For polarisations $e^{(i)}_m\hat e^{(i)}_n$ and momenta $k^{(1)}=-k^{(2)}=k$,
\begin{equation}
\frac{1}{a\cdot k}\left\langle V^{+2}_{1}\,V^{+3}_{a,2}\,\tfrac12\big[(\lambda\gamma^{0}\theta)+(\hat\lambda\gamma^{0}\hat\theta)\big]\right\rangle
=2k^0\,\big(e^{(1)}\!\cdot e^{(2)}\big)\big(\hat e^{(1)}\!\cdot\hat e^{(2)}\big) ,
\label{eq:2pt_closed_flat}
\end{equation}
which is $-\frac12(2i)^2$ times the kinematic factor; the left-moving variables are ordered to the left of the right-moving ones. With the plus sign in \eqref{eq:Wmikh} the two terms would cancel, as they must. The integral over $q$ and $x$ gives $\delta^{d-1}(\vec k^{(1)}+\vec k^{(2)})$ as for the open string, so the result is the free two-point function, proportional to $k^0$.

\subsection{The \texorpdfstring{$\ads$}{AdS5xS5} isometries
and the ghost-number-three vertex}\label{sec:slice}

The bulk integrand of a two-point function has the factor $(x_1\cdot X)^{-k}(x_2\cdot X)^{-k}$ times a function on $S^5$. 
This product is
invariant under the one-parameter group of $SO(1,5)$ boosts that fix the two boundary points $x_1$ and $x_2$. This is
the $AdS$ version of time translation. In radial quantisation around $x_1$ (with $x_2$ at infinity) it is the dilatation, whose charge is the
dimension $\Delta=k$. The $AdS_5$ integral therefore diverges along the boost orbits, while the $S^5$ integral is
finite. 
A coordinate that the boost shifts by a constant is
\begin{equation}
\phi_0=\frac12\log\frac{x_1\cdot X}{x_2\cdot X},\qquad \phi_0\to\phi_0+\beta ,
\end{equation}
and we take, as in \eqref{eq:mbrst},
\begin{equation}
\cV^{+1}=2\pi\,\delta(\phi_0)\,Q \cdot \phi_0 \, .
\label{eq:Vads}
\end{equation}
Using the BRST transformations given in 
\eqref{Qbulkpoints} and the properties of the polarization vectors of 
\eqref{eq:frame}, one derives 
\begin{equation}
Q \cdot \phi_0
=-\frac14\Big(\ch{\Theta\Lambda x_1X}-\ch{\Theta\Lambda x_2X}\Big)\equiv P .
\label{eq:P}
\end{equation}
$P$ has one fermion and one pure spinor, as expected. The last building block for the computation of the two-point functions is the ghost number three 
vertex operators. 
Similarly to the flat space case, 
it is hard to use directly \eqref{V3c}
since the $b$ and $c$ ghosts are composite operators. 
The $\ads$ vertices 
$V_a^{+3}$ were 
recently discussed in \cite{ferreira2026supersymmetrybicomplexpurespinor}. It was argued there that
they can be obtained as a product of 
$V^{+2}$ and a suitable 
$V^{+1}_a$. 
In our notation, they are given by 
\begin{equation}
j^{+1}_A=\Str\big(t_A \,g\,(\lambda_L-\lambda_R)\,g^{-1}\big) \, ,\qquad V^{+3}_A=
j^{+1}_A V^{+2}.
\label{V3V1}
\end{equation}
with $g$ the coset representative and $\lambda_{L,R}$ regarded as elements of the odd part of $\mathfrak{psu}(2,2|4)$.
The closure of $j^{+1}_A$ follows from  
\begin{equation}
Q \cdot g = g (\lambda_L +
\lambda_R+ \Sigma) \, , \quad 
(Q \cdot g^{-1} ) = - (\lambda_L + \lambda_R +\Sigma) g^{-1} \, , \quad
Q \cdot \lambda_s  = - \Sigma \lambda_s - \lambda_s \Sigma  \, ,
\end{equation} 
where $\Sigma$ is an $USp(2,2)\times USp(4)$ compensator when $g$ is gauged fixed $g=g_{\rm{new}}$.
The
natural $PSU(2,2|4)$ generators $t_A$ for $j_A$ in
\eqref{V3V1} are those that preserve the two boundary points: the boost $D_{12}\propto x_1\wedge x_2$ and the
$R$-symmetry rotation $J_{12}\propto y_1\wedge y_2$. 
With $g_{\rm{new}}$ in 
\eqref{eq: NewCosetRepresentative},
the fermionic expansion of $g(\lambda_L-\lambda_R)g^{-1}$ terminates after a few terms,
\begin{align}
j_D&=A[X\Lb Y\Theta]+A[\Psi Y\Lambda X]+A[\Psi\Theta X\Lb Y\Theta] ,\label{eq:jD}\\
j_J&=B[\Theta X\Lb Y]+B[Y\Lambda X\Psi]+B[\Theta X\Lb Y\Theta\Psi] .
\end{align}
where we have used the notation
$A[w]=\ch{x_1x_2w}-\ch{x_2x_1w}$, $B[w]=\ch{y_1y_2w}-\ch{y_2y_1w}$.
Each of the expressions above has terms with one and with three fermions; in the variables $\Lambda_{L,R}$, $\Theta_{L,R}$, $j_D$ has 48 terms. 
We additionally 
established three properties numerically\footnote{More precisely, we set the parameters of the problem to  rational numbers with large numerators and denominators, and then check our claims exactly.}: 
\begin{itemize}[leftmargin=*]
\item \emph{Closure.} Among the six structures of the block of $g(\lambda_L\mp\lambda_R)g^{-1}$ projected on
$x_1\wedge x_2$, the BRST-closed combinations on pure spinors span a two-dimensional space: $j_D$ and the exact $L{+}R$
combination $Q(A[\Psi\Theta])$.

\item \emph{Non-exactness.} $j_D$ is not in the span of $Q$ of all ghost-number-zero candidates with two and four
fermions.
\item \emph{Uniqueness.} In a general ansatz of 124 chains of ghost number one per generator, modulo $Q$ of 38 chains of ghost
number zero, the ghost-number-one cohomology is one-dimensional at 
$D_{12}$ and $J_{12}$.
\end{itemize}
We have now all the ingredients for computing an $\ads$ two-point function 
as the generalization of \eqref{eq:2ptpresc}. 
As in flat space, the result is divided by the charge $\pm k$ of the state. 
We also normalize the generators
$t_D$ and $t_J$.

\subsection{Zero-mode integrals}
\label{sec:2pt-integrals}
The two-point amplitude is then given by
\bea
{\cal A}^{k_1,k_2}_2={}& \frac{\mathcal K_2}{\frac12 (k_1+k_2)}\,\frac1{x_1\cdot x_2}\int_{AdS_5}\!dX\,\delta(\phi_0)\int_{S^5}\!dY\;
\Big\langle V(1)\;j_D\,V(2)\;P\Big\rangle_{X,Y},
\eea{eq:amp2}
with ${\mathcal K_2}$ an overall normalisation and where $\langle \cdot \rangle_{X,Y}$ denotes the zero mode integration without the $X$ and $Y$ integrals.
Only terms $\propto \Theta_L^5\Theta_R^5$ contribute, i.e.\ products of the form $V_a(1)\,j_D\,V_b(2)\,P\,(\fc_L)^{j_L}(\fc_R)^{j_R}$, which we call parts, with 
\begin{equation}
a+b+c_J+1+2(j_L+j_R)=10,
\end{equation}
where $c_J\in\{1,3\}$ as $j_D$ has terms with one or three $\Theta$'s and $P$ contributes one $\Theta$.
Two symmetries relate these parts to each other. The first is the exchange of the two points $1 \leftrightarrow 2$. The second symmetry exchanges the two sectors and arises from the $\mathbb Z_4$ automorphism of $\mathfrak{psu}(2,2|4)$,
\begin{equation}
\Omega:\qquad \Psi\to i\Psi ,\quad \Lambda\to i\Lambda ,\quad \Theta\to-i\Theta ,\quad \Lb\to-i\Lb ,
\label{eq:Omega}
\end{equation}
with $X$ and $Y$ fixed, which commutes with the BRST operator. By \eqref{thetaLR} and \eqref{eq:fcov} it maps
\begin{equation}
(\Theta_L,\Lambda_L,\fc_L)\to(\Theta_R,\Lambda_R,\fc_R) ,\qquad (\Theta_R,\Lambda_R,\fc_R)\to-(\Theta_L,\Lambda_L,\fc_L) ,
\end{equation}
so it exchanges the sectors up to signs. A part with $j_R$ corrections in the right sector has five $\Theta_R$'s, $3+j_R$ $\Lambda_R$'s and $j_R$ factors $\fc_R$, so these signs cancel, and the only sign of the zero-mode integral under $\Omega$ is the $-1$ from reordering $\Theta_L^5\Theta_R^5$. A monomial is multiplied by $i^q$, where $q$ is the number of $\Psi$'s and $\Lambda$'s minus the number of $\Theta$'s and $\Lb$'s. All terms of $V$ and of $j_D$ have $q=2$ mod 4, and those of $P$ have $q=0$, so $V(1)\,j_D\,V(2)\,P$ is odd as well. Hence the part with $(j_L,j_R)$ is equal to the one with $(j_R,j_L)$.
With these symmetries it is enough to consider the cases with $a \leq b$ and $j_L \geq j_R$, leaving us with the 22 terms in Table \ref{tab:2pt} for which we have to compute the zero-mode integral.
\begin{table}[t]
\centering
\begin{tabular}{cll}
\toprule
$j$ & $(a,b,c_J)$ & sector assignments $(j_L,j_R)$\\
\midrule
0 & $(0,6,3)$, $(0,8,1)$, $(2,4,3)$, $(2,6,1)$, $(4,4,1)$ & $(0,0)$\\
1 & $(0,4,3)$, $(0,6,1)$, $(2,2,3)$, $(2,4,1)$ & $(1,0)$\\
2 & $(0,2,3)$, $(0,4,1)$, $(2,2,1)$ & $(2,0)$, $(1,1)$\\
3 & $(0,0,3)$, $(0,2,1)$ & $(3,0)$, $(2,1)$\\
4 & $(0,0,1)$ & $(4,0)$, $(3,1)$, $(2,2)$\\
\bottomrule
\end{tabular}
\caption{The 22 independent parts  $V_a(1)\,j_D\,V_b(2)\,P\,(\fc_L)^{j_L}(\fc_R)^{j_R}$ contributing to the two-point amplitude organised by number of $\Theta_s$'s in each factor.}
\label{tab:2pt}
\end{table}

Doing these zero mode integrals is a formidable computation. A naive counting of terms in $V_a(1)\,j_D\,V_b(2)\,P\,(\fc_L)^{j_L}(\fc_R)^{j_R}$ which saturate $\Lambda_L^{3+j_L}\Lambda_R^{3+j_R} \Theta_L^5 \Theta_R^5$ yields $2.5 \cdot 10^8$. The number of terms is further increased by the differential $Z$-operators and the double insertion of the 2685-term $T$-tensor. The presence of long traces involving many copies of 10 different variables means that there are many genuinely different terms. For this reason we use a heavily optimized C++ program that represents the computations in terms of numerical matrices over a prime field, which is described in appendix \ref{app:numeric}. With this program, the computation of all the zero-mode integrals for the two-point function took about 7000 core hours on a cluster.

For simplicity we did this computation in the frame
\begin{equation}
x_i\cdot X=y_i\cdot Y=1 .
\label{eq:frame}
\end{equation}
With this choice each term is a polynomial of maximal degree three in $x_{12}=x_1\cdot x_2$ and $y_{12}=y_1\cdot y_2$ and degree four in $k$.
Finally we restore the full $x_i\cdot X$, $y_i\cdot Y$ dependence by homogeneity of the vertex \eqref{eq:vertex} and do the integral over a slice of AdS
\begin{gather}
\int_{AdS_5}\!dX\;\delta(\phi_0)\,(x_1\cdot X)^{-c_1}(x_2\cdot X)^{-c_2}=I_{AdS}(c)\,(x_1\cdot x_2)^{-c},\\
I_{AdS}(c)=2^c\pi^2\,\frac{\Gamma(c-2)}{\Gamma(c)}=\frac{2^c\pi^2}{(c-1)(c-2)} ,
\label{eq:slice}
\end{gather}
with $c=\frac12(c_1+c_2)$, as well as the sphere integral \cite{Lee:1998bxa}
\begin{equation}
\int_{S^5}\!dY\,(y_1\cdot Y)^a(y_2\cdot Y)^b=\delta_{ab}\,I_S(a)\,(y_1\cdot y_2)^a ,\qquad
I_S(a)=\frac{\pi^3\,a!}{(a+2)!\,2^{a-1}}=\frac{2^{1-a}\pi^3}{(a+1)(a+2)} .
\label{eq:IS}
\end{equation}
Each monomial in the frame \eqref{eq:frame} contributes
\bea
x_{12}^q\,y_{12}^r \to {}& \frac{(y_1 \cdot Y)^{k_1} (y_2 \cdot Y)^{k_2}}{(x_1 \cdot X)^{k_1} (x_2 \cdot X)^{k_2}} \left( \frac{x_1 \cdot x_2}{(x_1 \cdot X)(x_2 \cdot X)} \right)^q \left( \frac{y_1 \cdot y_2}{(y_1 \cdot Y)(y_2 \cdot Y)} \right)^r \\
\to {} & \delta_{k_1 k} \delta_{k_2k} \, I_{AdS}(k+q)\,I_S(k-r)\;\frac{(y_1\cdot y_2)^k}{(x_1\cdot x_2)^k} .
\eea{eq:2ptmono}
After these integrals, each part of Table \ref{tab:2pt} is a rational function in $k$, with simple or double poles at $k=-2,-1,0,1,2$. 
We include the result for each of these parts, as well as the formula for combining them, in an ancillary Mathematica file.

\subsection{Result}

After combining all the parts, including the additional parts related to the computed ones by symmetry, the result is
\begin{equation}
\boxed{
{\cal A}^{k,k}_2
= N_k \frac{(y_1\cdot y_2)^k}{(x_1\cdot x_2)^k} , \qquad
N_k =-\frac{\pi^5}{32} \frac{{\mathcal K_2}}{k}
= \frac{2\,\mathcal K_2\, \mathcal C_0}{k}.}
\label{eq:2ptpoly}
\end{equation}
All the poles in $k$ except the expected one at $k=0$ cancel with the numerators.
We also did this computation for the other choice of ghost-number-three vertex, for which we replaced
\beq
\frac{j_D}{x_1 \cdot x_2} \quad \to \quad \frac{j_J}{y_1 \cdot y_2},
\eeq
in \eqref{eq:amp2}. The final result \eqref{eq:2ptpoly} was the same.

\section{The three-point amplitude}\label{sec:3pt}

For the three-point amplitude the conformal Killing group can be fixed completely, and the amplitude is, up to an overall normalisation $\mathcal K_3$, simply the zero-mode integral of the product of three unintegrated vertex operators
\begin{equation}
{\cal A}^{k_1,k_2,k_3}_3 = \mathcal K_3\,
\vev{V(1)\,V(2)\,V(3)}.
\label{eq:3ptstructure}
\end{equation}

\subsection{Zero-mode integrals}
\label{sec:3pt-integrals}
As for the two-point amplitude, the main challenge is the large amount of terms to integrate.
It is enough to consider the parts $V_a(1)\,V_b(2)\,V_c(3)\,(\fc_L)^{j_L}(\fc_R)^{j_R}$ with
\begin{equation}
a+b+c+2j=10 ,\qquad j=j_L+j_R ,
\end{equation}
and, by symmetry we can restrict to $a\leq b \leq c$, $j_L \geq j_R$. The exchange of the sectors is again the symmetry \eqref{eq:Omega}: the product of three vertex operators is odd, which cancels the sign from reordering $\Theta_L^5\Theta_R^5$, so the terms with $(j_L,j_R)$ and $(j_R,j_L)$ are equal. The resulting 25 cases are listed in Table \ref{tab:3pt} and together, these products produce about $2.3 \cdot 10^8$ monomials. Applying the zero-mode prescription to all of them took about 20000 core hours on a cluster and produced polynomials in $k_i$, $x_{ij}$ and $y_{ij}$ in the frame \eqref{eq:frame}.
\begin{table}[t]
\centering
\begin{tabular}{cll}
\toprule
$j$ & parts $(a,b,c)$ & sector assignments $(j_L,j_R)$\\
\midrule
0 & $(0,0,10)$, $(0,2,8)$, $(0,4,6)$, $(2,2,6)$, $(2,4,4)$ & $(0,0)$\\
1 & $(0,0,8)$, $(0,2,6)$, $(0,4,4)$, $(2,2,4)$ & $(1,0)$\\
2 & $(0,0,6)$, $(0,2,4)$, $(2,2,2)$ & $(2,0)$, $(1,1)$\\
3 & $(0,0,4)$, $(0,2,2)$ & $(3,0)$, $(2,1)$\\
4 & $(0,0,2)$ & $(4,0)$, $(3,1)$, $(2,2)$\\
5 & $(0,0,0)$ & $(5,0)$, $(4,1)$, $(3,2)$\\
\bottomrule
\end{tabular}
\caption{The 25 independent parts $V_a(1)\,V_b(2)\,V_c(3)\,(\fc_L)^{j_L}(\fc_R)^{j_R}$ contributing to the three-point amplitude organised by number of $\Theta$'s in each factor.}
\label{tab:3pt}
\end{table}

The final step is the integration over the bulk point $(X, Y)$, which amounts to the integrals \cite{Freedman:1998tz,Lee:1998bxa}
\begin{align}
\int_{AdS_5}\!dX\,\prod_{i=1}^3(x_i\cdot X)^{-k_i}&=I_{AdS}(k_1,k_2,k_3)\prod_{i<j}(x_i\cdot x_j)^{-\alpha_{ij}} ,\label{eq:IAdS3}\\
\int_{S^5}\!dY\,\prod_{i=1}^3(y_i\cdot Y)^{k_i}&=I_S(k_1,k_2,k_3)\prod_{i<j}(y_i\cdot y_j)^{\alpha_{ij}} ,\label{eq:IS3}
\end{align}
with $\Sigma=k_1+k_2+k_3$, $\alpha_{12}=\alpha_3$, $\alpha_{13}=\alpha_2$, $\alpha_{23}=\alpha_1$, 
$2\alpha_1=(k_2+k_3-k_1),2\alpha_2=(k_1+k_3-k_2),2\alpha_3=(k_1+k_2-k_3)$, and
\begin{equation}
I_{AdS}=\frac{\pi^2\,2^{\Sigma/2}}{2}\,\frac{\Gamma(\tfrac\Sigma2-2)\,\Gamma(\alpha_1)\Gamma(\alpha_2)\Gamma(\alpha_3)}
{\Gamma(k_1)\Gamma(k_2)\Gamma(k_3)} ,\qquad
I_S=\frac{\pi^3}{(\tfrac\Sigma2+2)!\,2^{\Sigma/2-1}}\,\frac{k_1!\,k_2!\,k_3!}{\alpha_1!\,\alpha_2!\,\alpha_3!} .
\end{equation}
The Gamma functions and factorials always cancel in the product of AdS and sphere integrals up to rational functions, for example
\begin{equation}
I_{AdS}(k_1,k_2,k_3)\,I_S(k_1,k_2,k_3)=\frac{\pi^5\,k_1k_2k_3}{\alpha_1\alpha_2\alpha_3\,(\tfrac\Sigma2-2)(\tfrac\Sigma2-1)\tfrac\Sigma2(\tfrac\Sigma2+1)(\tfrac\Sigma2+2)} .
\label{eq:IAIS3}
\end{equation}
A monomial $\prod x_{ij}^{q_{ij}} y_{ij}^{r_{ij}}$ in the frame \eqref{eq:frame} shifts the weights, $k_1\to k_1+q_{12}+q_{13}$ in $I_{AdS}$, $k_1\to k_1-r_{12}-r_{13}$ in $I_S$, and so on, so that each part in Table \ref{tab:3pt} becomes some rational function in $k_1$, $k_2$ and $k_3$, which we include and combine in the ancillary Mathematica file.

\subsection{Result}

When summing all these rational functions, the entire rational dependence on the $k_i$ collapses into a constant
\begin{equation}
\boxed{\;C_{k_1k_2k_3}= \mathcal K_3\, \mathcal C_0, \qquad \mathcal C_0 = -\frac{\pi^5}{64}  .}
\label{eq:3ptresult}
\end{equation}
The operator-normalisation independent ratio of three- and two-point amplitudes then gives
\begin{equation}
\frac{C_{k_1k_2k_3}^2}{N_{k_1}N_{k_2}N_{k_3}}=\frac{\mathcal K_3^2}{8\,\mathcal K_2^3\,\mathcal C_0}\;k_1k_2k_3
=\frac{k_1k_2k_3}{N^2} .
\label{eq:wsratio}
\end{equation}
In the last step we used the normalisations of the two- and three-point amplitudes. They do not depend on the weights
and can therefore be fixed in the flat space limit, as we explain in appendix \ref{app:flat}:
\begin{equation}
\mathcal K_3= 2^{16}\,\frac{R^8}{4\kappa_{10}^2}\Big(\frac{2\pi\alpha'}{R^2}\Big)^3 ,\qquad
\mathcal K_2=-\frac18\, 2^{16}\,\frac{R^8}{4\kappa_{10}^2}\Big(\frac{2\pi\alpha'}{R^2}\Big)^2 .
\label{eq:K23}
\end{equation}
Here $R^8/4\kappa_{10}^2=N^2/16\pi^5$ is the string coupling, and there is one factor of $2\pi\alpha'/R^2$ for each vertex
operator, because it is $V/2\pi\alpha'$ that creates a fluctuation of the metric. The number $2^{16}$
accounts for the relative normalisation of the $\ads$ and flat space variables. The factor $\frac18$ in $\mathcal K_2$
comes from the normalisation of $j_D$, and its sign is fixed by the positivity of the two-point function. With
$\lambda=R^4/\alpha'^2$ this gives $N_k=64\pi^2N^2/(\lambda k)$ and $C_{k_1k_2k_3}=-2^9\pi^3N^2/\lambda^{3/2}$.

This can be compared to and matches exactly with $\mathcal N=4$ SYM with gauge group $SU(N)$, where the correlators of the half-BPS operators $\mathcal O_k(x,y)=\tr(y\cdot\Phi(x))^k$ are, at
leading order in $1/N$ and with the scalar propagator normalised to one \cite{Lee:1998bxa}
\bea
\vev{\mathcal O_k\,\mathcal O_k}&=k\,N^k\left(\frac{y_1\cdot y_2}{x_{12}^2}\right)^{k},\\
\vev{\mathcal O_{k_1}\mathcal O_{k_2}\mathcal O_{k_3}}&=k_1k_2k_3\,N^{\Sigma/2-1}\prod_{i<j}\left(\frac{y_i\cdot y_j}{x_{ij}^2}\right)^{\alpha_{ij}} .
\eea{SYM_result}
This implies the exact same ratio as in \eqref{eq:wsratio} as expected from AdS/CFT and non-\-re\-normalisation theorems.

\section{Discussion}\label{sec:discussion}

In this paper we formulated a BRST-invariant prescription for zero-mode integration in type IIB string theory on $AdS_5 \times S^5$. The prescription applies to genus-zero amplitudes with arbitrary multiplicity and external states. We implemented it explicitly for two and three-point functions of half-BPS states and reproduced the properly normalised three-point coefficients predicted by the AdS/CFT correspondence. By resolving the zero-mode integration problem, this prescription removes a central obstacle to the direct worldsheet computations of amplitudes on $AdS_5 \times S^5$. This opens several directions for future work. 

\begin{itemize}[leftmargin=*]
\item Four-point functions and the AdS Virasoro-Shapiro amplitude. The most natural extension of our results is the computation of four-point amplitudes of supergravity states. This requires an integrated vertex operator and the evaluation of the non-zero-mode worldsheet correlator in an $\alpha'$ expansion. This combined with the zero mode prescription given in this paper would allow a direct worldsheet test of  the proposed curvature expansion calculated by indirect means in \cite{Alday:2022uxp,Alday:2022xwz,Alday:2023jdk,Alday:2023mvu}. It would also be interesting to understand the worldsheet origin of the single-valued functions appearing in these predictions.
\item Massive string states. An important open problem for type IIB strings in $AdS_5 \times S^5$ is to calculate the spectrum of the first massive states, in a large radius expansion and at genus zero. This should be compared to the scaling dimension of the Konishi operator in planar ${\cal N}=4$ SYM, known from integrability methods \cite{Gromov:2009zb}. In the pure spinor formalism this can be achieved either by explicitly constructing their vertex operators or via factorisation limits of four point amplitudes. 
\item Worldsheet Ward identities. It should be possible to establish directly within the pure spinor formalism why two and three-point worldsheet correlators of half-BPS states receive no $\alpha'$ corrections. Understanding the contributions of non-zero modes, and the role of BRST symmetry and spacetime supersymmetry in their cancellation, would provide a worldsheet explanation of the protection assumed in the present computation. For higher point amplitudes, it would be interesting to derive the supersymmetric Ward identities, mimicking those of ${\cal N}=4$ SYM. 
\item Simplifying limits. It would be interesting to understand how different limits arise from the pure spinor computation. Two instructive limits are the supergravity limit and the flat space limit. Both limits involve $\alpha'/R^2 \to 0$, but the kinematical variables - involved in four and higher point amplitudes -  are tuned differently. Another interesting limit is the semiclassical limit, where some of the external operators become heavy. 
\item Our prescription for the zero-mode integration factorises into $\mathcal N_L\mathcal N_R$. This leads to a natural prescription for the zero-mode integration of disk amplitudes on $AdS_5\times S^5$. This provides a framework for studying, within the pure spinor formalism, the string configurations dual to Wilson loops \cite{Maldacena:1998im} and planar scattering amplitudes \cite{Alday:2007hr} in ${\cal N}=4$ SYM at strong coupling. Another interesting application would be the computation of correlators of supersymmetric Wilson loops with local half-BPS operators, combining our prescription with the analysis of supersymmetric string configurations in \cite{Martins:2024zud}.
\item Relation to string field theory. An advantage of the pure spinor
formalism is that it can consistently treat finite RR fluxes. RR
fluxes can also be treated perturbatively in the framework of
closed superstring field theory based on the NSR formalism. This was
done for $AdS_3 \times S^3 \times M_4$ in \cite{Cho:2018nfn}, and more
recently for $AdS_5 \times S^5$ in \cite{Cho:2025coy}. These techniques
can in principle be used to compute $AdS_5 \times S^5$ string amplitudes
in a small curvature expansion. It would be very interesting to find
the relation between the two formalisms.
\item Other backgrounds. Many of the ingredients used to construct
the zero-mode integration measure in this paper carry over to more
general backgrounds. It would be very interesting to extend the
amplitude computation to other backgrounds relevant for holography
that admit pure spinor or related covariant formulations. These
include, for example, $AdS_4 \times CP^3$
\cite{Fre:2008qc,DAuria:2008vov} and
$AdS_3 \times S^3 \times M_4$ \cite{Adam:2007ws,Daniel:2024yfr}.
In this context, it would be very interesting to write our prescription in terms of superspace covariant derivatives on $\ads$. Such an expression would naturally apply to other curved backgrounds.\footnote{We thank Nathan Berkovits for this suggestion.}
\end{itemize}

\section*{Acknowledgements}
We would like to thank Nathan Berkovits for useful discussions. 
This work has made use of the Hamilton HPC Service of Durham University. LFA is supported by the STFC grant ST/T000864/1. TF thanks the CNRS for support via the IRP project NP-Strong. TH is supported by the STFC grant ST/X000591/1. 
TF, LM and DZ would like to thank Serrapilheira
Institute Grant (Serra–R-2012-38185),
FUNPEC/UFRN/FRONTEIRAS DA CIENCIA, and Simons Foundation Grant (1023171 -RC) for financial support.  TF thanks IPhT and ENS where part of this work was done for the warm hospitality.  
For the purpose of open access, the authors have applied a CC BY public copyright licence to
any Author Accepted Manuscript (AAM) version arising
from this submission.

\appendix

\section{The zero-mode prescription as a differential operator}\label{app:prescription}

In this appendix we derive the zero-mode prescription as a differential operator. 
We showed in subsections \ref{sec:minimal} and \ref{sec:non-minimal} that both the minimal and the non-minimal formalisms define linear functionals $F_j[h^{\mu_1\ldots\mu_j}]$ which act on a function $h$ that is symmetric in its spinor indices, a polynomial of degree $3+j$ in $\lambda$ and degree $5$ in $\theta$. The functional has the following properties:
\begin{enumerate}[label=(\roman*),leftmargin=*]
\item \emph{Invariant under the $SO(10)$ Lorentz symmetry}: by definition.
\item \emph{Well defined on the pure spinor cone}: it has to vanish for $h \in \mathcal I$, with $\mathcal I$ being the ideal of polynomials in $\lambda$ generated by the ten $\lambda \gamma^m \lambda$ which vanish on the pure spinor cone.
\item \emph{Scaling}: $F_j[\lambda^{(\mu_1} H^{\mu_2\ldots\mu_j)}]
= - (10+j)  F_{j-1}[H^{\mu_1\ldots\mu_{j-1}}]$, as shown in \eqref{eq:m_scaling} and \eqref{eq:nm_scaling}.
\end{enumerate}
We start by using Lorentz invariance to show that such a functional is unique. Then we construct a factorised differential operator that satisfies all three properties.
The functional derived here has previously been derived in \cite{Gomez:2013sla} in a tensor form, but not as factorised differential operators.

\subsection{Uniqueness}\label{app:uniqueness}

We label irreducible representations of $\mathfrak{so}(10)$ by their highest weights, $V(\Lambda)$, and use the
fundamental weights $\omega_1,\ldots,\omega_5$ in the standard (Bourbaki) labelling \cite{FultonHarris1991}. In the basis of the
five commuting rotations in orthogonal planes, $\omega_i$ has its first $i$ entries equal to one and the others zero for
$i\le3$, $\omega_4=(\frac12,\frac12,\frac12,\frac12,-\frac12)$ and $\omega_5=(\frac12,\frac12,\frac12,\frac12,\frac12)$. So
$V(\omega_1)$ is the vector, and $V(\omega_5)=S$ and $V(\omega_4)=S^*$ are the two chiral spinors. We take $\lambda$ and $\theta$ in $S$. The polynomials of degree $n$ on the cone of pure spinors of one chirality form the irreducible
representation $V(n\omega_5)$, and those on the cone of the opposite chirality $V(n\omega_4)=V(n\omega_5)^*$.
By Lorentz invariance, the functional must be an invariant in
$\text{Sym}^j(S^*)\otimes V((j+3)\omega_5)\otimes\Lambda^5S$, where the first factor accounts for $h$'s $j$ symmetrised open indices, the second for the dependence of $h$ on $\lambda$, and $\Lambda^5S$ for the five $\theta$'s. 
In the non-minimal formalism all indices of $h$ are contracted with $\bar\lambda$'s, so we could replace the first factor with $V(j\omega_4)$, but we will consider the more general case.
The number of such invariants equals the
multiplicity of $\text{Sym}^j(S)$ in $V((j+3)\omega_5)\otimes\Lambda^5S$. We computed it with the Brauer--Klimyk formula
\cite{FultonHarris1991},
\begin{equation}
\mathrm{mult}\big(V(\nu)\subset V(\Lambda)\otimes U\big)=\sum_{\mu}m_U(\mu)\sum_{w\in W}\mathrm{sgn}(w)\,
\delta_{w(\Lambda+\mu+\rho)-\rho,\,\nu} ,
\end{equation}
which only requires the weights $\mu$ of $U=\Lambda^5S$ with their multiplicities $m_U(\mu)$; $W$ is the Weyl group of $D_5$ and $\rho$ the Weyl vector. We computed multiplicities for $j \leq 9$ with the result
\begin{equation}
\begin{tabular}{cll}
\toprule
$j$ & $\mathrm{Sym}^j(S^*)$ & singlets per piece \\
\midrule
0 & $1$ & 1 \\
1 & $V(\omega_4)$ & 1 \\
2 & $V(2\omega_4)\oplus V(\omega_1)$ & 1, 0 \\
3 & $V(3\omega_4)\oplus V(\omega_1+\omega_4)$ & 1, 0 \\
4 & $V(4\omega_4)\oplus V(\omega_1+2\omega_4)\oplus V(2\omega_1)$ & 1, 0, 0 \\
5 & $V(5\omega_4)\oplus V(\omega_1+3\omega_4)\oplus V(2\omega_1+\omega_4)$ & 1, 0, 0 \\
$j$ & $\bigoplus_{i}V(i\,\omega_1+(j-2i)\omega_4)$ & 1, 0, \ldots, 0 \\
\bottomrule
\end{tabular}
\end{equation}
For $j=0$ this is the uniqueness of the $T$ tensor. So $F_j[h^{\mu_1\ldots\mu_j}]$ is unique (at least) for every $j\leq 9$ and property (iii) fixes the relative factors. Next we construct such a functional.

\subsection{An operator on the pure spinor cone}\label{app:property_ii}

Let us consider a polynomial $h^\mu$, which might have additional spectator indices that we suppress, of degree $d$ in $\lambda$. By inspecting property (iii), we are led to construct a differential operator that contracts one index of $h^\mu$ and reduces its degree by one. For property (ii) we have to show that the operator maps $\mathcal I$ to $\mathcal I$. As it turns out the operator
\begin{equation}
Z^{(d)}[h]=\partial_\mu h^\mu-\frac{1}{4(d+2)}\,K[h] ,\qquad K[h]=(\gamma^m\lambda)_\mu\,\gamma_m^{\alpha\beta}\,\partial_\alpha\partial_\beta h^\mu ,
\end{equation}
has this property, as we now show.
We write $E^m\equiv\lambda\gamma^m\lambda$ for the polynomials generating the ideal $\mathcal I$ and  $\simeq$ for equality modulo $\mathcal I$.
Let $h^\mu=E^q u_q^\mu$ with $u_q^\mu$ of degree $d-2$. Then
\beq
\partial_\mu h^\mu=2(\gamma^q\lambda)_\mu u_q^\mu+E^q\partial_\mu u_q^\mu\simeq2(\gamma^q\lambda)_\mu u_q^\mu.
\eeq
For the second term,
\begin{equation}
\partial_\alpha\partial_\beta\big(E^qu\big)=2\gamma^q_{\alpha\beta}\,u+2(\gamma^q\lambda)_\alpha\partial_\beta u+2(\gamma^q\lambda)_\beta\partial_\alpha u+E^q\partial_\alpha\partial_\beta u ,
\end{equation}
and with the identities
\begin{equation}
\gamma^m_{\alpha\beta}\gamma_n^{\alpha\beta}=16\,\delta^m_n ,\qquad
\gamma^m_{\alpha\beta}\gamma^{n\,\beta\gamma}+\gamma^n_{\alpha\beta}\gamma^{m\,\beta\gamma}=2\eta^{mn}\delta_\alpha^\gamma ,
\label{eq:gid}
\end{equation}
this implies
\begin{equation}
K[h]\simeq32\,(\gamma^q\lambda)_\mu u_q^\mu+4\,(\gamma^m\lambda)_\mu\,(\lambda\gamma^q\gamma_m)^\beta\,\partial_\beta u_q^\mu .
\end{equation}
Here we use the Clifford relation from \eqref{eq:gid} on the second term
\begin{equation}
(\lambda\gamma^q\gamma_m)^\beta=2\delta^q_m\lambda^\beta-(\lambda\gamma_m\gamma^q)^\beta ,
\end{equation}
and the second piece gives $(\gamma^m\lambda)_\mu(\gamma_m\lambda)_\rho\gamma^{q\,\rho\beta}$, which lies in $\mathcal I$ as one can show by contracting the Fierz identity $\gamma^m_{(\alpha\beta}\gamma_{m\,\gamma)\delta}=0$ with $\lambda^\alpha\lambda^\gamma$. With $\lambda\cdot\partial\,u=(d-2)u$,
\begin{equation}
K[h]\simeq\big(32+8(d-2)\big)(\gamma^q\lambda)_\mu u_q^\mu=8(d+2)\,(\gamma^q\lambda)_\mu u_q^\mu ,
\qquad\text{so}\qquad Z^{(d)}[h]\simeq0 .
\end{equation}
All terms dropped along the way are explicit multiples of the $E^m$, so $Z^{(d)}[h]\in\mathcal I$, also when $h$ carries
spectator indices. The coefficient $-1/(4(d+2))$ in front of $K[h]$ is fixed by this requirement.

\subsection{The full functional}\label{app:full}
For symmetric $h^{\mu_1\ldots\mu_j}$ of degree $j+3$ define
\begin{equation}
Z_j[h]=Z^{(4)}_{\mu_j}\cdots Z^{(j+2)}_{\mu_2}Z^{(j+3)}_{\mu_1}\,h^{\mu_1\ldots\mu_j} ,
\end{equation}
The full functional is then
\beq
F_j[h^{\mu_1\ldots\mu_j}] = \kappa_j Z_j[h] \big|_{\lambda^3 \theta^5 \to T}\,,
\label{functional}
\eeq
where in the final step terms of order $\lambda^3 \theta^5$ are replaced by the tensor $T$ as in the flat space amplitude prescription, and any other terms do not contribute to the integral.
The right hand side is Lorentz invariant, and it vanishes for $h\in\mathcal I$, because each $Z^{(d)}$ maps $\mathcal I$ to $\mathcal I$ and the contraction with $T$ vanishes on $\mathcal I$. It therefore has properties (i) and (ii), and by the uniqueness shown in subsection \ref{app:uniqueness} it is proportional to the functional defined by the minimal and non-minimal formalisms. Only the constant $\kappa_j$ remains to be determined, for which it is enough to evaluate both sides on one function.
We choose $h^{\mu_1\ldots\mu_j}=\lambda^{\mu_1}\cdots\lambda^{\mu_j}g$ with $g$ of degree three in $\lambda$. On the left hand side, applying property (iii) $j$ times gives
\begin{equation}
F_j[\lambda^{\mu_1}\cdots\lambda^{\mu_j}g]=\prod_{m=1}^j\big(-(10+m)\big)\,F_0[g] .
\label{eq:Flambdas}
\end{equation}
On the right hand side, every operator $Z^{(d)}$ in $Z_j$ acts on a function of the form $h^\mu=\lambda^\mu H$, with $H$ of degree $d-1$, where $H$ contains the remaining factors of $\lambda$ with their spectator indices.
Then $\partial_\mu h^\mu=(16+d-1)\,H$. In $K[h]$ the term with both derivatives on $H$ contains $(\gamma^m\lambda)_\mu\lambda^\mu=E^m$,
and the two terms with one derivative on $\lambda^\mu$ give $2\,\lambda^\nu\gamma^m_{\nu\mu}\gamma_m^{\mu\beta}\partial_\beta H=20(d-1)\,H$.
Hence
\begin{equation}
Z^{(d)}[\lambda^\mu H]\simeq c_d\,H ,\qquad c_d=15+d-\frac{5(d-1)}{d+2}=\frac{(d+5)(d+7)}{d+2} .
\label{eq:cd}
\end{equation}
Therefore $Z_j[\lambda^{\mu_1}\cdots\lambda^{\mu_j}g]\simeq c_{j+3}\cdots c_4\,g$, and comparing with \eqref{eq:Flambdas} gives
\begin{equation}
\kappa_j=\prod_{m=1}^j\frac{-(10+m)}{c_{m+3}}=\prod_{m=1}^j\frac{-(m+10)(m+5)}{(m+8)(m+10)}
=\frac{(-1)^j336}{(j+6)(j+7)(j+8)} .
\label{kappadef}
\end{equation}
With this normalisation, the functional \eqref{functional} satisfies properties (i)-(iii).

\section{Numerical implementation}\label{app:numeric}

In this appendix we describe how we applied the zero-mode integral prescription \eqref{eq:presc} to each of the terms listed in Tables \ref{tab:2pt} and \ref{tab:3pt} which contribute to the two-point and three-point amplitudes. In principle this is straightforward algebra in four steps:
(i) multiply out the
factors and keep the terms with the right number of zero modes, $(3+j_L,3+j_R,5,5)$ in
$(\Lambda_L,\Lambda_R,\Theta_L,\Theta_R)$; (ii) apply the operators $Z^L_{j_L}$, $Z^R_{j_R}$ of \eqref{eq:presc},
which compute the picture changing corrections; (iii) replace $\Lambda_s^3 \Theta_s^5$ in each sector ($s \in \{L,R\}$) with the $T$ tensor \eqref{eq:T2k}; (iv) integrate over the bulk point analytically (\cref{sec:2pt-integrals,sec:3pt-integrals}).
What makes this difficult is the large number of terms arising combinatorically.
As mentioned before, step (i) yields over $2\cdot 10^8$ terms for each of the two-point and three-point amplitudes. Depending on the number of corrections, step (ii) multiplies this by another large factor, and (iii) sums over $105^2$ pairings of indices.

We started by writing symbolic code which performs steps (i) and (ii) in Mathematica and step (iii) in FORM \cite{Vermaseren:2000nd}, however this approach turned out to be too slow. We then used Claude Code (Opus 5.5) to implement a C++ program which performs steps (i)-(iii) more efficiently, and which was validated against the symbolic reference implementation on many examples, including all the integrals over exact expressions mentioned in section \ref{sec:prescription} and parts of $\langle VVV \rangle$ with up to five corrections.

The two main concepts that make the numerical implementation efficient are: (a) evaluation of terms at numerical kinematic points, in integers modulo a large prime, with reconstruction of rational numbers and kinematics dependence in the end; and (b) storage of terms in a format that makes equivalence of terms due to symmetries manifest so that they can be merged after every step, to keep the number of terms under control as far as possible.
In the remainder of this appendix we explain this in more detail.

\subsection{Arithmetic modulo a prime}

Mapping symbolic computations to numerical arithmetic over a finite field is a standard technique for optimizing large computations, which is used for example in multi-loop computations \cite{vonManteuffel:2014ixa,Peraro:2016wsq}.
We perform the computations in $\Fp$, the integers modulo a prime $p$, and take $p=4611686018427387817$, the largest prime below $2^{62}$ with $p = 1\bmod4$. In $\Fp$ rational numbers are expressed as integers, the size of our chosen prime ensures that every number including intermediate results fits into a 64-bit machine word, and the condition $p = 1\bmod4$ ensures that $-1$ has a square root in $\Fp$, which represents $i$. Choosing a large $p$ ensures that the reconstruction is usually unique, obviating the need for additional runs. We also pick numerical values for the weights $k_i$ and polarisations $x_i$ and $y_i$, which we explain next.

The weights $k_i$ appear only in the coefficients of the vertex parts $V_{m_i}$ as polynomials with degree at most $m_i/2$. Each coefficient is therefore carried through the computation as a list of its values at $\prod_i(m_i/2+1)\le18$ values of the weights. The expensive work, bringing terms into canonical form and combining them, is done only once for all values. 

We further choose numerical values for $x_i$, $y_i$, $X$, $Y$ in the frame \eqref{eq:frame} and reconstruct the dependence of the final result on the six invariants $x_i \cdot x_j$,  $y_i \cdot y_j$ from several such evaluations. Bounds on the degree of the polynomials can be found by inspecting how many powers of $x_i$ and $y_i$ the inputs contain. For example, $V_0$ and $f_\text{cov}$ are independent of $x_i$, $y_i$, so any three-point computation with two powers of $V_0$ is independent of the invariants and one evaluation is enough. Generally $V_{m_i}$ contains at most $m_i/2+1$ powers of $x_i$ and $y_i$. The case with the largest polynomial is $V_2 V_4 V_4$ with up to 17 monomials each in $x_i$ and $y_i$, requiring up to $17^2 = 289$ points.

For the reconstruction of the kinematic polynomials and rational coefficients, we evaluate each term at the points required for kinematic reconstruction plus two extra points. Rational numbers $a/b$ with small enough numerator and denominator can be recovered from $a b^{-1} \bmod p$ by rational reconstruction \cite{Wang,Monagan}. We restrict to denominators with prime factors up to 13, as we don't expect any large prime factors to appear. In addition to the kinematic runs, an additional check run at a second prime must produce the reconstructed result. Otherwise a second set of runs is combined with the first one by the Chinese remainder theorem.

\subsection{Efficient merging of terms}

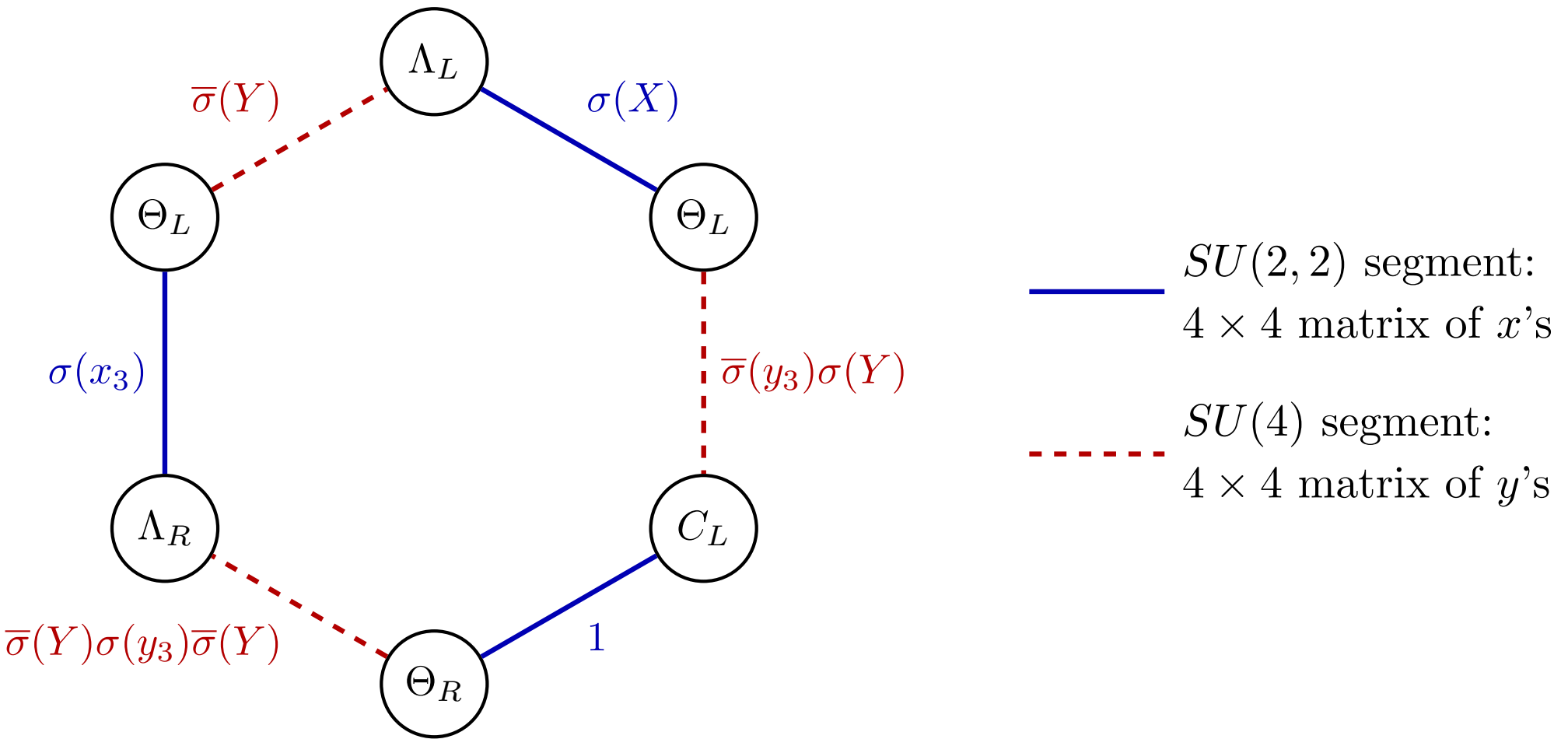
\begin{figure}[t]
\centering
\begin{tikzpicture}[scale=1.15,
  slot/.style={circle,draw,thick,minimum size=9mm,inner sep=0pt,font=\small},
  R/.style={very thick,blue!70!black}, J/.style={very thick,red!70!black,dashed}]
\foreach \n/\lab [count=\q from 0] in {a/{$\Lambda_L$},b/{$\Theta_L$},c/{$C_L$},d/{$\Theta_R$},e/{$\Lambda_R$},f/{$\Theta_L$}}
  \node[slot] (\n) at ({90-60*\q}:2.3) {\lab};
\draw[R] (a) -- node[above right,font=\small]{$\sigma(X)$} (b);
\draw[J] (b) -- node[right,font=\small]{$\bar\sigma(y_3)\sigma(Y)$} (c);
\draw[R] (c) -- node[below right,font=\small]{$1$} (d);
\draw[J] (d) -- node[below left,font=\small]{$\bar\sigma(Y)\sigma(y_3)\bar\sigma(Y)$} (e);
\draw[R] (e) -- node[left,font=\small]{$\sigma(x_3)$} (f);
\draw[J] (f) -- node[above left,font=\small]{$\bar\sigma(Y)$} (a);
\begin{scope}[xshift=4.4cm]
\draw[R] (0,0.6) -- (1,0.6) node[right,black,align=left]{$SU(2,2)$ segment:\\ $4\times4$ matrix of $x$'s};
\draw[J] (0,-0.6) -- (1,-0.6) node[right,black,align=left]{$SU(4)$ segment:\\ $4\times4$ matrix of $y$'s};
\end{scope}
\end{tikzpicture}
\caption{A ring. The value of a term is the product over its rings of the traces of the alternating matrix products,
with the spinors inserted, times the coefficient and the Grassmann sign of the order of the $\Theta$'s.}
\label{fig:ring}
\end{figure}

After step (i) each term is a product of traces of the spinor variables $\Lambda_s, \Theta_s, C_s$ with $s \in \{L,R\}$  and the bosonic variables $x_i$, $X$, $y_i$, $Y$.
Here $C_s$ is a placeholder for the free spinor index of a correction $f_s$, written as $C_s\cdot f_s$, which is removed by the operators $Z_j$ of \eqref{functional}.
Consider an arbitrary example trace
\beq
\begin{aligned}
{}&[\Lambda_L X \Theta_L y_3 Y C_L \Theta_R Y y_3 Y \Lambda_R x_3 \Theta_L Y ] = \\
{}&\Lambda^{R_1}_{LI_1} X_{R_1 R_2} \Theta^{R_2}_{LI_2} y_3^{I_2 I_3} Y_{I_3 I_4}
C^{I_4}_{LR_3} \Theta^{R_3}_{RI_5} Y^{I_5 I_6} y_{3 I_6 I_7} Y^{I_7 I_8} \Lambda^{R_4}_{RI_8} x_{3 R_4 R_5} \Theta^{R_5}_{L I_9} Y^{I_9 I_1}\,.
\end{aligned}
\eeq
We can think of each such a trace as a ring with vertices corresponding to the spinor variables.
Each spinor carries one $SU(2,2)$ and one $SU(4)$ index, so the vertices are connected by alternating segments depending on the $SU(2,2)$ variables $x_i$, $X$ and $SU(4)$ variables $y_i$, $Y$. The ring corresponding to the example term is depicted in Figure \ref{fig:ring}.
Given the number of variables and possible contractions, there is obviously a large number of possible rings. For this reason it is important that we can efficiently identify equivalent terms and merge expressions to eliminate duplicates. We do this on two levels.

The first level of simplification is that at a kinematic point, every segment is simply a $4 \times 4$ matrix over $\Fp$ and identities between Pauli matrices hold automatically. We normalise each distinct matrix, such that its first nonzero entry is one, assign a label to it and store its coefficient. In this way we automatically work with the minimal number of required segments.

The second level of simplification is to bring each ring into a canonical form, taking into account its cyclic and reflection symmetry and the simplified segments. To this end we assign a fixed ordering to the spinor types and segment labels and use the symmetries to minimise the sequence that labels the ring. For example, the sequence $(1213)$ is smaller than its cyclic and reflection images $(1312)$, $(2131)$  and $(3121)$. The coefficient contains an overall factor and the Grassmann sign from commuting $\Theta$'s to bring each term into the canonical form. The canonical form also detects terms that vanish due to the symmetries.

All operations of steps (ii) and (iii) cut rings open at some spinors and reconnect the rings with new segment insertions.  Step (ii) uses the operator \eqref{eq:ZAdS} to remove all the extra $\Lambda_s$ which arise from higher picture changing operator corrections. In step (iii) we insert the tensor $T$ once for each sector, removing all remaining spinors. In the end we are left with a trace over a $4\times 4$ matrix and finally a number. As these operations can be done in any order, the sector with fewer corrections is treated first. Segments as well as rings are usually merged after every $Z$ or $T$ insertion. For some parts of the computation, where intermediate terms combine little, steps can be fused so that intermediate terms don't need to be stored. The program is multithreaded, and the stored values of the $T$ contractions are shared between the threads.

\section{The flat space limit}\label{app:flat}
The constants $\mathcal K_2$ and $\mathcal K_3$ in \eqref{eq:amp2} and \eqref{eq:3ptstructure} are properties of the
prescriptions and do not depend on the inserted vertex operators. They can therefore be determined from wave functions
that vary rapidly on the scale of the radius $R$. For these the amplitudes are those of flat space in the local frame at
the bulk point, where they are known: for the fluctuation $h$ of the metric $g=\eta+h$ the three-point amplitude is the
cubic vertex of $\frac1{2\kappa_{10}^2}\int\sqrt gR$, and the two-point amplitude is the norm of the state
\cite{Erbin:2019uiz}.

In flat space we use the conventions of section \ref{sec:review}: the closed-string vertex operator of a massless state
with polarisation $e_m\hat e_n$ is the product of two open-string vertex operators $\lambda^\alpha A_\alpha$ with the
superfield \eqref{TheAvertex}, and $Qx^m=\frac12(\lambda\gamma^m\theta+\hat\lambda\gamma^m\hat\theta)$. The relation to the
$\ads$ variables follows from the action of the BRST operator on the bulk point. For a tangent vector $v$ at $X$,
\begin{equation}
v\cdot QX=\tfrac18\,v_m\big(\Lambda_L\gamma^m\Theta_L+\Lambda_R\gamma^m\Theta_R\big)+Q(\ldots) ,
\end{equation}
with the gamma matrices \eqref{eq:gammaAdS}, so that the flat space variables are $\Lambda_{L,R}/2$ and
$\Theta_{L,R}/2$. Since the bracket contains three pure spinors and five fermions of each sector, the bracket
in the $\ads$ variables is $2^{-16}$ times the flat space one, both normalised by \eqref{eq:shortcut_measure_flat}. The
integrated vertex operator contains $\partial x^m\bar\partial x^n\,e_m\hat e_n\,e^{ik\cdot x}$ with unit coefficient,
while a fluctuation of the metric changes the action by $\frac1{2\pi\alpha'}\int d^2z\,h_{mn}\,\partial x^m\bar\partial x^n$
\cite{Berkovits:2001ue}. So $V$ creates the fluctuation $h_{mn}=2\pi\alpha'\,e_m\hat e_n\,e^{ik\cdot x}$.

For three gravitons, $\hat e_i=e_i$, the flat space bracket is the square of the three-gluon amplitude \eqref{eq:3g},
$\vev{V_1V_2V_3}_{\rm flat}=(\mathcal A^{(123)})^2=-\mathrm{YM}_3^2$, where $\mathrm{YM}_3$ is the square bracket
of \eqref{eq:3g}. The cubic vertex of $\sqrt gR$ for $h_{mn}=\sum_ie_{i,m}e_{i,n}e^{ik_i\cdot x}$ is
$\frac12\mathrm{YM}_3^2$. Equating the two amplitudes in absolute value,
\begin{equation}
\mathcal K_3\cdot2^{-16}\cdot\mathrm{YM}_3^2=\frac{(2\pi\alpha')^3}{2\kappa_{10}^2}\cdot\frac12\,\mathrm{YM}_3^2 ,
\end{equation}
which is $\mathcal K_3$ of \eqref{eq:K23} in units $R=1$. Its sign depends on the sign convention for the vertex
operators and we choose it positive.

The flat space two-point amplitude was computed in section \ref{sec:2pt}: the bracket divided by the charge is
\eqref{eq:2pt_closed_flat}, in absolute value $2k^0(e_1\cdot e_2)^2$ for $\hat e_i=e_i$. The norm of the state is the quadratic term of the action,
$\frac{1}{4\kappa_{10}^2}\,2k^0\,(e_1\cdot e_2)^2$ for the fluctuation above. In $\ads$ the operator \eqref{eq:Vads} becomes
the flat space one with $\phi=x^0$, and $j_DV$ becomes the vertex \eqref{eq:Wmikh} with charge $a\cdot k=8k\,(x_1\cdot x_2)$, whereas
\eqref{eq:amp2} divides by $k\,(x_1\cdot x_2)$. Hence
\begin{equation}
|\mathcal K_2|\cdot2^{-16}\cdot8\cdot2k^0\,(e_1\cdot e_2)^2
=\frac{(2\pi\alpha')^2}{4\kappa_{10}^2}\,2k^0\,(e_1\cdot e_2)^2 ,
\end{equation}
which is $\mathcal K_2$ of \eqref{eq:K23}. The sign of $\mathcal K_2$ is not a convention. The norm of a state is
positive, and evaluating the bracket of \eqref{eq:amp2} at the bulk point for a state and its conjugate we find that it
is negative, so $\mathcal K_2<0$.

\section{\texorpdfstring{$PSU(2,2|4)$}{PSU(2,2|4)} Algebra}
\label{app: algebra}

The $PSU(2,2|4)$ algebra presented in (\ref{eq: rr_flux_acting_algebra}), when written in $SU(2,2)\times SU(4)$ covariant form, becomes
\begin{equation}
\{{t}^{I}_{S},{t}^{R}_{J}\}=\delta^{I}_{J} t^{R}_{S} + \delta^{R}_{S} t^{I}_{J}.   
\end{equation}
\begin{equation}
[t^{I}_{J},t^{K}_{L}]=\delta^{K}_{J} t^{I}_{L} - \delta^{I}_{L} t^{K}_{J}\,,\qquad [t^{R}_{S},t^{T}_{U}]=\delta^{T}_{S} t^{R}_{U} - \delta^{R}_{U} t^{T}_{S}\,,\nonumber
\end{equation}
\begin{equation}
[t^{R}_{S},{t}^{I}_{T}]=+\delta^{R}_{T}{t}^{I}_{S}-\frac{1}{4}\delta^{R}_{S}t^{I}_{T}\,, \qquad [t^{R}_{S},{t}^{T}_{I}]=-\delta^{T}_{S}{t}^{R}_{I}+\frac{1}{4}\delta^{R}_{S}t^{T}_{I}\,,\nonumber    
\end{equation}
\begin{equation}
[t^{I}_{J},{t}^{K}_{R}]=+\delta^{K}_{J}{t}^{I}_{R}-\frac{1}{4}\delta^{I}_{J}t^{K}_{R}\,, \qquad [t^{I}_{J},{t}^{R}_{K}]=-\delta^{I}_{K}{t}^{R}_{J}+\frac{1}{4}\delta^{I}_{J}t_{K}^{R}\,\nonumber    .
\end{equation}
In \cite{Fleury:2021ieo}, these $SU(2,2)\times SU(4)$ covariant generators are decomposed as
\begin{equation}
\begin{pmatrix}
 t_{R}^{S} & t_{R}^{I}\\
 t_{R}^{I}& t_{I}^{J} 
\end{pmatrix}=
\begin{pmatrix}
 M_{\alpha}^{\beta}+\frac{\Delta\delta_{\alpha}^{\beta}}{2}&  K_{\alpha}\,^{\dot\beta} &q_{\alpha}^{i}&  q_{\alpha}^{\overline i}\\
  P_{\dot\alpha}\,^{\beta}  &M_{\dot\alpha}\,^{\dot\beta}-\frac{\Delta \delta_{\dot\alpha}^{\dot\beta}}{2}&q_{\dot\alpha}^{i}  &q_{\dot\alpha}^{\overline i}\\
 q^{\alpha}_{i}&   q^{\dot\alpha}_{i}&M_{i}^{j}+\frac{\Delta\delta_{ i}^{ j}}{2}&  P_{ i}\,^{\overline j} \\
  q^{\alpha}_{\overline i} &q^{\dot\alpha}_{\overline i} &K_{\overline i}\,^{ j}  &M_{\overline i}\,^{\overline j}-\frac{\Delta \delta_{\overline i}^{\overline j}}{2}
\end{pmatrix}  \,,
\end{equation}
and the algebra becomes
\begin{equation}
\begin{aligned} 
\label{algebra}
& \qquad \qquad \qquad \qquad \qquad \qquad \qquad   \{q^{R}_{J},q^{I}_{S}\}=   \\
& =\delta^{I}_{J}\left[\delta^{R}_{\alpha}\delta_{S}^{\beta}\left(M^{\alpha}_{\beta}+\frac{1}{2}\Delta\delta^{\alpha}_{\beta}\right)  +\delta^{R}_{\dot\alpha}\delta_{S}^{\dot\beta}\left(M^{\dot\alpha}_{\dot\beta}-\frac{1}{2}\Delta \delta^{\dot\alpha}_{\dot\beta}\right)+\delta^{R}_{\dot\alpha}\delta_{S}^{\beta} K^{\dot\alpha}_{\beta}+\delta^{R}_{\alpha}\delta_{S}^{\dot\beta} P^{\alpha}_{\dot\beta}\right]  \\   
& +\delta^{R}_{S}\left[\delta^{I}_{i}\delta_{J}^{j}\left(M^{i}_{j}+\frac{1}{2}J\delta^{i}_{j}\right)  +\delta^{I}_{\bar i}\delta_{J}^{\bar j}\left(M^{\bar i}_{\bar j}-\frac{1}{2}J \delta^{\bar i}_{\bar j}\right)+\delta^{I}_{\bar i}\delta_{J}^{j} P^{\bar i}_{j}+\delta^{I}_{i}\delta_{J}^{\bar j} K^{i}_{\bar j}\right] \, , 
\end{aligned} 
\end{equation}
and
\begin{equation}
\label{firstlinecommutator}
\begin{aligned}
[\Delta, P^{\alpha}_{\dot\alpha}]= P^{\alpha}_{\dot\alpha}\,,\qquad & [\Delta,K^{\dot\alpha}_{\alpha}]= -K^{\dot\alpha}_{\alpha}\,, \\
 [\Delta,{q}^{R}_{I}]=\frac{1}{2}\delta^{R}_{\alpha}{q}^{\alpha}_{I}-\frac{1}{2}\delta^{R}_{\dot\alpha}{q}^{\dot\alpha}_{I}\,,\qquad & [\Delta,{q}^{I}_{R}]=-\frac{1}{2}\delta^{\alpha}_{R}{q}^{I}_{\alpha}+\frac{1}{2}\delta^{\dot\alpha}_{R}{q}^{I}_{\dot\alpha}\,, \\
[J,P^{\overline i}_{i}]= -P^{\overline i}_{i}\,,\qquad  & [J, K^{i}_{\overline i}]=K^{i}_{\overline i}\,, \\
[J,{q}^{I}_{R}]=\frac{1}{2}\delta^{I}_{i}{q}^{i}_{R}-\frac{1}{2}\delta^{I}_{\overline i}{q}^{\overline i}_{R}\,,\qquad & [J,{q}^{R}_{I}]=-\frac{1}{2}\delta^{i}_{I}{q}^{R}_{i}+\frac{1}{2}\delta^{\overline i}_{I}{q}^{R}_{\overline i}\, .
\end{aligned} 
\end{equation}
\begin{equation}
\begin{aligned}
[M^{\alpha}_{\beta},M^{\gamma}_{\delta}]=\delta^{\gamma}_{\beta}M^{\alpha}_{\delta}-\delta^{\alpha}_{\delta}M^{\gamma}_{\beta}\,,\qquad & [M^{\dot\alpha}_{\dot\beta},M^{\dot\gamma}_{\dot\delta}]=\delta^{\dot\gamma}_{\dot\beta}M^{\dot\alpha}_{\dot\delta}-\delta^{\dot\alpha}_{\dot\delta}M^{\dot\gamma}_{\dot\beta}    \,, \\
[M^{\alpha}_{\beta},{q}^{R}_{I}]=\delta^{R}_{\beta}{q}^{\alpha}_{I}-\frac{1}{2}\delta^{\alpha}_{\beta}{q}^{R}_{I
}\,,  \qquad  & [M^{\alpha}_{\beta},{q}^{I}_{R}]=-\delta^{\alpha}_{R}{q}^{I}_{\beta}+\frac{1}{2}\delta^{\alpha}_{\beta}{q}^{I}_{R}\,, \\
[M^{\dot\alpha}_{\dot\beta},{q}^{R}_{I}]=\delta^{R}_{\dot\beta}{q}^{\dot\alpha}_{I}-\frac{1}{2}\delta^{\dot\alpha}_{\dot\beta}{q}^{R}_{I}  \,,\qquad & [M^{\dot\alpha}_{\dot\beta},{q}^{I}_{R}]=-\delta^{\dot\alpha}_{R}{q}^{I}_{\dot\beta}+\frac{1}{2}\delta^{\dot\alpha}_{\dot\beta}{q}^{I}_{R}\,, \\
[M^{\alpha}_{\beta},P^{\gamma}_{\dot\gamma}]=\delta^{\gamma}_{\beta}P^{\alpha}_{\dot\gamma}-\frac{1}{2}\delta^{\alpha}_{\beta}P^{\gamma}_{\dot\gamma}\,,\qquad   &  [M^{\dot \alpha}_{\dot \beta},P^{\gamma}_{\dot\gamma}]=-\delta^{\dot\alpha}_{\dot\gamma}P^{\gamma}_{\dot\beta}+\frac{1}{2}\delta^{\dot \alpha}_{\dot \beta}P^{\gamma}_{\dot\gamma} \,,
\\
[M^{\dot\alpha}_{\dot\beta},K^{\dot\gamma}_{\gamma}]=\delta^{\dot\gamma}_{\dot\beta}K^{\dot\alpha}_{\gamma}-\frac{1}{2}\delta^{\dot\alpha}_{\dot\beta}K^{\dot\gamma}_{\gamma}\,,\qquad    & [M^{ \alpha}_{ \beta},K^{\dot\gamma}_{\gamma}]=-\delta^{\alpha}_{\gamma}K^{\dot\gamma}_{\beta} +\frac{1}{2}\delta^{ \alpha}_{ \beta}K^{\dot\gamma}_{\gamma}\,,
\end{aligned}
\end{equation}
\begin{equation}
\begin{aligned} 
[M^{i}_{j},M^{k}_{l}]=\delta^{k}_{j}M^{i}_{l}-\delta^{i}_{l}M^{k}_{j}\,,\qquad & [M^{\bar i}_{\bar j},M^{\bar k}_{\bar l}]=\delta^{\bar k}_{\bar j}M^{\bar i}_{\bar l}-\delta^{\bar i}_{\bar l}M^{\bar k}_{\bar j} \, ,    \\
[M^{i}_{j},{q}^{I}_{R}]=\delta^{I}_{j}{q}^{i}_{R}-\frac{1}{2}\delta^{i}_{j}{q}^{I}_{R
}    \,,\qquad &  [M^{i}_{j},{q}^{R}_{I}]=-\delta^{i}_{I}{q}^{R}_{j}+\frac{1}{2}\delta^{i}_{j}{q}^{R}_{I}\,, \\
[M^{\bar i}_{\bar j},{q}^{I}_{R}]=\delta^{I}_{\bar j}{q}^{\bar i}_{R}-\frac{1}{2}\delta^{\bar i}_{\bar j}{q}^{I}_{R} \,,\qquad & [M^{\bar i}_{\bar j},{q}^{R}_{I}]=-\delta^{\bar i}_{I}{q}^{R}_{\bar j}+\frac{1}{2}\delta^{\bar i}_{\bar j}{q}^{R}_{I}\,,   
\\
[M^{i}_{ j},K^{k}_{\bar k}]=\delta^{k}_{ j}K^{ i}_{\bar k}-\frac{1}{2}\delta^{ i}_{ j}K^{k}_{\bar k}\,,\qquad   
& [M^{\bar   i}_{\bar   j},K^{k}_{\bar k}]=-\delta^{\bar  i}_{\bar k}K^{k}_{\bar  j}+\frac{1}{2}\delta^{\bar   i}_{\bar   j}K^{k}_{\bar k} \, , \\
[M^{\bar  i}_{\bar  j},P^{\bar k}_{k}]=\delta^{\bar k}_{\bar  j}P^{\bar  i}_{k}-\frac{1}{2}\delta^{\bar  i}_{\bar  j}P^{\bar k}_{k}\,,\qquad   &  [M^{  i}_{  j},P^{\widehat k}_{k}]=-\delta^{ i}_{k}P^{\widehat k}_{ j} +\frac{1}{2}\delta^{  i}_{  j}P^{\widehat k}_{k} \, . 
\end{aligned} 
\end{equation}
\begin{equation}
\begin{aligned} 
& [K^{\dot\alpha}_{\beta},q^{R}_{I}]=\delta^{R}_{\beta}q^{\dot\alpha}_{I}\,,\qquad [K^{\dot\alpha}_{\beta},q^{I}_{R}]=-\delta^{\dot\alpha}_{R}q^{I}_{\beta}\,, \\
& [P^{\alpha}_{\dot\beta},q^{R}_{I}]=\delta^{R}_{\dot\beta}q^{\alpha}_{I}\,,\qquad [P^{\alpha}_{\dot\beta},q^{I}_{R}]=-\delta^{\alpha}_{R}q^{I}_{\dot\beta}\,,    \\
& [K^{\dot\alpha}_{\alpha},P^{\beta}_{\dot\beta}]  = \delta^{\beta}_{\alpha} M^{\dot\alpha}_{\dot\beta} - \delta^{\dot\alpha}_{\dot\beta}M^{\beta}_{\alpha} - \delta^{\dot\alpha}_{\dot\beta}\delta^{\beta}_{\alpha} \Delta \, ,     
\label{TheKPqalgebra}
\end{aligned}
\end{equation}
\begin{equation}
\begin{aligned} 
& [K^{i}_{\bar j},q^{I}_{R}]=\delta^{I}_{\bar j}q^{i}_{R}\,,\qquad [K^{i}_{\bar j},q^{R}_{I}]=-\delta^{i}_{I}q^{R}_{\bar j}\,,    \\
& [P^{\bar i}_{j},q^{I}_{R}]=\delta^{I}_{j}q^{\bar i}_{R}\,,\qquad [P^{\bar i}_{j},q^{R}_{I}]=-\delta^{\bar i}_{I}q^{R}_{j}\,,  \\  
& [K^{i}_{\bar i},P^{\bar j}_{j}]=\delta^{\bar j}_{\bar i} M^{i}_{j} - \delta^{i}_{j} M^{\bar j}_{\bar i} + \delta^{\bar j}_{\bar i}\delta^{i}_{j} J \, .     
\label{TheKPqalgebraI}
\end{aligned}
\end{equation}

\section{On the original proposal for the zero-mode measure on \texorpdfstring{$AdS_5 \times S^5$}{AdS5xS5}}

\label{sec:previous_proposal}

The original prescription for $\ads$ pure spinor superstring amplitudes did not introduce picture-lowering operators or extensions of the complex to include non-minimal variables \cite{Berkovits:2008ga}. The proposal was based on the observation that, unlike in flat background, one can combine left and right pure spinor variables to form a scalar $(\lambda_L\eta\lambda_R)$ that is BRST closed but not exact. Here, $\eta$ denotes the Ramond-Ramond bi-spinor of the $\ads$ background. Using this observation, it was argued that one could extend the complex by including inverse powers of $(\lambda_L\eta\lambda_R)$, and that this extension would not trivialize the cohomology. 

In flat background, if we allow inverse powers of $(\lambda_L\eta\lambda_R)$ in the minimal complex, since $(\lambda_L\eta\lambda_R)$ is exact, specifically $(\lambda_L\eta\lambda_R) = Q_{\text{flat}} (\lambda_L\eta\theta_R)$, one could write any closed expression $V$ as an exact term via $V = Q_{\text{flat}} \left(\frac{(\lambda_L\eta\theta_R)}{(\lambda_L\eta\lambda_R)}V\right)$, thus trivializing the cohomology. Therefore, in flat background we do not allow inverse powers of $(\lambda_L\eta\lambda_R)$ in the minimal complex, and this presents a complication, as it is not possible to construct expressions with negative ghost number by working solely with polynomials of pure spinor variables. One way to obtain expressions with negative ghost number in the minimal formalism is through picture-lowering operators, which break Lorentz covariance. To restore Lorentz covariance, one needs to extend the complex to become non-minimal \cite{Berkovits:2005bt}. 

Expressions of negative ghost number are important not just for multi-loop scattering amplitudes, where the b-ghosts pairs with the Beltrami differentials, but also for the tree-level amplitudes, because each pure spinor component carries $+1$ ghost number and has a zero mode on the sphere. The integral of these zero modes introduces positive ghost number to the worldsheet measure that must be offset by the introduction of picture-lowering operators, or the extension to the non-minimal complex. For example, for open strings in flat background, the pure spinor integration $\int [D\lambda]$ contributes with $+8$ ghost number
\begin{equation}
\int [D^{11}\lambda] \lambda^{\alpha}\lambda^{\beta}\lambda^{\gamma} =  \int (\epsilon T)^{((\alpha \beta \gamma))}_{[ \rho_1 \ldots \rho_{11}]} 
 \, d \lambda^{\rho_1} \wedge\ldots  \wedge d\lambda^{\rho_{11}}\,,   
\end{equation}
where the tensor $(\epsilon T)^{((\alpha \beta \gamma))}_{[ \rho_1 \ldots \rho_{11}]}$ is the one appearing in \eqref{epsT}. Therefore, to obtain the appropriate $+3$ ghost number of the sphere, one needs to introduce $11$ picture-lowering operators. These $11$ picture-lowering operators also play the role of a regulator to these integrals, as the pure spinor space contains noncompact directions corresponding to the ghost-number scaling of the pure spinor variables. 

The proposal made in \cite{Berkovits:2008ga} was that, because $(\lambda_L\eta\lambda_R)$ is not exact, one could extend the complex to include inverse powers of $(\lambda_L\eta\lambda_R)$, thus bypassing the need for picture-lowering operators or non-minimal variables. Instead of regularizing the pure spinor integration using picture-lowering operators, the proposal was to treat $\lambda_L$ and $\lambda_R\eta$ as complex conjugates and fix $(\lambda_L\eta\lambda_R) = \rho\neq 0$,
thus eliminating the noncompact directions of the pure spinor integration. Since these noncompact directions correspond to the left and right ghost-number symmetries, it was argued that the value of $\rho$ could be absorbed by shifts of the string coupling. The integration over the remaining compact directions of the pure spinor space was defined as
\begin{equation}
\label{eq:316}
\int [D^{10}\lambda_{L}D^{10}\lambda_{R}]\;\lambda_L^{\alpha}\lambda_L^{\beta}\lambda_L^{\gamma}\lambda_R^{\widehat\alpha}\lambda_R^{\widehat\beta}\lambda_R^{\widehat\gamma}=T^{((\alpha\beta\gamma))((\widehat\alpha\widehat\beta\widehat\gamma))}\,,
\end{equation}
where $T^{((\alpha\beta\gamma))((\widehat\alpha\widehat\beta\widehat\gamma))}$ is constructed from products of Ramond-Ramond bi-spinors $\eta^{\alpha\widehat\alpha}\eta^{\beta\widehat\beta}\eta^{\gamma\widehat\gamma}$ followed by symmetrization and gamma-trace removal on $(\alpha\beta\gamma)$ and $(\widehat\alpha\widehat\beta\widehat\gamma)$. Notice that, on the right-hand side of \eqref{eq:316}, the integration over the $10+10$ compact directions of the pure spinor space has already been performed. Consequently, in the amplitude prescription proposed in \cite{Berkovits:2008ga}, the superspace integration and the pure spinor integration are decoupled. Indeed, the mixing between these two integrations comes precisely from the picture-lowering operators, or the non-minimal regulator, which are not used in this proposal.

As it turns out, there is a flaw in the argument proposed in \cite{Berkovits:2008ga}. 
Although it is true that $(\lambda_L\eta\lambda_R)$ is not exact, its square, $(\lambda_L\eta\lambda_R)^2$, is. That is, there is a $W$ such that $QW = (\lambda_L\eta\lambda_R)^2$. Therefore, allowing inverse powers of $(\lambda_L\eta\lambda_R)^2$ in the complex does trivialize the cohomology, since any closed object $V$, satisfying $QV=0$, can be written as $V=Q \left(\frac{W}{(\lambda_L\eta\lambda_R)^2}V\right)$.
In particular, fixing $(\lambda_L\eta\lambda_R) = \rho \neq 0 $ also trivializes the cohomology, since any closed object $V$, satisfying $QV=0$, can be written as an exact $V = Q\left(\frac{W}{\rho^2}V\right)$. Consequently, the pure spinor integration of (\ref{eq:316}) proposed in \cite{Berkovits:2008ga} is inconsistent with the BRST symmetry of the pure spinor formalism. 

To show that $(\lambda_L\eta\lambda_R)^2$ is exact, let us use the following parametrization of the $\ads$ superspace:
\begin{equation}
g=g_{\mathrm b}e^{\psi} e^{\theta},
\qquad g_{\mathrm b}=\begin{pmatrix}v&0\\0&u\end{pmatrix}.
\label{eq:param_new}
\end{equation}
The BRST transformation decomposes as 
\begin{equation}
Q = \delta + \mathcal{R}\,,  
\label{eq: brst_decomposition}
\end{equation}
where $\delta$ defines a complex
\begin{equation}
\delta  \psi = \lambda\,,\qquad \delta \theta = \overline\lambda - \theta \lambda\theta\,,\qquad \delta \lambda = 0\,,\qquad \delta \overline\lambda = \overline\lambda \lambda \theta - \theta\lambda\overline\lambda\,,    
\end{equation}
and $\mathcal{R}$ act as $SU(2,2)\times SU(4)$ generators
\begin{align}
\mathcal{R}\psi = -\omega_v \psi - \psi \omega_u\,,\qquad \mathcal{R}\theta = -\omega_u \theta - \theta \omega_v\,,\\
\mathcal{R}\lambda = -\omega_v \lambda + \lambda \omega_u\,,\qquad \mathcal{R}\overline\lambda = -\omega_u \overline\lambda + \overline\lambda \omega_v\,,
\end{align}
with fermionic parameters
\begin{equation}
\omega_v = v^{-1} Qv  \,,\qquad \omega_u = u^{-1} Q u\,.  
\end{equation}
To obtain the coordinates of \cite{Fleury:2026xpg}, solve $e^{\Psi}e^{\Theta}g_{\mathrm b}=g_{\mathrm b}e^{\psi} e^{\theta}$ which results in the following dressing relations
\begin{equation}
\Lambda=u\lambda v^{-1},\qquad \overline\Lambda=v\overline\lambda u^{-1},\qquad
\Psi=u\psi v^{-1},\qquad \Theta =v\theta u^{-1}.
\label{eq:dressing}
\end{equation}
The variables $\lambda$ and $\overline\lambda$ are related with the usual left and right pure spinor by holomorphic and anti-holomorphic combinations
\begin{equation}
\lambda_L=\lambda+\overline\lambda ^\vee,\qquad \lambda_R=i(\lambda-\overline\lambda ^\vee),
\qquad
\lambda=\frac{\lambda_L-i\lambda_R}{2},\quad
\overline\lambda =\frac{(\lambda_L+i\lambda_R)^\vee}{2},
\label{eq:lr}
\end{equation}
where $m^\vee=-\varepsilon m^{\mathsf T}\varepsilon $ denotes the raising and lowering of indices by the symplectic metrics of $USp(2,2)$ and $USp(4)$. The symplectic structure appears both in the spacetime parametrizations $v\in \frac{SU(2,2)}{USp(2,2)}$, $u\in \frac{SU(4)}{USp(4)}$, and in the pure spinor constraints 
\begin{align}
\lambda \overline\lambda +(\lambda \overline\lambda )^\vee-\tfrac12\tr(\lambda \overline\lambda )\mathbb{I}=0\,,\nonumber\\
\overline\lambda \lambda +(\overline\lambda \lambda )^\vee-\tfrac12\tr(\overline\lambda \lambda )\mathbb{I}=0\,,\nonumber\\
\langle \lambda\lambda ^\vee+\overline\lambda ^\vee \overline\lambda \rangle_{\text{traceless}}=0\,,\nonumber\\
\langle \lambda ^\vee \lambda +\overline\lambda\overline\lambda ^\vee\rangle_{\text{traceless}}=0\,,
\label{eq:ps_constraints}
\end{align}
where $\langle \dots\rangle_{\text{traceless}}$ denotes the traceless part. The scalar built from pure spinors and Ramond-Ramond bi-spinor then takes the form
\begin{equation}
(\lambda_L\eta\lambda_R) = \frac{i}{2}\tr(\lambda_L\lambda_R^\vee)=\frac12\left[\tr(\overline\lambda^\vee \overline\lambda)-\tr(\lambda \lambda^\vee)\right].
\label{eq:vd}
\end{equation}
Our strategy to show that $(\lambda_L\eta\lambda_R)^2$ is exact will be the following. We first use the pure spinor constraints to write $(\lambda_L\eta\lambda_R)^2$ as a polynomial of traces and determinants that do not involve the symplectic metrics $\varepsilon$ (or the $^\vee$) of $USp(2,2)$ and $USp(4)$, thus reducing the BRST transformation of (\ref{eq: brst_decomposition}) to the ones generated by $\delta$ since $\mathcal{R}$ act as $SU(2,2)\times SU(4)$ generators. Combining the pure spinor constraints (\ref{eq:ps_constraints}) with equation (\ref{eq:vd}), one can show that
\bea
(\lambda_L\eta\lambda_R)^2 ={}& \frac{1}{2}\left[\tr(\lambda\lambda^{\vee})\right]^2 -   \tr\left[(\lambda\lambda^{\vee})^2\right]   
+ \frac{1}{2}\left[\tr(\overline \lambda\overline \lambda^{\vee})\right]^2 -   \tr\left[(\overline \lambda\overline \lambda^{\vee})^2\right] \\
&+ 2 \tr\left[(\overline\lambda\lambda)^2\right] - \left[\tr(\overline\lambda\lambda)\right]^2\,,
\eea{eq: dilaton square}
where by the identity relating symplectic structures and determinants
\begin{equation}
4 \det (a) = \frac{1}{2}\left[\tr\left(aa^\vee\right)\right]^2 - \tr\left[\left(aa^\vee\right)^2\right]\,, 
\end{equation}
results in an expression involving only polynomials of traces and determinants
\begin{align}
(\lambda_L\eta\lambda_R)^2 = 4\det(\lambda) + 4\det(\overline\lambda)+ 2 \tr\left[(\overline\lambda\lambda)^2\right] - \left[\tr(\overline\lambda\lambda)\right]^2\,,
\label{eq: llsquare_trace}
\end{align}
and do not involve the symplectic metrics (or the $^\vee$) of $USp(2,2)$ and $USp(4)$. Therefore, it is sufficient to show that the above expression is $\delta$-exact. It turns out that the $\delta$ complex of unconstrained polynomials has trivial cohomology on positive-degree due to the homotopy operator
\begin{equation}
h\psi = 0\,,\qquad h\theta = 0\,,\qquad  h\lambda = \psi\,,\qquad h\overline\lambda =\theta - \theta\psi\theta\,,   
\end{equation}
where $\{h, \delta\} = N$ counts the variables $(\lambda,\chi,\theta,\psi)$, with $\chi = \overline\lambda -\theta\lambda\theta$, and the normalized operator $hN^{-1}$ inverts $\delta$:
\begin{equation}
\{\delta, hN^{-1}\} = 1\,.
\end{equation}
Since we do not need to use the pure spinor constraint to show that the polynomial on the right-hand side of (\ref{eq: llsquare_trace}) is $\delta$ closed, it follows from the contracting homotopy above that it is also $\delta$ exact. Therefore, we conclude that $(\lambda_L\eta\lambda_R)^2$ is $Q$ exact, thus obstructing the alternative proposal of \cite{Berkovits:2008ga} for the pure spinor scattering amplitude in $\ads$.

\section{The Haar supermeasure}
\label{Haarsupermeasure}

In the context of integration on groups, the Haar measure is the unique measure function that is invariant under left and right multiplication. This can be explicitly constructed using the Maurer-Cartan forms. Similarly, in the case of integration on coset spaces $G/H$, there also exists a measure with the property of being invariant under the (left) action of $G$ on the coset space. But this time under the condition that the representation matrix of $H$ acting on directions orthogonal to its algebra has unit determinant.

The invariant measure function for the $G/H$ coset space is given by the determinant of the vielbein $\det (e)$, and we will see that the natural generalization to our super-coset space setting ${\rm{sdet}}(E^A_M)$ (used in \cite{Berkovits:2008ga} as an extra factor in the zero-mode integration) has also the invariant property under $PSU(2,2|4)$.

To see this, consider $(\mathcal Z^M)$ as coordinates and $\mathcal Z'^M = f^M(\mathcal Z)$ as the functional form of the global $PSU(2,2|4)$ transformations acting on those coordinates, then the vielbein will transform as
  \begin{equation}
  E^A_M(\mathcal Z) = \frac{\partial \mathcal Z'^N}{\partial\mathcal Z^M}\ E_N^B(\mathcal Z'(\mathcal Z))\ h_B^A(\mathcal Z)
  \end{equation}
which implies that the combination $[d\mathcal Z]\ {\rm sdet}E(\mathcal Z)$ is $PSU(2,2|4)$-invariant if and only if
\begin{equation}
[d\mathcal Z]\ {\rm sdet}[E](\mathcal Z) = [d\mathcal Z']\ {\rm sdet}[E](\mathcal Z') \qquad \Leftrightarrow \qquad {\rm sdet}[h^A_B] = 1
\end{equation}
This last condition is easily seen to be true since ${\rm sdet}[h^A_B] = \det[h^a_b] / \det[h^\alpha_\beta] \det[h^{\widehat\alpha}_{\widehat\beta}]$ and the determinant of the vector and spinor representations of $SO(1,4)\times SO(5)$ is one.

For the coset representative $g_{\rm{new}}$ given in \eqref{eq: CosetIdeal}, whose variables are $\Psi^S_I, \Theta_S^I$ and $X^{\alpha}_{\dot\alpha}$, $Y^{\bar{i}}_i$, $Z$ and $W$, the explicit expression of the invariant measure is
\begin{equation}
{\rm sdet}(E^A_M) = \exp(4Z) \exp (-4W) \, . 
\end{equation}
The prescription \eqref{eq:presc} 
integrates $d X_{RS} \, d Y_{IJ}$ and the sdet above 
gives precisely the Jacobian of the transformation from the variables of $g_{\rm{new}}$, 
\begin{equation}
dX_{RS} \, dY_{IJ} = \exp(4Z) \exp (-4W) \, 
d X^{\alpha}_{\dot\alpha} \, d Z \, d Y^{\bar{i}}_{i} \, d W \, . 
\end{equation}

\bibliographystyle{JHEP}
\bibliography{main}
\end{document}